\documentclass[10pt]{article}
\usepackage[numbers]{natbib}
\usepackage{times}
\usepackage{graphicx}
\usepackage{fullpage}
\usepackage{amsmath, amssymb, amstext}

\newcommand{\Ep}{\ensuremath{\epsilon_p}}

\newcommand{\Epdot}[1]{\ensuremath{\dot{\epsilon}_{#1}}}

\newcommand{\That}{\ensuremath{\hat{T}}}

\newcommand{\Partial}[2]{\ensuremath{\frac{\partial #1}{\partial #2}}}

\begin{document}

  \title{The Mechanical Threshold Stress model for various tempers
         of AISI 4340 steel}
  \author{Biswajit Banerjee
          \footnote{E-mail: banerjee@eng.utah.edu.
          Phone: (801) 585-5239 Fax: (801) 585-0039}
          \\
          Department of Mechanical Engineering, University of Utah, 
          Salt Lake City, UT 84112, USA}
  \maketitle
  \begin{abstract}
  Numerical simulations of high-strain-rate and high-temperature 
  deformation of pure metals and alloys require realistic plastic
  constitutive models.  Empirical models include the widely used
  Johnson-Cook model and the semi-empirical Steinberg-Cochran-Guinan-Lund 
  model.  Physically based models such as the Zerilli-Armstrong model, 
  the Mechanical Threshold Stress model, and the Preston-Tonks-Wallace 
  model are also coming into wide use.  In this paper, we determine the
  Mechanical Threshold Stress model parameters for various tempers of 
  AISI 4340 steel using experimental data from the open 
  literature.  We also compare stress-strain curves and Taylor impact test
  profiles predicted by the Mechanical Threshold Stress model with those 
  from the Johnson-Cook model for 4340 steel.  Relevant temperature-
  and pressure-dependent shear modulus models, melting temperature models,
  a specific heat model, and an equation of state for 4340 steel are
  discussed and their parameters are presented.
  \end{abstract}

\section{Introduction}\label{sec:intro}
  The present work was motivated by the need to simulate numerically
  the deformation and fragmentation of a heated AISI 4340 steel 
  cylinder loaded by explosive deflagration.  Such simulations
  require a plastic constitutive model that is valid over temperatures
  ranging from 250 K to 1300 K and over strain rates ranging from 
  quasistatic to the order of $10^5$ /s.  The Mechanical Threshold Stress 
  (MTS) model (\citet{Follans88,Kocks01}) is a physically-based model that 
  can be used for the range of temperatures and strain rates of interest
  in these simulations.  In the absence of any MTS models specifically
  for 4340 steels, an existing MTS model for HY-100 steel
  (\citet{Goto00,Goto00a}) was initially explored as a surrogate for 4340
  steel.  However, the HY-100 model failed to produce results that were
  in agreement with experimental stress-strain data for 4340 steel.  This
  paper attempts to redress that issue by providing the MTS parameters for 
  a number of tempers of 4340 steel (classified by their Rockwell C 
  hardness number).  The MTS model is compared with the Johnson-Cook
  (JC) model (\citet{Johnson83,Johnson85}) for 4340 steel and the relative
  advantages and disadvantages of these models are discussed.

  The MTS model requires a temperature and pressure dependent elastic
  shear modulus. We describe a number of shear modulus models and the
  associated melting temperature models.  Conversion of plastic work into
  heat is achieved through a specific heat model that takes the 
  transformation from the bcc ($\alpha$) phase to the fcc ($\gamma$)
  phase into account.  The associated Mie-Gr{\"u}neisen equation of state
  for the pressure is also discussed.

  The organization of this paper is as follows.  For completeness we
  provide brief descriptions of the models used in this paper in Section
  ~\ref{sec:models}.  Parameters for the submodels required by
  the MTS model (for example, the shear modulus model) are determined and 
  validated in Section~\ref{sec:subModel}.  Details of the procedure used
  to determine the MTS model parameters are given in 
  Section~\ref{sec:MTSParam}.  Predictions from the MTS model are 
  compared with those from the Johnson-Cook model in 
  Section~\ref{sec:MTSComp}.  These comparisons include both stress-strain
  curves and Taylor impact tests.  Conclusions and final remarks are 
  presented in Section~\ref{sec:conclude}.

\section{Models}\label{sec:models}
  In this section, we describe the form of the MTS plastic flow 
  stress model and the associated submodels for the specific heat,
  melting temperature, shear modulus, and the equation of state that have
  have used for the computations in this paper.  The submodels are used 
  during the stress update step in elastic-plastic numerical computations 
  at high strain rates and high temperatures.  The submodels
  discussed in this paper are:
  \begin{enumerate}
     \item {\bf Specific Heat}: the Lederman-Salamon-Shacklette model.
     \item {\bf Melting Temperature}: the Steinberg-Cochran-Guinan (SCG) model
           and the Burakovsky-Preston-Silbar (BPS) model.
     \item {\bf Shear Modulus}: the Varshni-Chen-Gray model (referred to as 
           the MTS shear modulus model in this paper), the 
           Steinberg-Cochran-Guinan (SCG) model, and the 
           Nadal-LePoac (NP) model.
     \item {\bf Equation of State}: the Mie-Gr{\"u}neisen model.
  \end{enumerate}
  More details about the models may be found in the cited references.  

  The following notation has been used uniformly in the equations that follow.
  \begin{align*}
    \Epdot{} &=~ \text{Strain rate} \\
    \Ep &=~ \text{Plastic strain} \\
    \mu &=~ \text{Shear modulus} \\
    \rho &=~ \text{Current mass density} \\
    \rho_0 &=~ \text{Initial mass density} \\
    \eta = \rho/\rho_0 &=~ \text{Compression} \\
    \sigma_y &=~ \text{Yield Stress} \\
    b   &=~ \text{Magnitude of the Burgers vector} \\
    k_b &=~ \text{Boltzmann constant} \\
    p   &=~ \text{Pressure (positive in compression)} \\
    C_p &=~ \text{Specific heat at constant pressure} \\
    C_v &=~ \text{Specific heat at constant volume} \\
    T   &=~ \text{Temperature} \\
    T_m &=~ \text{Melting temperature}
  \end{align*}
  Other symbols that appear in the text are identified following the 
  relevant equations.
  
  \subsection{Mechanical Threshold Stress Model}
  The Mechanical Threshold Stress (MTS) model (\citet{Follans88,Goto00a})  
  gives the following form for the flow stress
  \begin{equation} \label{eq:MTSSigmay}
    \sigma_y(\Ep,\Epdot{},T) = 
      \sigma_a + (S_i \sigma_i + S_e \sigma_e)\frac{\mu(p,T)}{\mu_0} 
  \end{equation}
  where
  $\sigma_a$ is the athermal component of mechanical threshold stress,
  $\sigma_i$ is the intrinsic component of the flow stress due to barriers 
  to thermally activated dislocation motion, $\sigma_e$ is the strain
  hardening component of the flow stress, $(S_i, S_e)$ are strain-rate and
  temperature dependent scaling factors, and $\mu_0$ is the shear modulus 
  at 0 K and ambient pressure.  The scaling factors $S_i$ and $S_e$
  have the modified Arrhenius form
  \begin{align}
    S_i & = \left[1 - \left(\frac{k_b~T}{g_{0i}b^3\mu(p,T)}
    \ln\frac{\Epdot{0i}}{\Epdot{}}\right)^{1/q_i}
    \right]^{1/p_i} \label{eq:MTSSi}\\
    S_e & = \left[1 - \left(\frac{k_b~T}{g_{0e}b^3\mu(p,T)}
    \ln\frac{\Epdot{0e}}{\Epdot{}}\right)^{1/q_e}
    \right]^{1/p_e} \label{eq:MTSSe}
  \end{align}
  where ($g_{0i}, g_{0e}$) are normalized activation energies, 
  ($\Epdot{0i}, \Epdot{0e}$) are constant reference strain rates, and
  ($q_i, p_i, q_e, p_e$) are constants.  The strain hardening component
  of the mechanical threshold stress ($\sigma_e$) is given by a
  modified Voce law
  \begin{equation}
    \frac{d\sigma_e}{d\Ep} = \theta(\sigma_e)
  \end{equation}
  where
  \begin{align}
    \theta(\sigma_e) & = 
       \theta_0 [ 1 - F(\sigma_e)] + \theta_{1} F(\sigma_e) 
    \label{eq:theta}\\
    \theta_0 & = a_{00} + a_{10} \ln \Epdot{} + a_{20} \sqrt{\Epdot{}} +
                 a_{30} T \label{eq:theta_0}\\
    \theta_1 & = a_{01} + a_{11} \ln \Epdot{} + a_{21} \sqrt{\Epdot{}} +
                 a_{31} T \\
    F(\sigma_e) & = 
      \cfrac{\tanh\left(\alpha \cfrac{\sigma_e}{\sigma_{es}}\right)}
      {\tanh(\alpha)}\\
    \ln(\cfrac{\sigma_{es}}{\sigma_{0es}}) & =
    \left(\frac{k_b T}{g_{0es} b^3 \mu(p,T)}\right)
    \ln\left(\cfrac{\Epdot{}}{\Epdot{0es}}\right) \label{eq:Sigma0esG0es}
  \end{align}
  and $\theta_0$ is the strain hardening rate due to dislocation accumulation,
  $\theta_{1}$ is a saturation hardening rate (usually zero),
  ($a_{0j}, a_{1j}, a_{2j}, a_{3j}, \alpha$) are constants ($j=0,1$),
  $\sigma_{es}$ is the saturation stress at zero strain hardening rate, 
  $\sigma_{0es}$ is the saturation threshold stress for deformation at 0 K,
  $g_{0es}$ is the associated normalized activation energy, and 
  $\Epdot{0es}$ is the reference maximum strain rate.  Note
  that the maximum strain rate for which the model is valid is usually 
  limited to approximately $10^7$/s.

  \subsection{Adiabatic Heating and Specific Heat Model}
  A part of the plastic work done is converted into heat and used to 
  update the temperature.  The increase in temperature ($\Delta T$) due to 
  an increment in plastic strain ($\Delta\epsilon_p$) is given by the equation
  \begin{equation}
    \Delta T = \cfrac{\chi\sigma_y}{\rho C_p} \Delta \epsilon_p
  \end{equation}
  where $\chi$ is the Taylor-Quinney coefficient, and $C_p$ is the specific
  heat.  The value of the Taylor-Quinney coefficient is taken to be 0.9
  in all our simulations (see \citet{Ravi01} for more details on how
  $\chi$ varies with strain and strain rate).

  The relation for the dependence of $C_p$ upon temperature that is 
  used in this paper has the form (\citet{Lederman74})
  \begin{align}
    C_p & = \begin{cases}
            A_1 + B_1~t + C_1~|t|^{-\alpha} & \text{if}~~ T < Tc \\
            A_2 + B_2~t + C_2~t^{-\alpha^{'}} & \text{if}~~ T > Tc 
          \end{cases} \label{eq:CpSteel}\\
    t & = \cfrac{T}{T_c} - 1 
  \end{align}
  where $T_c$ is the critical temperature at which the phase transformation
  from the $\alpha$ to the $\gamma$ phase takes place, and 
  $A_1, A_2, B_1, B_2, \alpha, \alpha^{'}$ are constants.

  \subsection{Melting Temperature Models}
  \subsubsection{Steinberg-Cochran-Guinan Model}
  The Steinberg-Cochran-Guinan (SCG) melting temperature model 
  (\citet{Steinberg80}) is based on a modified Lindemann law and has the form
  \begin{equation} \label{eq:TmSCG}
    T_m(\rho) = T_{m0} \exp\left[2a\left(1-\frac{1}{\eta}\right)\right]
              \eta^{2(\Gamma_0-a-1/3)}; 
  \end{equation}
  where $T_{m0}$ is the melt temperature at $\eta = 1$, $a$ is the 
  coefficient of a first order volume correction to the Gr{\"u}neisen 
  gamma ($\Gamma_0$).  

  \subsubsection{Burakovsky-Preston-Silbar Model}
  The Burakovsky-Preston-Silbar (BPS) model is based on dislocation-mediated
  phase transitions (\citet{Burakovsky00}).  The BPS model has the form
  \begin{align}
    T_m(p) & = T_m(0)
      \left[\cfrac{1}{\zeta} + 
            \cfrac{1}{\zeta^{4/3}}~\cfrac{\mu_0^{'}}{\mu_0}~p\right]~; 
    \quad
    \zeta = \left(1 + \cfrac{K_0^{'}}{K_0}~p\right)^{1/K_0^{'}} 
    \label{eq:TmBPS}\\
    T_m(0) & = \cfrac{\kappa\lambda\mu_0~v_{WS}}{8\pi\ln(z-1)~k_b}
               \ln\left(\cfrac{\alpha^2}{4~b^2\rho_c(T_m)}\right)~;
    \quad \lambda = b^3/v_{WS}
  \end{align}
  where $\zeta$ is the compression, $\mu_0$ is the shear modulus at 
  room temperature and zero pressure, $\mu_0^{'} = \partial\mu/\partial p$ 
  is the pressure derivative of the shear modulus at zero pressure, 
  $K_0$ is the bulk modulus at room temperature and zero pressure, 
  $K_0^{'} = \partial K/\partial p$ is the pressure derivative of the 
  bulk modulus at zero pressure, $\kappa$ is a constant, 
  $v_{WS}$ is the Wigner-Seitz volume, $z$ is the crystal coordination 
  number, $\alpha$ is a constant, and $\rho_c(T_m)$ is the critical density 
  of dislocations.  Note that $\zeta$ in the BPS model is derived from 
  the Murnaghan equation of state with pressure as an input and may be 
  different from $\eta$ in numerical computations.

  \subsection{Shear Modulus Models}
  \subsubsection{MTS Shear Modulus Model}
  The Varshni-Chen-Gray shear modulus model has been used in conjunction with
  the MTS plasticity models by \citet{Chen96} and \citet{Goto00a}. Hence, we 
  refer to this model as the MTS shear modulus model.  The MTS shear
  modulus model is of the form (\citet{Varshni70,Chen96}) 
  \begin{equation} \label{eq:MTSShear}
    \mu(T) = \mu_0 - \frac{D}{exp(T_0/T) - 1}
  \end{equation}
  where $\mu_0$ is the shear modulus at 0 K, and $D, T_0$ are material
  constants.  There is no pressure dependence of the shear modulus in the
  MTS shear modulus model.

  \subsubsection{Steinberg-Cochran-Guinan Model}
  The Steinberg-Guinan (SCG) shear modulus model 
  (\citet{Steinberg80,Zocher00}) is pressure dependent and has the form
  \begin{equation} \label{eq:SCGShear}
    \mu(p,T) = \mu_0 + \Partial{\mu}{p} \frac{p}{\eta^{1/3}} +
         \Partial{\mu}{T}(T - 300) ; \quad
    \eta = \rho/\rho_0
  \end{equation}
  where, $\mu_0$ is the shear modulus at the reference state($T$ = 300 K, 
  $p$ = 0, $\eta$ = 1).  When the temperature is above $T_m$, the shear 
  modulus is instantaneously set to zero in this model.

  \subsubsection{Nadal-Le Poac Model}\label{sec:NPShear}
  A modified version of the SCG model has been developed by 
  \citet{Nadal03} that attempts to capture the sudden drop in the
  shear modulus close to the melting temperature in a smooth manner.
  The Nadal-LePoac (NP) shear modulus model has the form
  \begin{equation} \label{eq:NPShear}
    \mu(p,T) = \frac{1}{\mathcal{J}(\That)}
      \left[
        \left(\mu_0 + \Partial{\mu}{p} \cfrac{p}{\eta^{1/3}} \right)
        (1 - \That) + \frac{\rho}{Cm}~k_b~T\right]; \quad
    C := \cfrac{(6\pi^2)^{2/3}}{3} f^2
  \end{equation}
  where
  \begin{equation}
    \mathcal{J}(\That) := 1 + \exp\left[-\cfrac{1+1/\zeta}
        {1+\zeta/(1-\That)}\right] \quad
       \text{for} \quad \That:=\frac{T}{T_m}\in[0,1+\zeta],
  \end{equation}
  $\mu_0$ is the shear modulus at 0 K and ambient pressure, $\zeta$ is
  a material parameter, $m$ is the atomic mass, and $f$ is the 
  Lindemann constant.

  \subsection{Mie-Gr{\"u}neisen Equation of State}
  The hydrostatic pressure ($p$) is calculated using a temperature-corrected 
  Mie-Gr{\"u}neisen equation of state of the form (\citet{Zocher00},
  see also \citet{Wilkins99}, p. 61)
  \begin{equation} \label{eq:EOSMG}
   p = \frac{\rho_0 C_0^2 (\eta -1)
              \left[\eta - \frac{\Gamma_0}{2}(\eta-1)\right]}
             {\left[\eta - S_{\alpha}(\eta-1)\right]^2} + \Gamma_0 E
  \end{equation}
  where $C_0$ is the bulk speed of sound, 
  $\Gamma_0$ is the Gr{\"u}neisen's gamma at the reference state, 
  $S_{\alpha} = dU_s/dU_p$ is a linear Hugoniot slope coefficient, 
  $U_s$ is the shock wave velocity, $U_p$ is the particle velocity, and
  $E$ is the internal energy per unit reference specific volume.  
  The internal energy is computed using
  \begin{equation}
    E = \frac{1}{V_0} \int C_v dT \approx \frac{C_v (T-T_0)}{V_0}
  \end{equation}
  where $V_0 = 1/\rho_0$ is the reference specific volume at the
  reference temperature $T_0$.

\section{Submodel Parameters and Validation}\label{sec:subModel}
  The accuracy of the yield stress predicted by the MTS model depends on
  the accuracy of the shear modulus, melting temperature, equation of state, 
  and specific heat models.  The following discussion shows why we
  have chosen to use a temperature-dependent specific heat model, the BPS 
  melting temperature model, the NP shear modulus model, and the 
  Mie-Gr{\"u}neisen equation of state model.  The relevant parameters of
  these models are determined and the models are validated against 
  experimental data.
  
  \subsection{Specific Heat Model for 4340 Steel}
\begin{figure}[t]
    \centering
    \scalebox{0.40}{\includegraphics{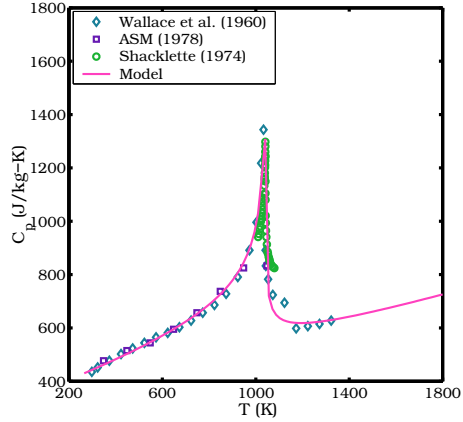}}
    \caption{Comparison of experimental data and model prediction of 
             specific heat for 4340 steel as a function of temperature.}
    \label{fig:CpSteel}
  \end{figure}

  The parameters for the specific heat model (equation~\ref{eq:CpSteel})
  were fit with a least squares technique using experimental data for 
  iron (\citet{Wallace60,Shacklette74}) and AISI 3040 steel (\cite{ASM78}).  
  The variation of specific heat with temperature predicted by the model 
  is compared with experimental data in Figure~\ref{fig:CpSteel}.  The 
  transition from the bcc $\alpha$ phase to the fcc $\gamma$ phase is 
  clearly visible in the figure.  The constants used in the calculations
  are shown in Table~\ref{tab:CpSteel}.  If we use a constant (room
  temperature) specific heat for 4340 steel, there will be an unrealistic 
  increase in temperature close to the phase transition which can cause
  premature melting in numerical simulations (and the associated numerical
  problems).  We have chosen to use the temperature-dependent specific heat
  model to avoid such issues.
  \begin{table}[t]
    \centering
    \caption{Constants used in specific heat model for 4340 steel.}
    \begin{tabular}{ccccccccc}
       \hline
       $T_c$ & $A_1$ & $B_1$ & $C_1$ & $\alpha$ & 
               $A_2$ & $B_2$ & $C_2$ & $\alpha^{'}$ \\
       (K) & (J/kg-K) & (J/kg-K) & (J/kg-K) &  & 
             (J/kg-K) & (J/kg-K) & (J/kg-K) &  \\
       \hline
       1040 & 190.14 & -273.75 & 418.30 & 0.20 & 
              465.21 &  267.52 &  58.16 & 0.35 \\
       \hline
    \end{tabular}
    \label{tab:CpSteel}
  \end{table}

  \subsection{Melting Temperature Model for 4340 Steel}
\begin{figure}[t]
    \centering
    \scalebox{0.45}{\includegraphics{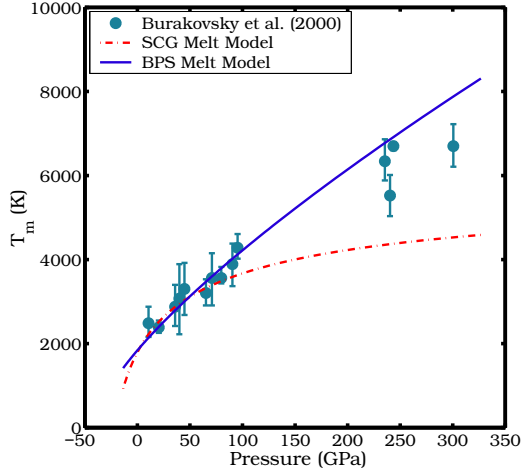}}
    \caption{Comparison of experimental data and model predictions of 
             melting temperature for 4340 steel as a function of pressure.}
    \label{fig:TmSteel}
  \end{figure}

  For the sake of simplicity, we do not consider a phase change in the 
  melting temperature model and assume that the iron crystals remain bcc 
  at all temperatures and pressures.  We also assume that iron has the
  same melting temperature as 4340 steel.  In Figure~\ref{fig:TmSteel} we
  have plotted experimental data (\citet{Burakovsky00,Williams87,Yoo93})
  for the melting temperature of iron at various pressures.  Melting
  curves predicted by the SCG model (Equation \ref{eq:TmSCG}) and the
  BPS model (Equation \ref{eq:TmBPS}) are shown as smooth curves on
  the figure.  The BPS model performs better at high pressures, but 
  both models are within experimental error below 100 GPa.  We have chosen
  to use the BPS melting temperature model because of its larger range
  of applicability.

  The parameters used for the models are shown in Table~\ref{tab:TmSteel}.  
  An initial density ($\rho_0$) of 7830 kg/m$^3$ has been used in the 
  model calculations.
  \begin{table}[t]
    \centering
    \caption{Parameters used in melting temperature models for 4340 steel.
             The SCG model parameters are from \citet{Gust82}.
             The bulk and shear moduli and their derivatives have been 
             obtained from \citet{Guinan74}.  The parameters for the BPS 
             model at zero pressure have been obtained from 
             \citet{Burakovsky00a,Burakovsky00b}, and the lattice constant 
             ($a$) is from \citet{Jansen84}.}  
    \begin{tabular}{ccccccccccc}
       \hline
       \hline
       \multicolumn{11}{l}{Steinberg-Cochran-Guinan (SCG) model} \\
       \hline
       $T_{m0} (K) $ & $\Gamma_0$ & $a$ \\
       \hline
       1793     & 1.67       & 1.67 \\
       \hline
       \hline
       \multicolumn{11}{l}{Burakovsky-Preston-Silbar (BPS) model} \\
       \hline
       $K_0$ (GPa) & $K_0^{'}$ & $\mu_0$ (GPa) & $\mu_0^{'}$ &
       $\kappa$ & $z$ & $b^2\rho_c(T_m)$ & $\alpha$ & $\lambda$ & 
       $v_{WS}$ ($\AA^3$) & $a$ ($\AA$) \\
       \hline
       166   & 5.29      & 81.9    & 1.8         &
       1        & 8   & 0.78             & 2.9      & 1.30      &
       $a^3/2$  & 2.865 \\ 
       \hline
    \end{tabular}
    \label{tab:TmSteel}
  \end{table}
  
  \subsection{Shear Modulus Models for 4340 Steel}\label{sec:Shear}
\begin{figure}[p]
    \centering
    \scalebox{0.45}{\includegraphics{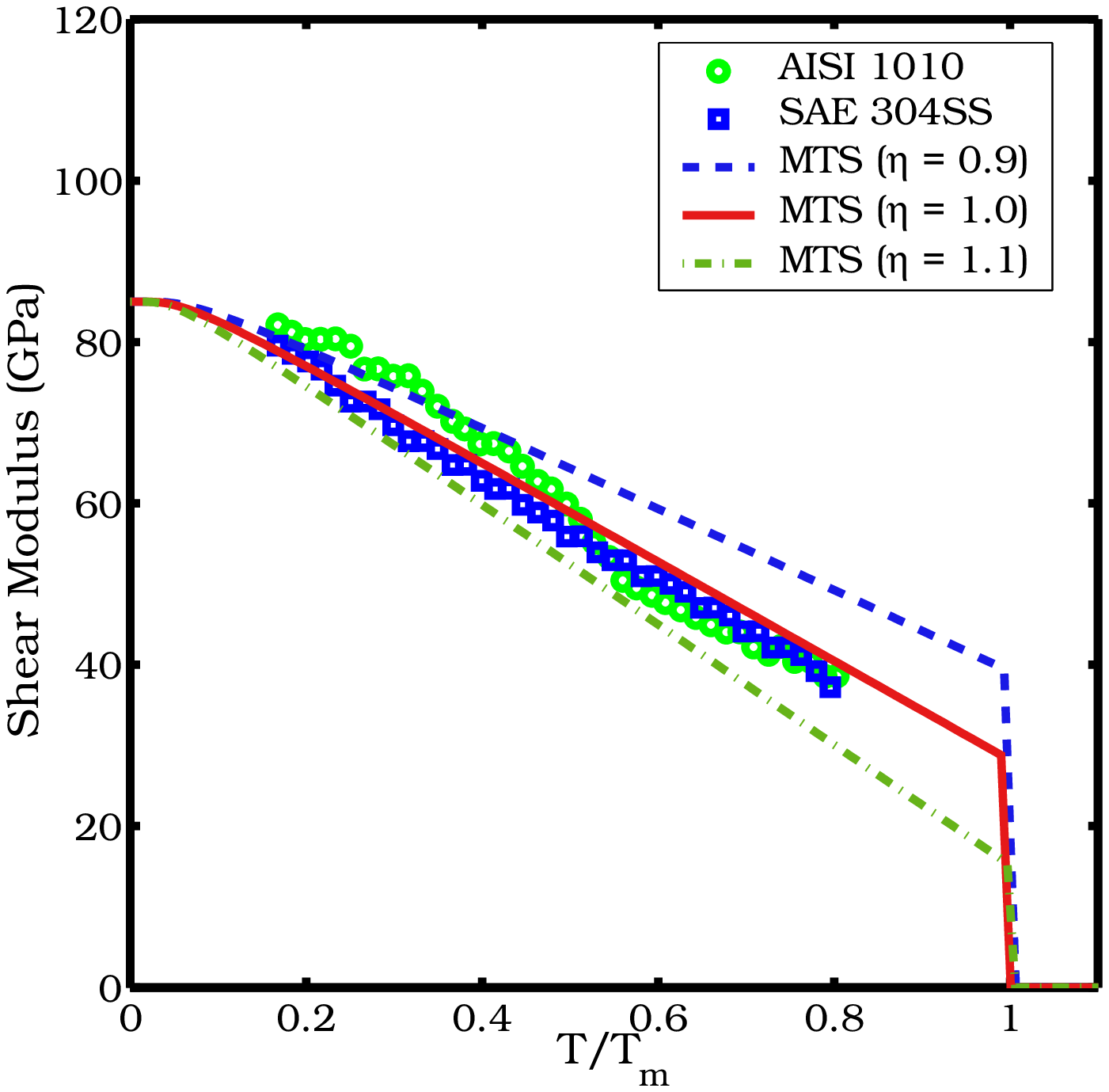}
                    \includegraphics{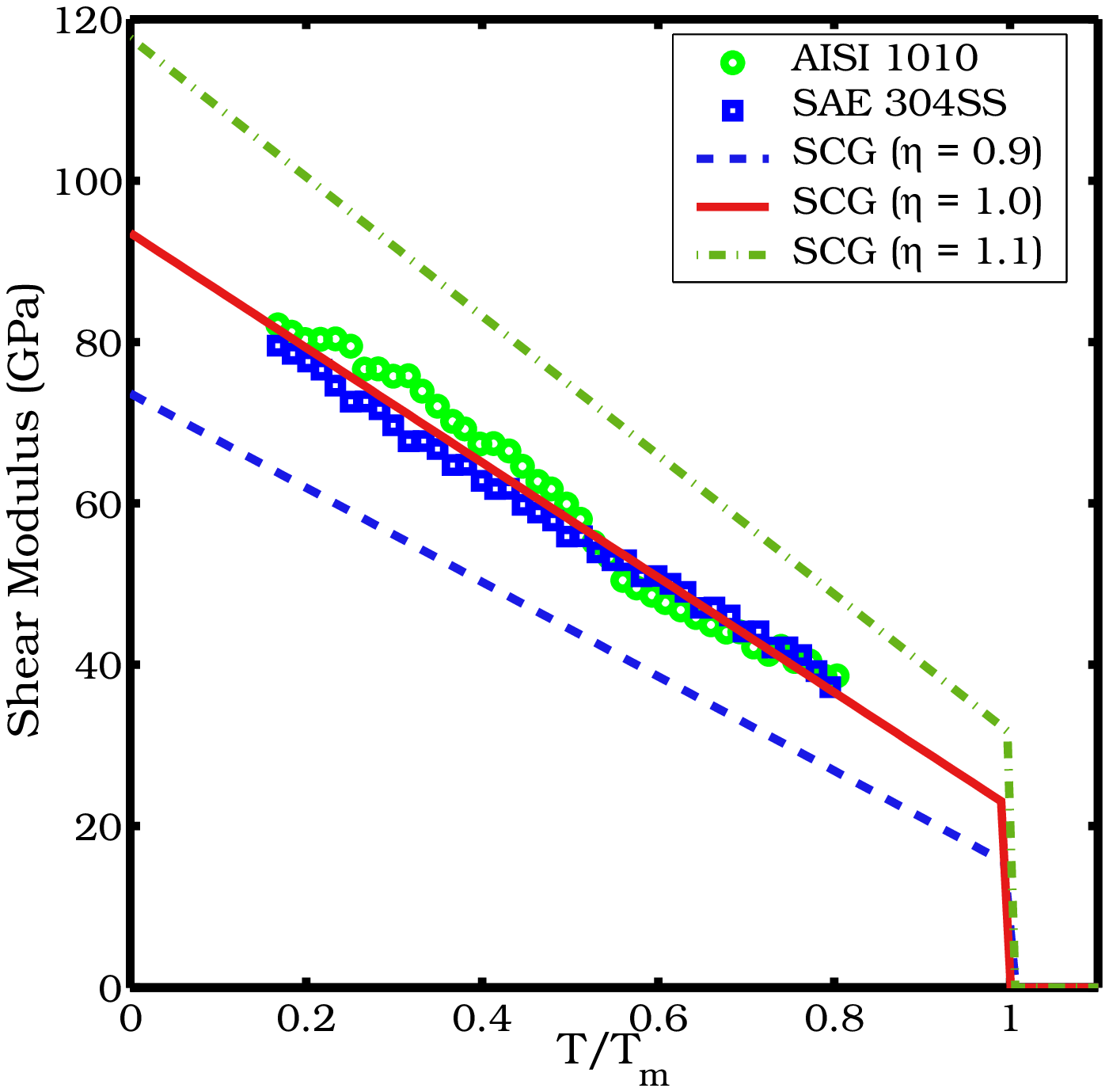}}\\
    (a) MTS Shear Model \hspace{2in} 
    (b) SCG Shear Model \\
    \vspace{12pt}
    \scalebox{0.45}{\includegraphics{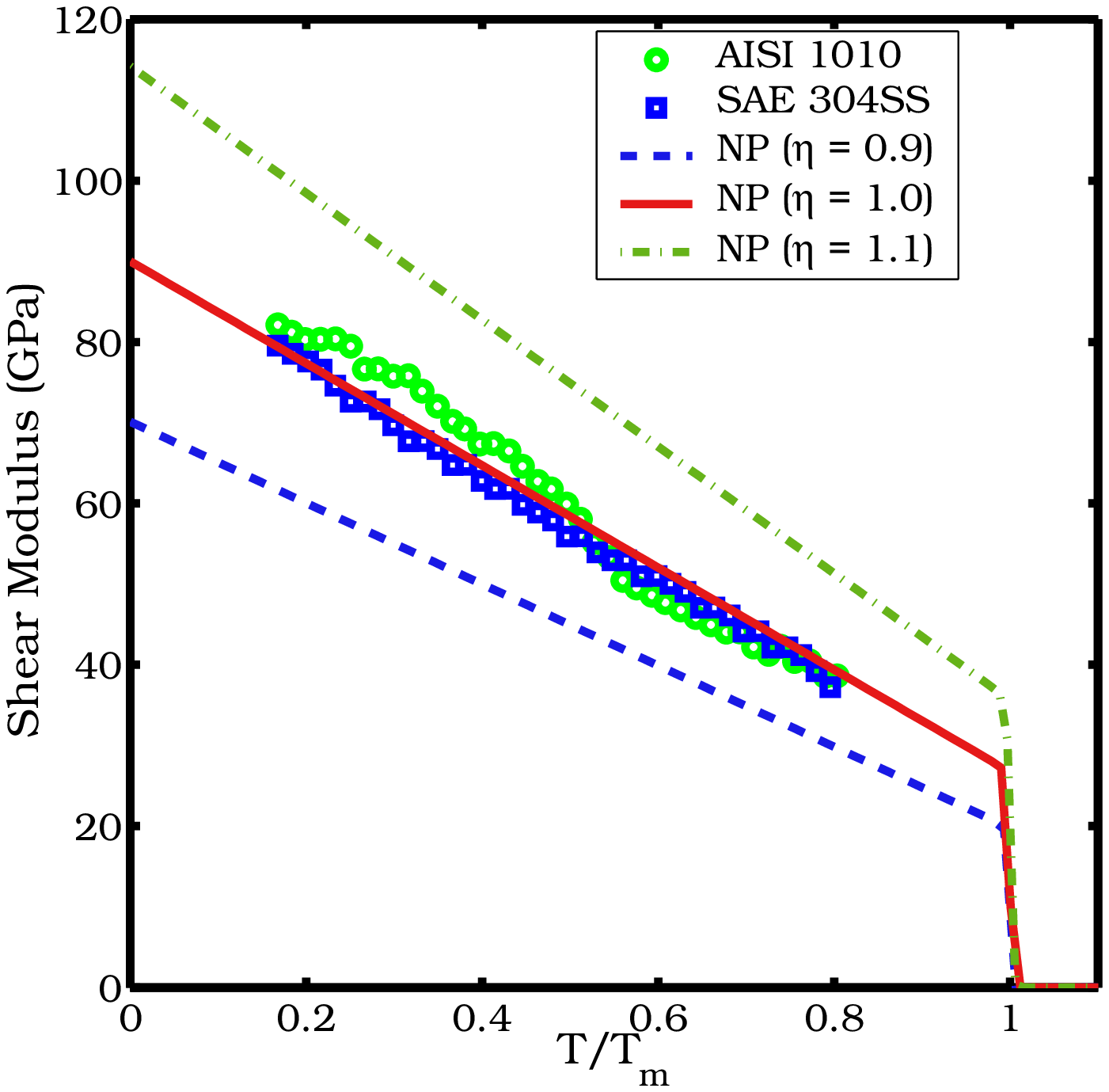}} \\
    (c) NP Shear Model \\
    \caption{Comparison of experimental data with model predictions of 
             shear modulus for 4340 steel.  The experimental data are for
             AISI 1010 steel and SAE 304 stainless steel 
             (\citet{Fukuhara93}) at standard pressure.}
    \label{fig:ModelShearSt}
  \end{figure}

  Figures~\ref{fig:ModelShearSt}(a), (b), and (c) show shear moduli predicted
  by the MTS shear modulus model, the SCG shear modulus model, and the NP
  shear modulus model, respectively.  Three values of compression 
  ($\eta$ = 0.9, 1.0, 1.1) are considered for each model.  The 
  pressure-dependent melting temperature has been determined using the BPS 
  model in each case.  The initial density is taken to be 7830 kg/m$^3$.
  The model predictions are compared with experimental data for AISI 1010 
  steel and SAE 304 stainless steel.  As the figure shows, both steels behave 
  quite similarly as far as their shear moduli are concerned.  We assume that 
  4340 steel also shows a similar dependence of shear modulus on temperature.

  The MTS model does not incorporate any pressure dependence of the shear 
  modulus.  The pressure dependence observed in 
  Figure~\ref{fig:ModelShearSt}(a) is due to the pressure dependence of
  $T_m$.  Both the SCG and NP shear modulus models are pressure dependent 
  and provide a good fit to the data.  Though the SCG model is computationally
  more efficient than and as accurate as the NP model, we have chosen to
  the NP shear modulus model for subsequent calculations for 4340 steel 
  because of its smooth transition to zero shear modulus at melt.

  The parameters used in the shear modulus models are shown in 
  Table~\ref{tab:ModelShearSt}.  The parameters for the MTS model have been
  obtained from a least squares fit to the data at a compression of 1.
  The values of $\mu_0$ and $\partial\mu/\partial p$ for the SCG model
  are from \citet{Guinan74}.  The derivative with respect to temperature has 
  been chosen so as to fit the data at a compression of 1.  The NP shear
  model parameters $\mu_0$ and $C$ have also been chosen to fit the data.
  A value of 0.57 for $C$ is suggested by \citet{Nadal03}.  However, that
  value leads to a higher value of $\mu$ at high temperatures than suggested
  by the experimental data.
  \begin{table}[t]
    \centering
    \caption{Parameters used in shear modulus models for 4340 steel.}
    \begin{tabular}{ccccc}
       \hline
       \hline
       \multicolumn{5}{l}{MTS shear modulus model} \\
       \hline
       $\mu_0$ (GPa) & $D$ (GPa) & $T_0$ (K) \\
       \hline
       85.0          & 10.0      & 298       \\
       \hline
       \hline
       \multicolumn{5}{l}{SCG shear modulus model} \\
       \hline
       $\mu_0$ (GPa) & $\partial\mu/\partial p$ 
                     & $\partial\mu/\partial T$ (GPa/K) \\
       \hline
       81.9          & 1.8                & 0.0387 \\
       \hline
       \hline
       \multicolumn{5}{l}{NP shear modulus model} \\
       \hline
       $\mu_0$ (GPa) & $\partial\mu/\partial p$ & $\zeta$ & $C$ & $m$ (amu) \\
       \hline
       90.0          & 1.8               & 0.04    & 0.080 & 55.947 \\
       \hline
    \end{tabular}
    \label{tab:ModelShearSt}
  \end{table}

  \subsection{Equation of State for 4340 Steel}
  The pressure in the steel is calculated using the Mie-Gr{\"u}neisen 
  equation of state (equation ~\ref{eq:EOSMG}) assuming a linear Hugoniot
  relation.  The Gr{\"u}neisen gamma ($\Gamma_0$) is assumed to be a 
  constant over the regime of interest. The specific heat at 
  constant volume is assumed to be the same as that at constant pressure
  and is calculated using equation (\ref{eq:CpSteel}).

\begin{figure}[t]
    \centering
    \scalebox{0.40}{\includegraphics{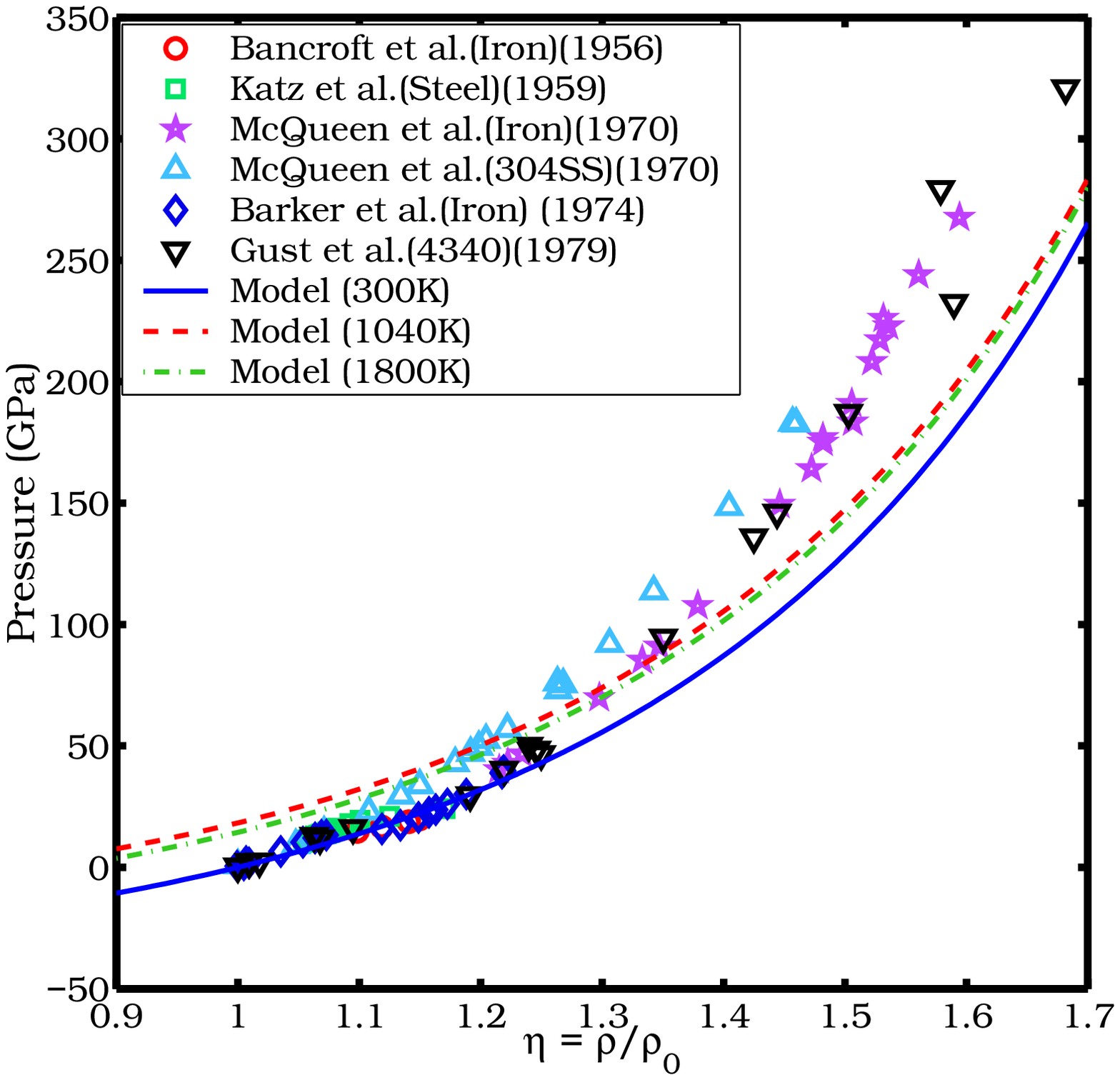}}
    \caption{Comparison of experimental data with model predictions of 
             equation of state for 4340 steel.}
    \label{fig:EOSMGSt}
  \end{figure}

  Figure~\ref{fig:EOSMGSt} compares model predictions with experimental data
  for iron (\citet{Bancroft56,McQueen70,Barker74}), mild steel
  (\citet{Katz59}), 300 series stainless steels (\citet{McQueen70}), and
  for AISI 4340 steel (\citet{Gust79}).  The high pressure experimental
  data are not along isotherms and show the temperature increase due to
  compression.  The equation of state provides a reasonable match to the
  experimental data at compressions below 1.2 which is reasonable for
  the simulations of interest in this paper.  Improved equations of state
  should be used for overdriven shocks.
  
  In the model calculations, the bulk speed of sound ($C_0$) is 3935 m/s
  and the linear Hugoniot slope coefficient ($S_{\alpha}$) is 1.578.  Both
  parameters are for iron and have been obtained from \citet{Brown00}.
  The Gr{\"u}neisen gamma value ($\Gamma_0$ = 1.69) has been interpolated 
  from values given by \citet{Gust79}.  An initial temperature ($T_0$) 
  of 300 K and an initial density of 7830 kg/m$^3$ have been used in the model
  calculations. 
  
\section{Determination of MTS Model Parameters} \label{sec:MTSParam}
  The yield strength of high-strength low-alloy (HSLA) steels such as
  4340 steel can vary dramatically depending on the heat treatment that it 
  has undergone.  This is due to the presence of bcc ferrite-bainite phases
  along with the dominant bcc martensite phase at room temperature.  At 
  higher temperatures (below the $\alpha$-$\gamma$ transition) the phases
  partially transform into the fcc austenite and much of the effect of
  heat treatment is expected to be lost.  Beyond the transition temperature,
  the alloy is mostly the fcc $\gamma$ phase that is expected to behave
  differently than the lower temperature phases.  Hence, purely empirical 
  plasticity models have to be recalibrated for different levels of 
  hardness of 4340 steel and for different ranges of temperature.  

  In the absence of relevant microstructural models for the various tempers
  of 4340 steel, we assume that there is a direct correlation between the 
  Rockwell C hardness of the alloy steel and the yield stress (see the  
  ASM \citet{ASM78}).  We determine the MTS parameters for four
  tempers of 4340 steel.  Empirical relationships are then determined 
  that can be used to calculate the parameters of intermediate tempers of 
  4340 steel via interpolation.

  The experimental data used to determine the MTS model parameters are
  from the sources shown in Table~\ref{tab:4340ExptSource}.  All the 
  data are for materials that have been oil quenched after austenitization.
  More details can be found in the cited references.  The 4340 VAR 
  (vacuum arc remelted) steel has a higher fracture toughness than the
  standard 4340 steel.  However, both steels have similar yield behavior
  (\citet{ASMH96}).
  \begin{table}[t]
    \centering
    \caption{Sources of experimental data for 4340 steel.}
    \begin{tabular}{ccccccccl}
       \hline
       Material & Hardness & Normalize & Austenitize & Tempering &Reference \\
                &          & Temp. (C) & Temp. (C)   & Temp. (C) &          \\
       \hline
       4340 Steel     & $R_c$ 30   &     &     &     & \citet{Johnson85} \\
       4340 Steel     & $R_c$ 38   & 900 & 870 & 557 & \citet{Larson61} \\
       4340 Steel     & $R_c$ 38   &     & 850 & 550 & \citet{Lee97} \\
       4340 VAR Steel & $R_c$ 45   & 900 & 845 & 425 & \citet{Chi89} \\
       4340 VAR Steel & $R_c$ 49   & 900 & 845 & 350 & \citet{Chi89} \\
       \hline
    \end{tabular}
    \label{tab:4340ExptSource}
  \end{table}

  The experimental data are either in the form of true stress versus
  true strain or shear stress versus average shear strain.  These curves
  were digitized manually with care and corrected for distortion. The
  error in digitization was around 1\% on average.  The shear
  stress-strain curves were converted into an effective tensile stress-strain
  curves assuming von Mises plasticity (see \citet{Goto00}).
  The elastic portion of the strain was then subtracted from the total
  strain to get true stress versus plastic strain curves.  The Young's
  modulus was assumed to be 213 MPa. 

  \subsection{Determination of $\sigma_a$}
  The first step in the determination of the parameters for the MTS 
  models is the estimation of the athermal component of the yield stress
  ($\sigma_a$).  This parameter is dependent on the Hall-Petch effect 
  and hence on the characteristic martensitic packet size.  The packet 
  size will vary for various tempers of steel and will depend on the 
  size of the austenite crystals after the $\alpha$-$\gamma$ phase 
  transition.  Since we do not have unambiguous grain sizes and other
  information needed to determine $\sigma_a$, we assume that this constant 
  is independent of temper and has a value of 50 MPa based on the value used 
  for HY-100 steel (\citet{Goto00}).  We have observed that a value of
  150 MPa leads to a better fit to the modified Arrhenius equation for
  $\sigma_i$ and $g_{0i}$ for the $R_c$ 30 temper.  However, this value is
  quite high and probably unphysical because of the relatively large grain 
  size at this temper.

  \subsection{Determination of $\sigma_i$ and $g_{0i}$}
  From equation (\ref{eq:MTSSigmay}), it can be seen that $\sigma_i$ can
  be found if $\sigma_y$ and $\sigma_a$ are known and $\sigma_e$ is zero.
  Assuming that $\sigma_e$ is zero when the plastic strain is zero,
  and using equation (\ref{eq:MTSSi}), we get the relation
  \begin{equation} \label{eq:FisherSigi}
    \left(\cfrac{\sigma_y-\sigma_a}{\mu}\right)^{p_i} =
    \left(\cfrac{\sigma_i}{\mu_0}\right)^{p_i} 
    \left[
      1 - \left(\cfrac{1}{g_{0i}}\right)^{1/q_i}
          \left[\cfrac{k_b T}{\mu b^3}
                \ln\left(\cfrac{\Epdot{0i}}{\Epdot{}}\right)
          \right]^{1/q_i}
    \right]
  \end{equation}
  Modified Arrhenius (Fisher) plots based on equation (\ref{eq:FisherSigi})
  are used to determine the normalized activation energy ($g_{0i}$) and 
  the intrinsic thermally activated portion of the yield stress ($\sigma_i$).
  The parameters $p_i$ and $q_i$ for iron and steels (based on the 
  effect of carbon solute atoms on thermally activated dislocation motion) 
  have been suggested to be 0.5 and 1.5, respectively 
  (\citet{Kocks75,Goto00}).  Alternative values can be obtained depending
  on the assumed shape of the activation energy profile or the obstacle 
  force-distance profile (\citet{Cottrell49,Caillard03}).  

  We have observed that the values suggested for HY-100 give us a 
  value of the normalized activation energy $g_{0i}$ for $R_c$ = 30 that 
  is around 40, which is not physical.  Instead, we have assumed a 
  rectangular force-distance profile which gives us values of 
  $p_i$ = 2/3 and $q_i = 1$ and reasonable values of $g_{0i}$.  We have 
  assumed that the reference strain rate is $\Epdot{0i} = 10^8$/s. 

\begin{figure}[p]
    \centering
    \scalebox{0.45}{\includegraphics{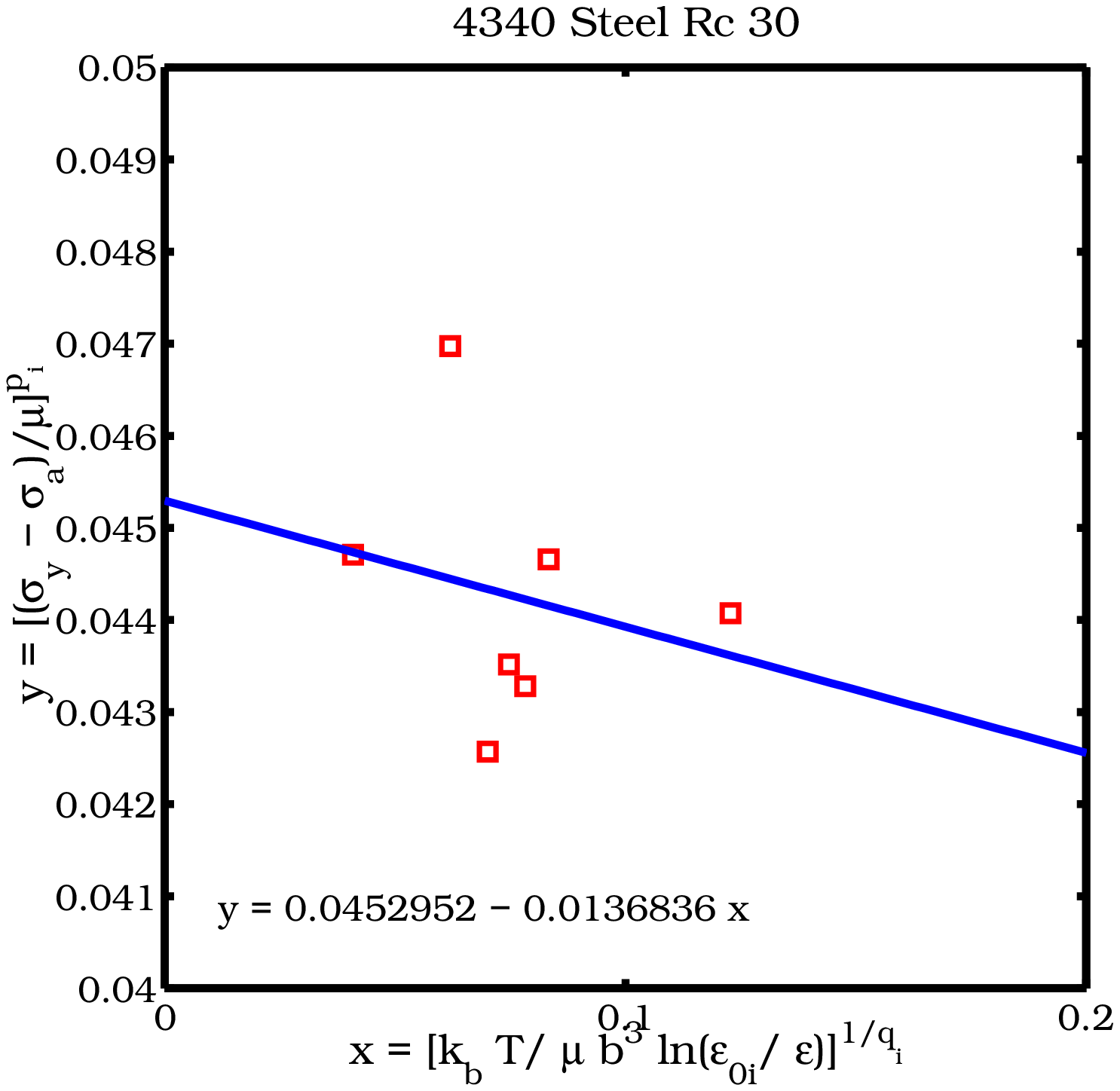}
                    \includegraphics{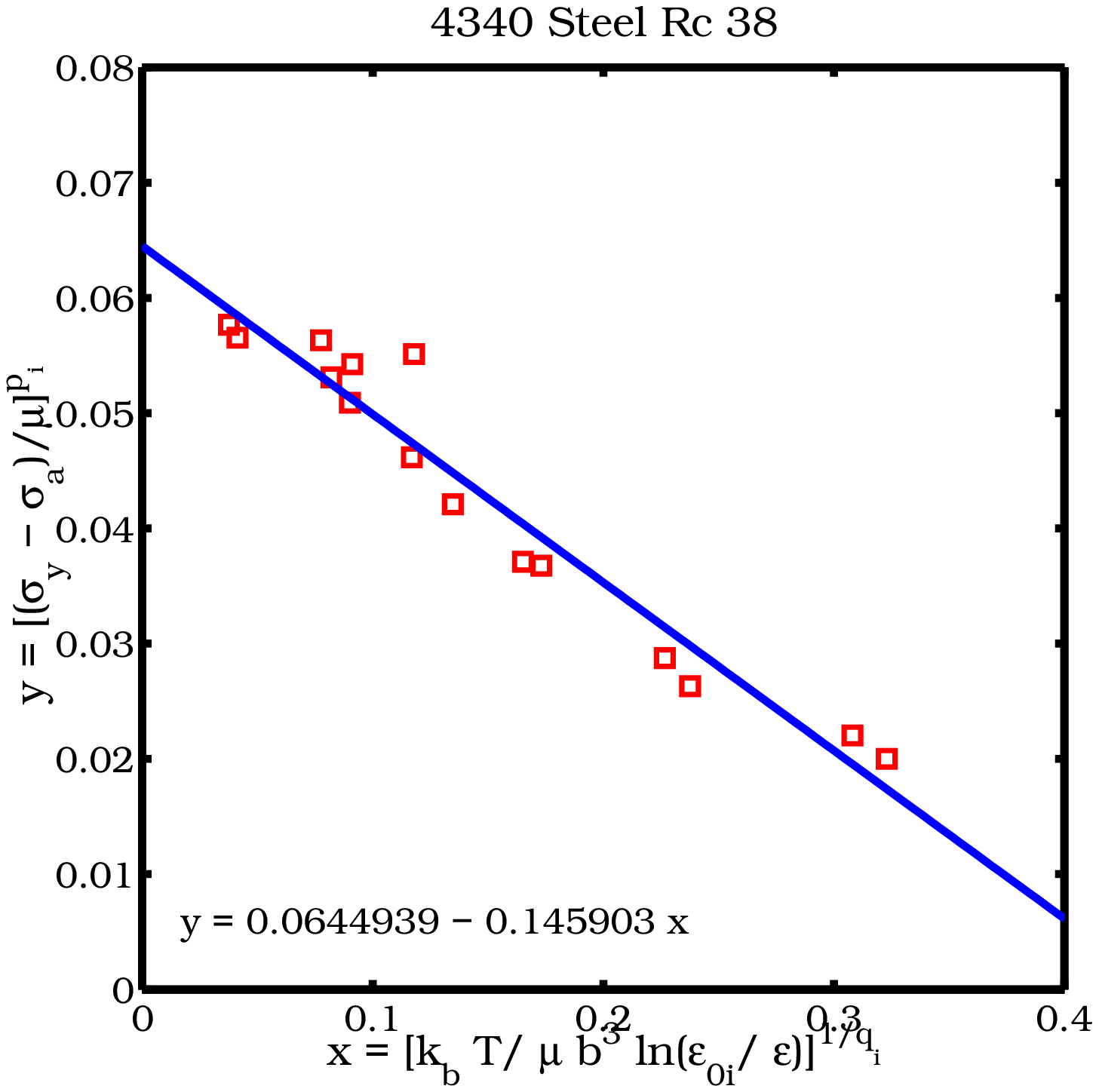}}\\
    (a) $R_c$ = 30 \hspace{2in} 
    (b) $R_c$ = 38 \\
    \vspace{20pt}
    \scalebox{0.45}{\includegraphics{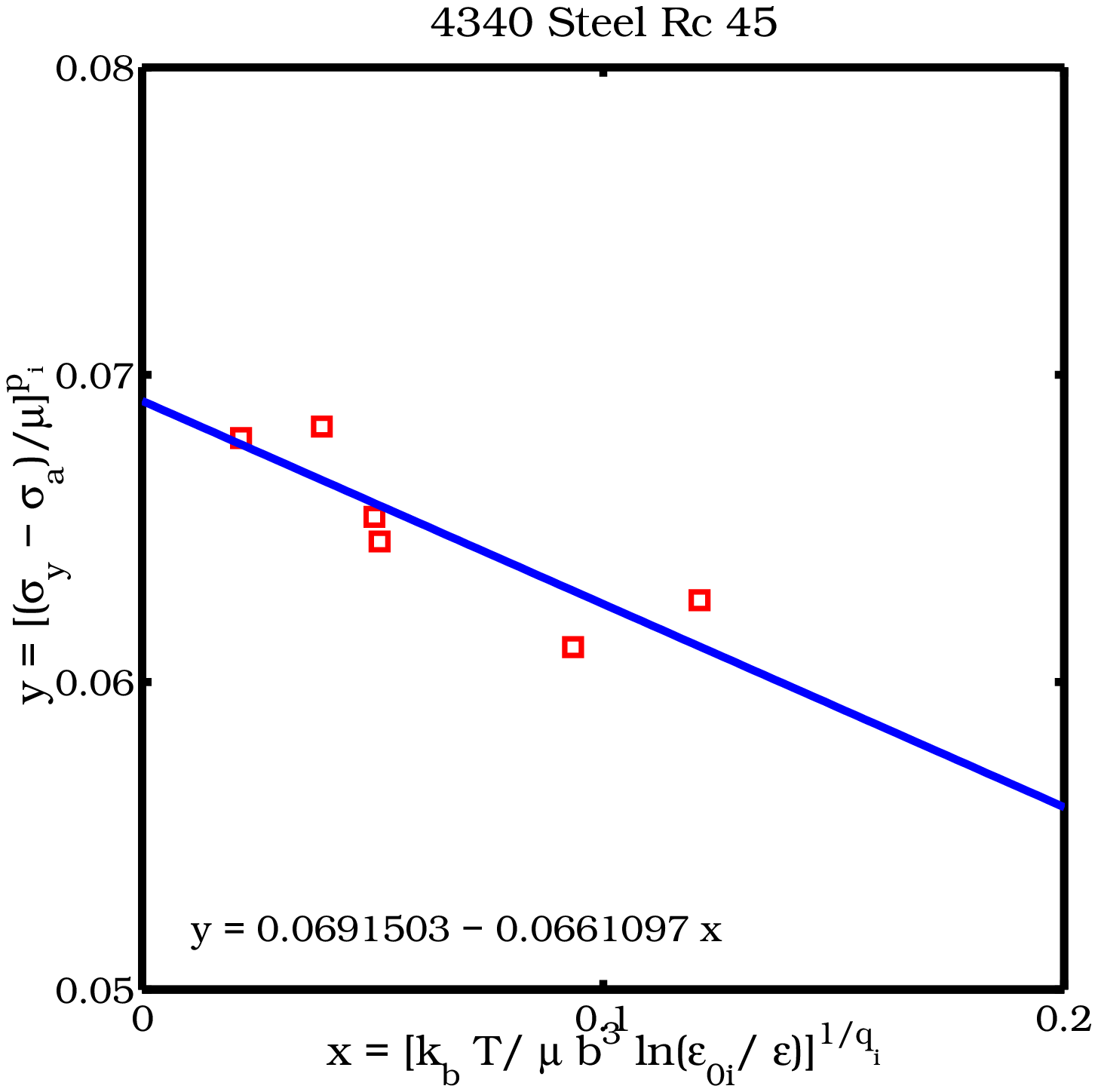}
                    \includegraphics{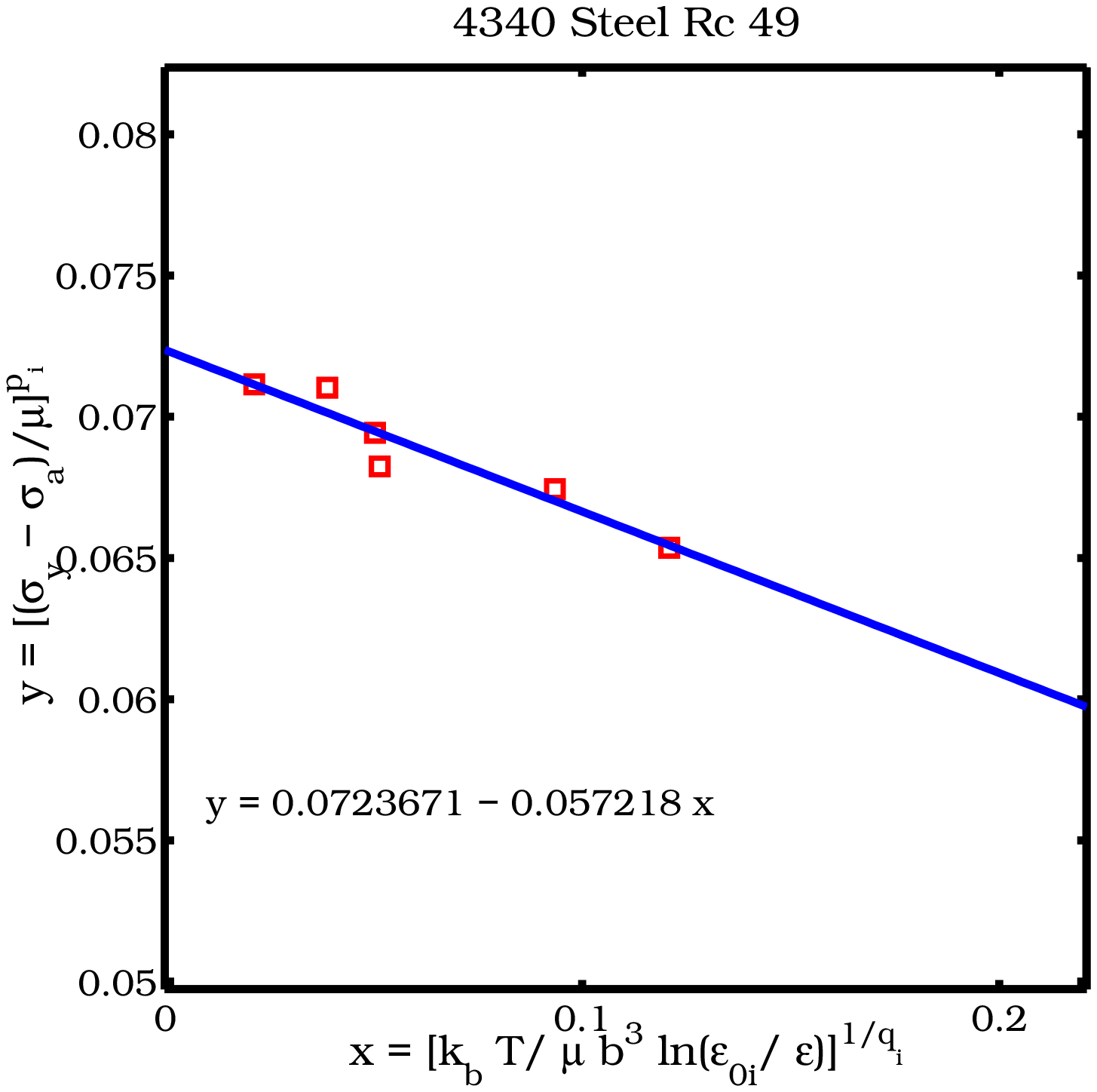}}\\
    (c) $R_c$ = 45 \hspace{2in} 
    (d) $R_c$ = 49 \\
    \caption{Fisher plots for the intrinsic component of the MTS
             model for various tempers of 4340 steel.}
    \label{fig:FisherSigi}
  \end{figure}

  The Fisher plots of the raw data (based on Equation~(\ref{eq:FisherSigi})) 
  are shown as squares in Figures~\ref{fig:FisherSigi}(a), (b), (c), and (d).
  Straight line least squares fits to the data are also shown in the figures.
  For these plots, the shear modulus ($\mu$) has been calculated using the NP 
  shear modulus model discussed in Sections \ref{sec:NPShear} and 
  \ref{sec:Shear}.  The yield stress at zero plastic strain ($\sigma_y$) 
  is the intersection of the stress-plastic strain curve with the stress 
  axis.  The value of the Boltzmann constant ($k_b$) is
  1.3806503e-23 J/K and the magnitude of the Burgers' vector ($b$) is
  assumed to be 2.48e-10 m.  The density of the material is assumed to
  be constant with a value of 7830 kg/m$^3$.  The raw data used in these
  plots are given in Appendix~\ref{app:FisherSigi}.

  The spread in the data for $R_c$ 30 (Figure~\ref{fig:FisherSigi}(a))
  is quite large and a very low $R^2$ value is obtained for the fit.
  This error is partially due to the inclusion of both tension and 
  shear test data (in the form of effective tensile stress) in the plot.  
  Note that significantly different yield stresses can be obtained from 
  tension and shear tests (especially at large strains) 
  (\citet{Johnson85,Goto00}).  However, this difference is
  small at low strains and is not expected to affect the intrinsic part
  of the yield stress much.  A more probable cause of the spread is 
  that the range of temperatures and strain rates is quite limited.
  More data at higher strain rates and temperatures are needed to get an 
  improved correlation for the $R_c$ 30 temper of 4340 steel.  

  Figure~\ref{fig:FisherSigi}(b) shows the fit to the Fisher plot data
  for 4340 steel of hardness $R_c$ 38.  The low strain rate data
  from \citet{Larson61} are the outliers near the top of the plot.  The
  hardness of this steel was estimated from tables given in ~\cite{ASM78} 
  based on the heat treatment and could be higher than $R_c$ 38.  However,
  the \citet{Larson61} data are close to the data from \citet{Lee97} as 
  can be seen from the plot.  A close examination of the high temperature
  data shows that there is a slight effect due to the $\alpha$ to $\gamma$
  phase transformation at high temperatures.

  The stress-strain data for 4340 steel $R_c$ 45 shows anomalous 
  temperature dependent behavior under quasistatic conditions. For instance, 
  the yield stress at 373 K is higher than that at 298 K.  The fit to the
  Fisher plot data for this temper of steel is shown in 
  Figure~\ref{fig:FisherSigi}(c).  The fit to the data can be improved if
  the value of $\sigma_a$ is assumed to be 150 MPa and $q_i$ is assumed to
  be equal to 2.  However, larger values of $\sigma_a$ can lead to 
  large negative values of $\sigma_e$ at small strains - which is 
  unphysical.  

  The fit to the Fisher data for the $R_c$ 49 temper is shown in 
  Figure~\ref{fig:FisherSigi}(d).  The fit is reasonably good.  More
  data at high strain rates and high temperatures are needed for both 
  the $R_c$ 45 and the $R_c$ 49 tempers of 4340 steel.

  The values of $\sigma_i$ and $g_{0i}$ for the four tempers of 4340
  are shown in Table~\ref{tab:SigmaiGoi}.  The value of $g_{0i}$ for
  the $R_c$ 38 temper is quite low and leads to values of the 
  Arrhenius factor ($S_i$) that are zero for temperatures more than 800 K.
  In the following section, we consider the effect of dividing the 
  $R_c$ 38 data into high and low temperature regions to alleviate this
  problem.
  \begin{table}[t]
    \centering
    \caption{Values of $\sigma_i$ and $g_{0i}$ for four tempers of 
             4340 steel.}
    \begin{tabular}{ccc}
       \hline
       Hardness ($R_c$) & $\sigma_i$ (MPa) & $g_{0i}$ \\
       \hline
       30 & 867.6  & 3.31 \\
       38 & 1474.1 & 0.44 \\
       45 & 1636.6 & 1.05 \\
       49 & 1752   & 1.26 \\
       \hline
    \end{tabular}
    \label{tab:SigmaiGoi}
  \end{table}

  \subsubsection{High temperature values of $\sigma_i$ and $g_{0i}$}
  More data at higher temperatures and high strain rates are
  required for better characterization of the $R_c$ 30, $R_c$ 45, and
  $R_c$ 49 tempers of 4340 steel.  In the absence of high temperature data,
  we can use data for the $R_c$ 38 temper at high temperatures to obtain
  the estimates of $\sigma_i$ and $g_{0i}$ for other tempers.  We explore
  two approaches of determining these parameters:
  \begin{enumerate} 
    \item {\bf Case 1:}
          Divide the temperature regime into three parts: $T_0 <$ 573 K;
          573 K $\le T_0 < $ 1040 K; $T_0 \ge $ 1040 K.  The values of
          $\sigma_i$ and $g_{0i}$ are calculated for each of these regimes
          from the $R_c$ 38 data.  The values of the two parameters for
          temperatures above 573 K are assumed to be applicable to all the
          tempers.  Note that the choice of 573 K for the cut-off temperature
          is arbitrary and loss of temper is likely to occur at a higher
          temperature.
    \item {\bf Case 2:}
          Divide the temperature regime into two parts: $T_0 < $ 1040 K
          and $T_0 \ge$ 1040 K.  In this case, we assume that the various
          tempers retain distinctive properties up to the phase transition
          temperature.  All the tempers are assumed to have identical 
          values of $\sigma_i$ and $g_{0i}$ above 1040 K.
  \end{enumerate} 

\begin{figure}[t]
    \centering
    \scalebox{0.45}{\includegraphics{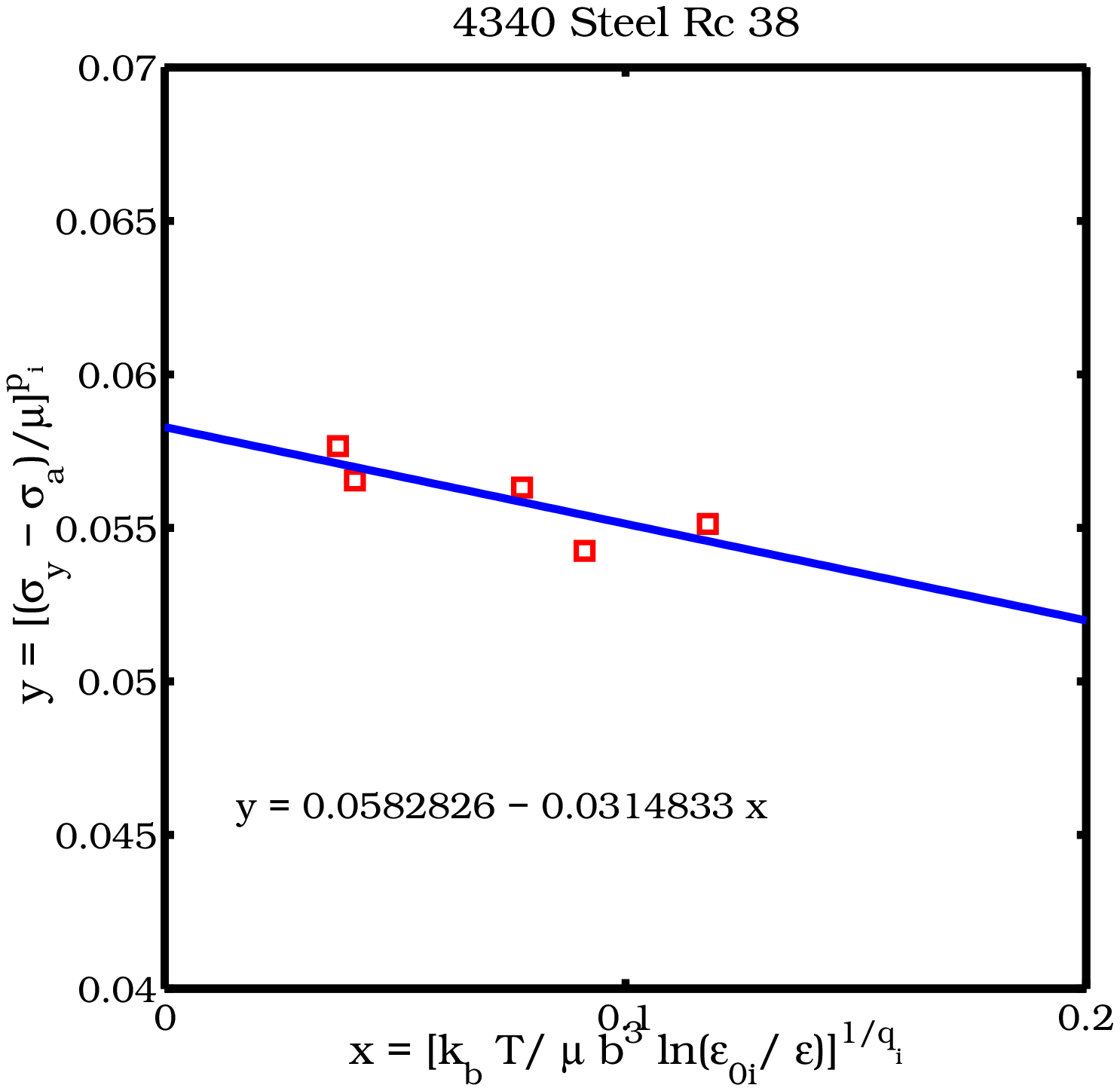}
                    \includegraphics{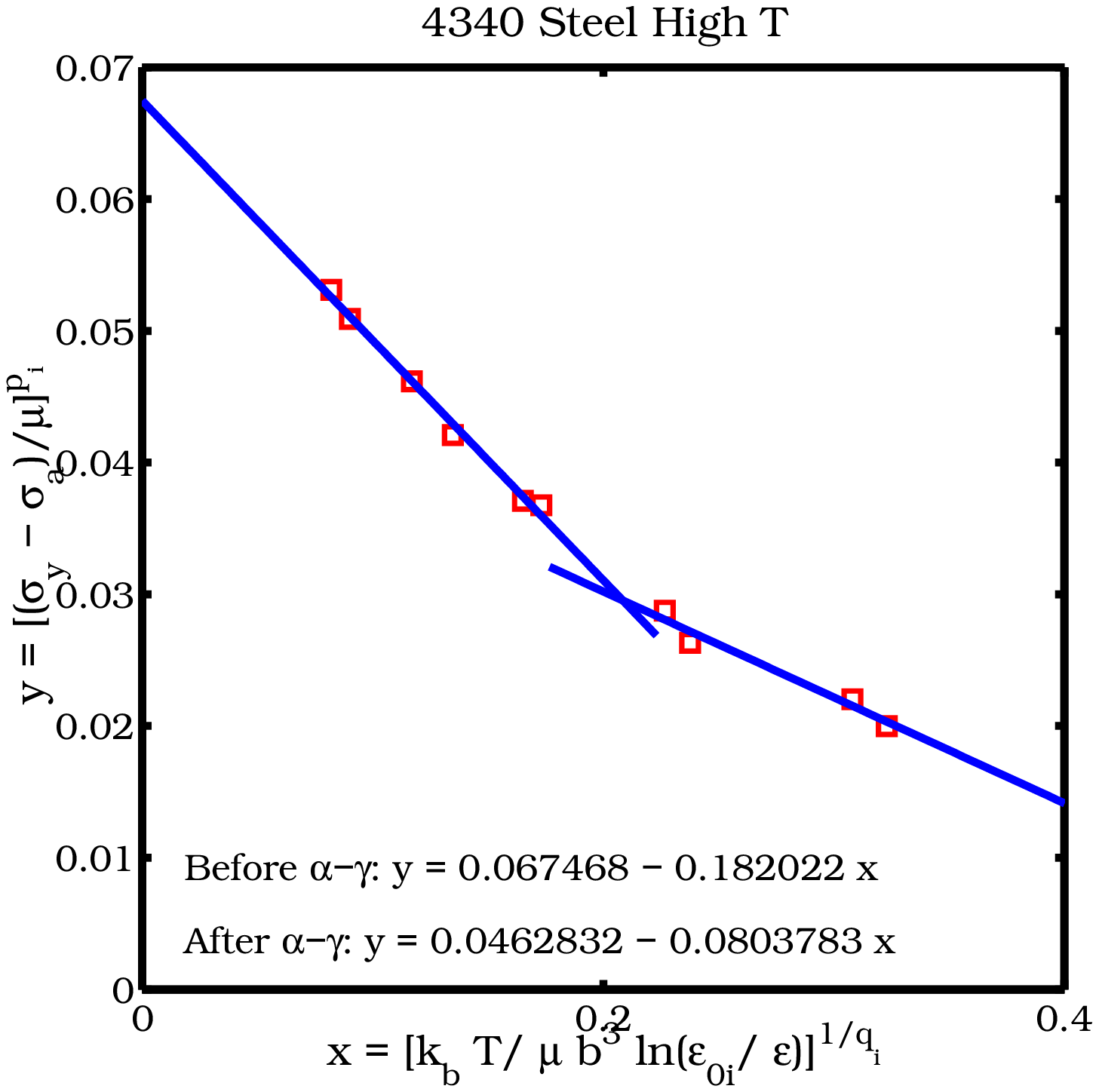}}\\
    (a) $R_c$ = 38 (T $<$ 573 K) \hspace{1.0in} 
    (b) $R_c$ = 38 (T $>$ 573 K) \\
    \caption{Fisher plots for the intrinsic component of the MTS
             model for $R_c$ 38 4340 steel assuming three temperature
             regimes.}
    \label{fig:SigiRc38}
  \end{figure}

  \paragraph{Case 1: Three temperature regimes.}
  The low and high temperature Fisher plots for $R_c$ 38 4340 steel are 
  shown in Figures~\ref{fig:SigiRc38} (a) and (b), respectively.
  A comparison of Figures~\ref{fig:FisherSigi}(b) and \ref{fig:SigiRc38}(a)
  shows that the low temperature Fisher plot has a distinctly lower slope
  that the plot that contains all the $R_c$ 38 data.  The values of 
  $\sigma_i$ and $g_{0i}$ at low temperatures for the $R_c$ 38 temper are
  1266 MPa and 1.85, respectively.   The high temperature plot 
  (Figure~\ref{fig:SigiRc38}(b)) shows that the slope of the Fisher 
  plot is quite steep between 573 K and 1040 K and decreases slightly after 
  the $\alpha$ to $\gamma$ phase transition.  The values of $\sigma_i$ 
  and $g_{0i}$ for temperatures between 573 K and 1040 K are 1577.2 MPa 
  and 0.371, respectively.  After the phase transition at 1040 K, these 
  quantities take values of 896.1 MPa and 0.576, respectively.  

\begin{figure}[t]
    \centering
    \scalebox{0.45}{\includegraphics{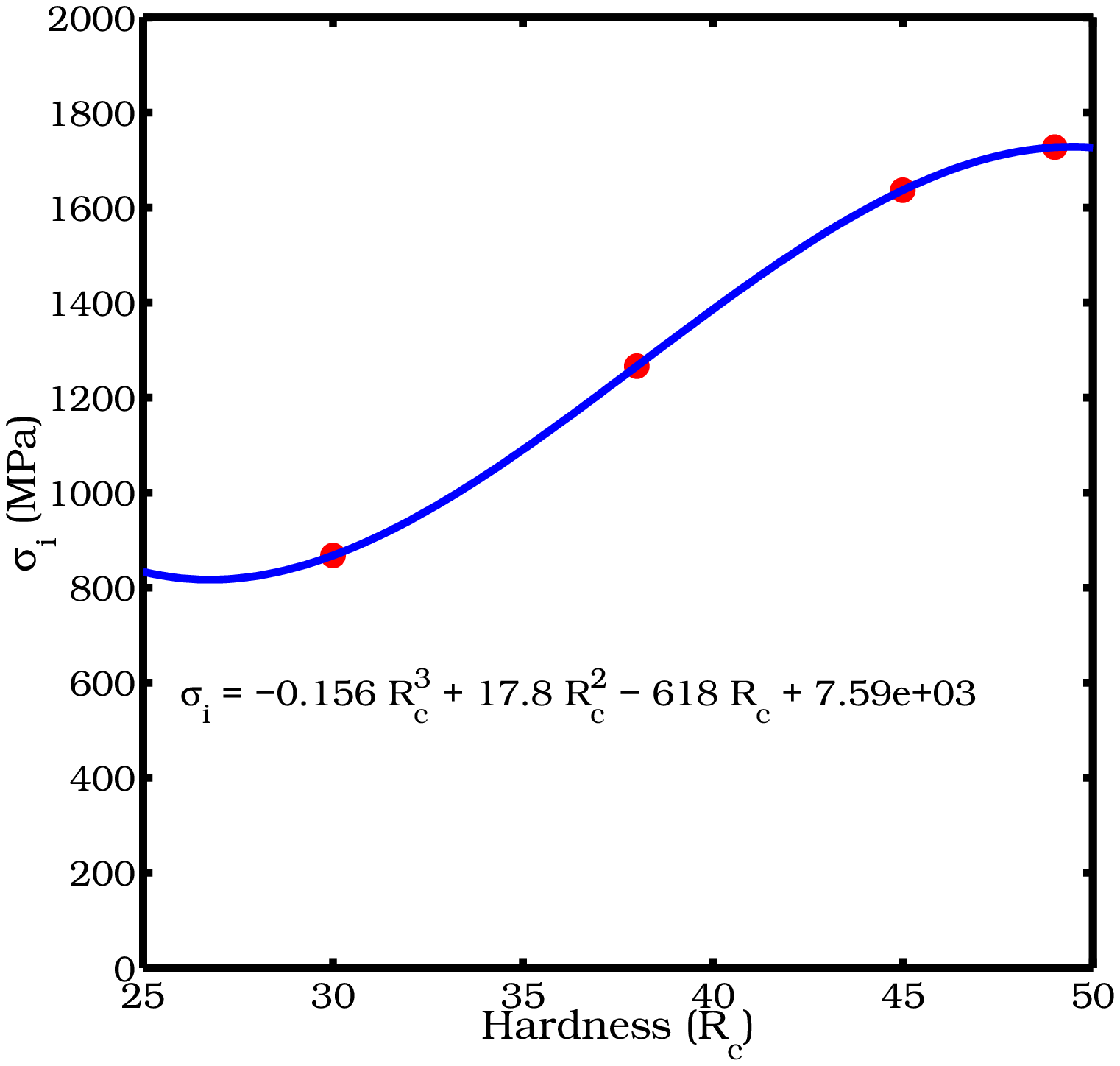}
                    \includegraphics{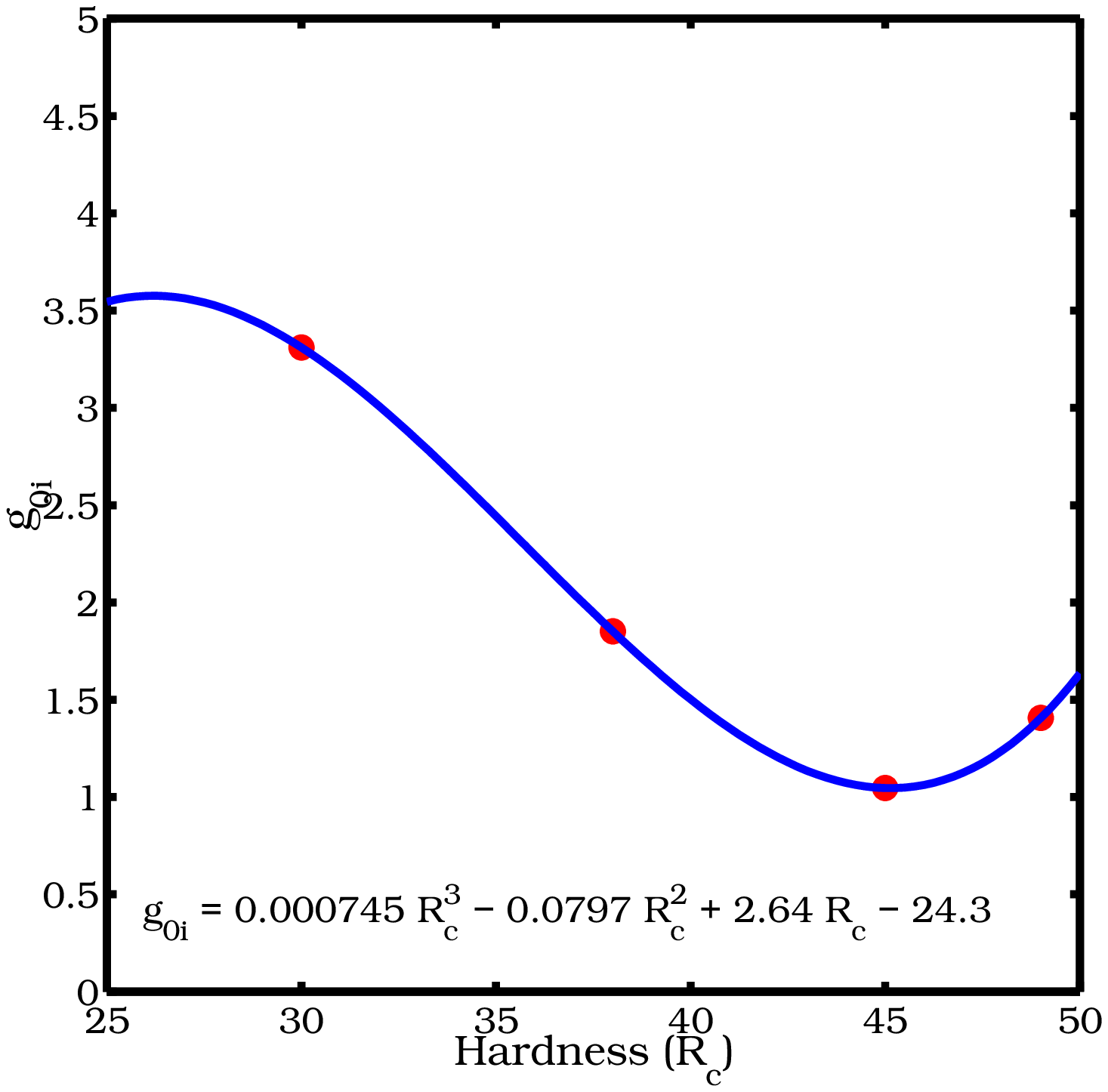}}\\
    (a) $\sigma_i$ \hspace{2in} 
    (b) $g_{0i}$ \\
    \caption{Values of $\sigma_i$ and $g_{0i}$ obtained from the 
             Fisher plots for various tempers of 4340 steel close
             to room temperature (for the three temperature regime).}
    \label{fig:SigmaiGoiLow}
  \end{figure}

  Plots of $\sigma_i$ and $g_{0i}$ as functions of the Rockwell hardness
  number (for temperatures below 573 K) are shown in 
  Figures~\ref{fig:SigmaiGoiLow}(a) and (b), respectively.  
  These plots show a smooth increase in the value of $\sigma_i$ and 
  a decrease in the normalized activation energy ($g_{0i}$) with 
  increasing hardness.
  The high temperature values of $g_{0i}$ 
  for the $R_c$ 38 give reasonable values of $S_i$ (non-zero) at temperatures
  above 800 K.  However, the lower temperature (less that 1040 K) values of 
  the two parameters give a poor fit to the experimental stress-strain data.
  This is probably due to the anomalous behavior of 4340 steel at 373 K
  and low strain rates. 

\begin{figure}[p]
    \centering
    \scalebox{0.45}{\includegraphics{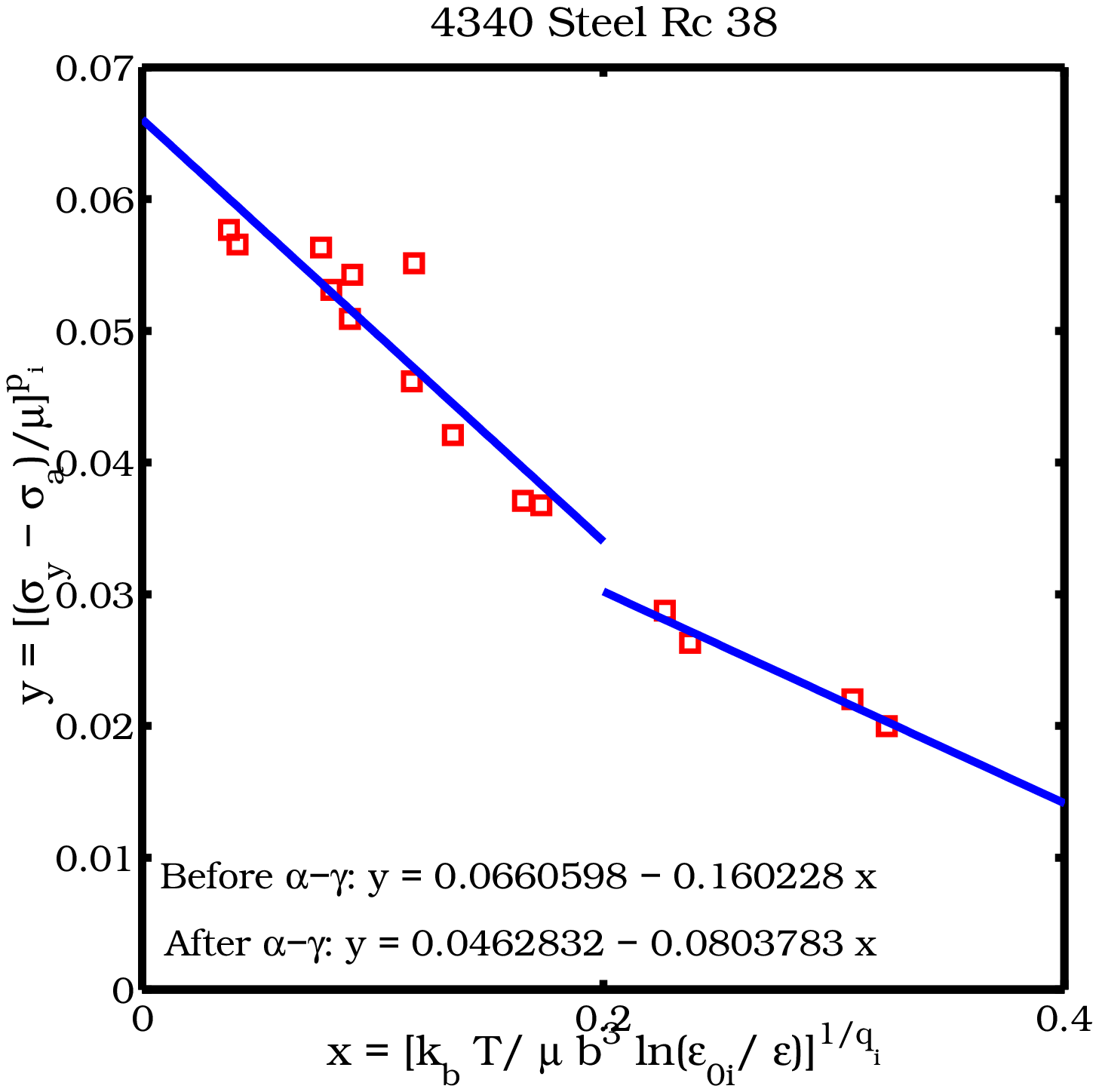}}\\
    \caption{Fisher plots for the intrinsic component of the MTS
             model for the $\alpha$ and $\gamma$ phases of
             $R_c$ 38 4340 steel assuming two temperature regimes.}
    \label{fig:SigiRc38HiLo}
  \end{figure}

  \paragraph{Case 2: Two temperature regimes.}
  The two-regime fits to the Fisher plot data for $R_c$ 38 are shown 
  in Figure~\ref{fig:SigiRc38HiLo}.
  The values of $\sigma_i$ and $g_{0i}$ for the $R_c$ 38 temper 
  (in the $\alpha$ phase) are 1528 MPa and 0.412, respectively, while 
  those for the $\gamma$ phase are 896 MPa and 0.576, respectively.  
  The fits show a jump in value at 1040 K that is not ideal for Newton 
  iterations in a typical elastic-plastic numerical code.  We suggest 
  that the $\gamma$ phase values of these parameters be used if there
  is any problem with convergence.

\begin{figure}[p]
    \centering
    \scalebox{0.45}{\includegraphics{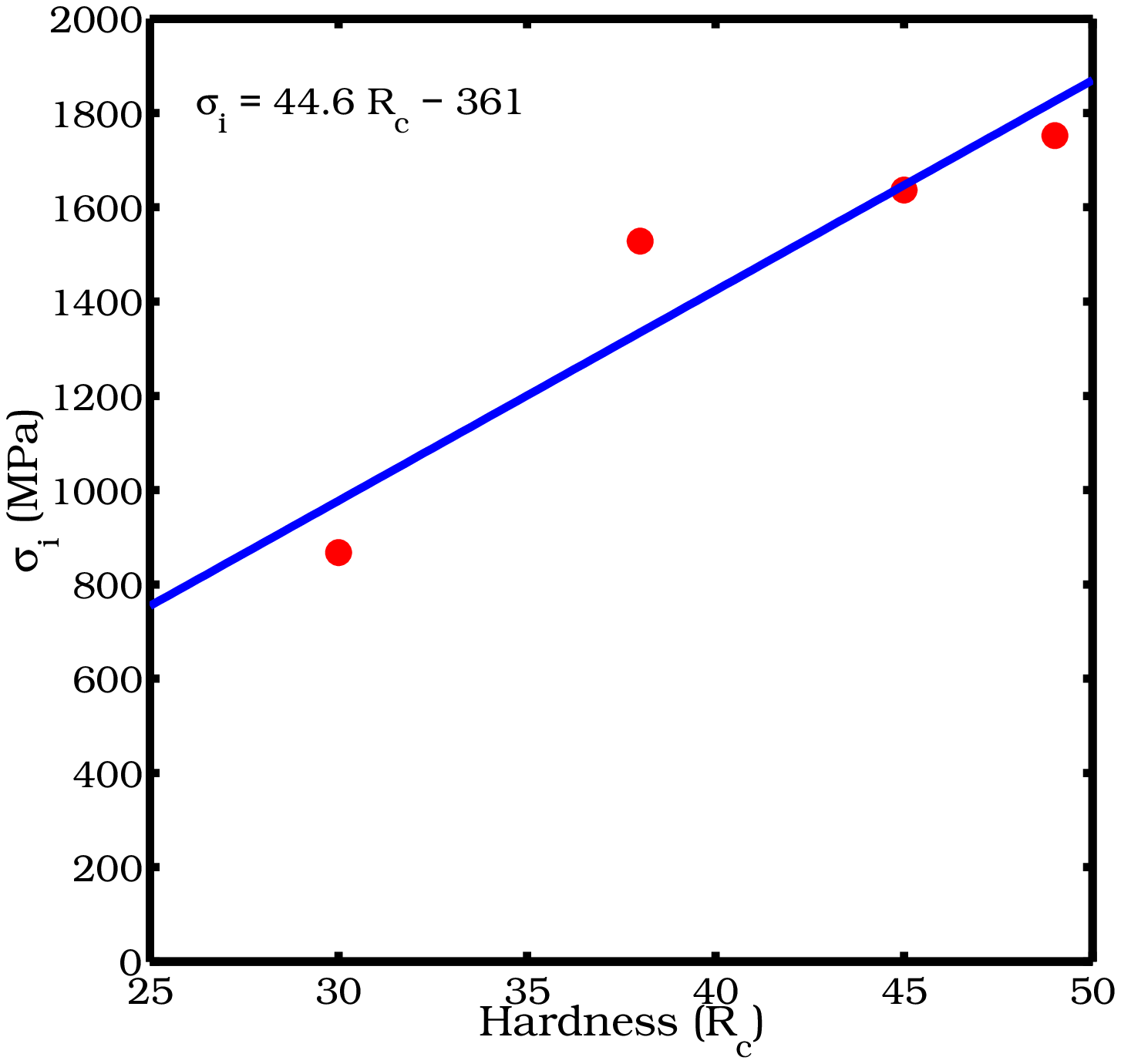}
                    \includegraphics{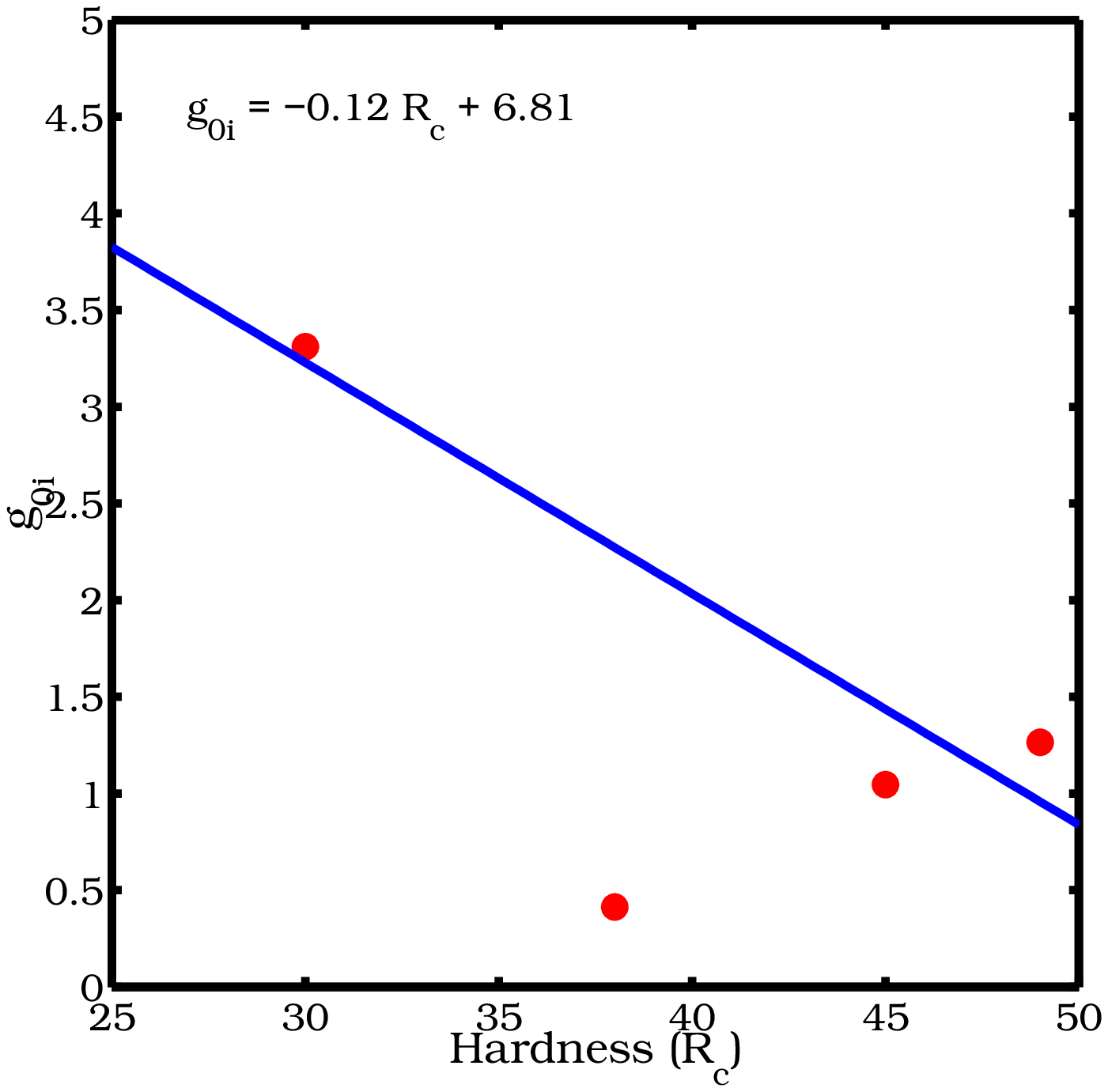}}\\
    (a) $\sigma_i = 44.628 R_c - 361.33$ \hspace{36pt} 
    (b) $g_{0i} = - 0.1195 R_c + 6.814$ \\
    \caption{Values of $\sigma_i$ and $g_{0i}$ obtained from the 
             Fisher plots for various tempers of the $\alpha$ phase
             of 4340 steel. The fit for $g_{0i}$ excludes the low 
             value for $R_c$ 38 4340 steel.}
    \label{fig:SigmaiGoialpha}
  \end{figure}

  Plots of $\sigma_i$ and $g_{0i}$ as functions of the Rockwell hardness
  number (for temperatures below 1040 K) are shown in 
  Figures~\ref{fig:SigmaiGoialpha}(a) and (b), respectively.  
  Straight line fits to the $\sigma_i$ and $g_{0i}$ versus $R_c$ data 
  can be used to estimate these parameters for intermediate tempers of 
  the $\alpha$ phase of 4340 steel.  These fits are shown in 
  Figure~\ref{fig:SigmaiGoialpha}.  

  The value of $\sigma_i$ increases with increasing hardness.  However, 
  the value of $g_{0i}$ does not decrease monotonically with hardness.
  More experimental data are needed to determine if the trend of $g_{0i}$
  is physical.  The value of $g_{0i}$ for the $R_c$ 38 temper appears to be
  unusually low.  However, these values lead to good fit to experimental
  data for $R_c$ 38 temper.  For that reason, we have used the two 
  temperature regime values of $\sigma_i$ and $g_{0i}$ for all 
  subsequent computations that use these parameters.

  \subsection{Determination of $\sigma_{0es}$ and $g_{0es}$}
  Once estimates have been obtained for $\sigma_i$ and $g_{0i}$, the
  value of $S_i\sigma_i$ can be calculated for a particular strain rate 
  and temperature.  From equation (\ref{eq:MTSSigmay}), we then get
  \begin{equation} \label{eq:Sigmae}
    \sigma_e = \cfrac{1}{S_e}
               \left[\cfrac{\mu_0}{\mu}\left(\sigma_y - \sigma_a\right)
                     - S_i\sigma_i\right] ~.
  \end{equation}
  Equation (\ref{eq:Sigmae}) can be used to determine the saturation value 
  ($\sigma_{es}$) of the structural evolution stress ($\sigma_e$). 
  Given a value of $\sigma_{es}$, equation (\ref{eq:Sigma0esG0es})
  can be used to compute $\sigma_{0es}$ and the
  corresponding normalized activation energy ($g_{0es}$) from the
  relation  
  \begin{equation}
    \ln(\sigma_{es}) = \ln(\sigma_{0es}) - 
    \frac{k_b T}{g_{0es} b^3 \mu}
    \ln\left(\cfrac{\Epdot{}}{\Epdot{0es}}\right) ~.
  \end{equation}

  The value of $\sigma_{es}$ can be determined either from a 
  plot of $\sigma_e$ versus the plastic strain or from a plot of the 
  tangent modulus $\theta(\sigma_e)$ versus $\sigma_e$.  
  The value of $S_e$ is required before $\sigma_e$ can be calculated.  
  Following \citet{Goto00},
  we assume that $\Epdot{0e}$, $\Epdot{0es}$, $p_e$, $q_e$, and $g_{0e}$
  take the values 10$^7$ /s, 10$^7$ /s, 2/3, 1, and 1.6, respectively.
  These values are used to calculate $S_e$ at various temperatures and
  strain rates.  The values of $\sigma_i$ and $g_{0i}$ used to compute
  $\sigma_e$ vary with hardness for temperatures below 1040 K, and are
  constant above that temperature as discussed in the previous section.
  Adiabatic heating is assumed for strain rates greater than 500 /s.
  
\begin{figure}[p]
    \centering
    \scalebox{0.45}{\includegraphics{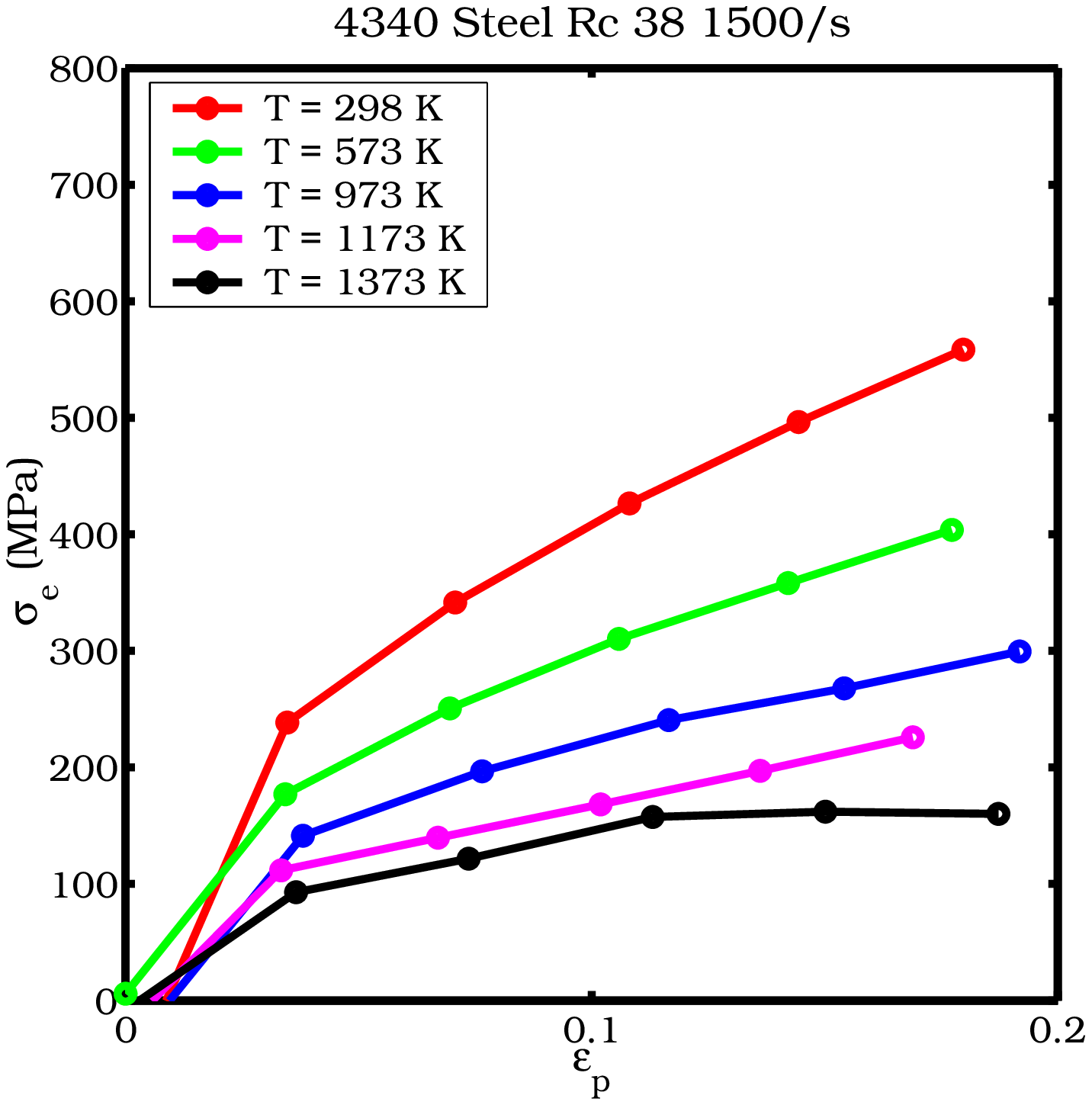}
                    \includegraphics{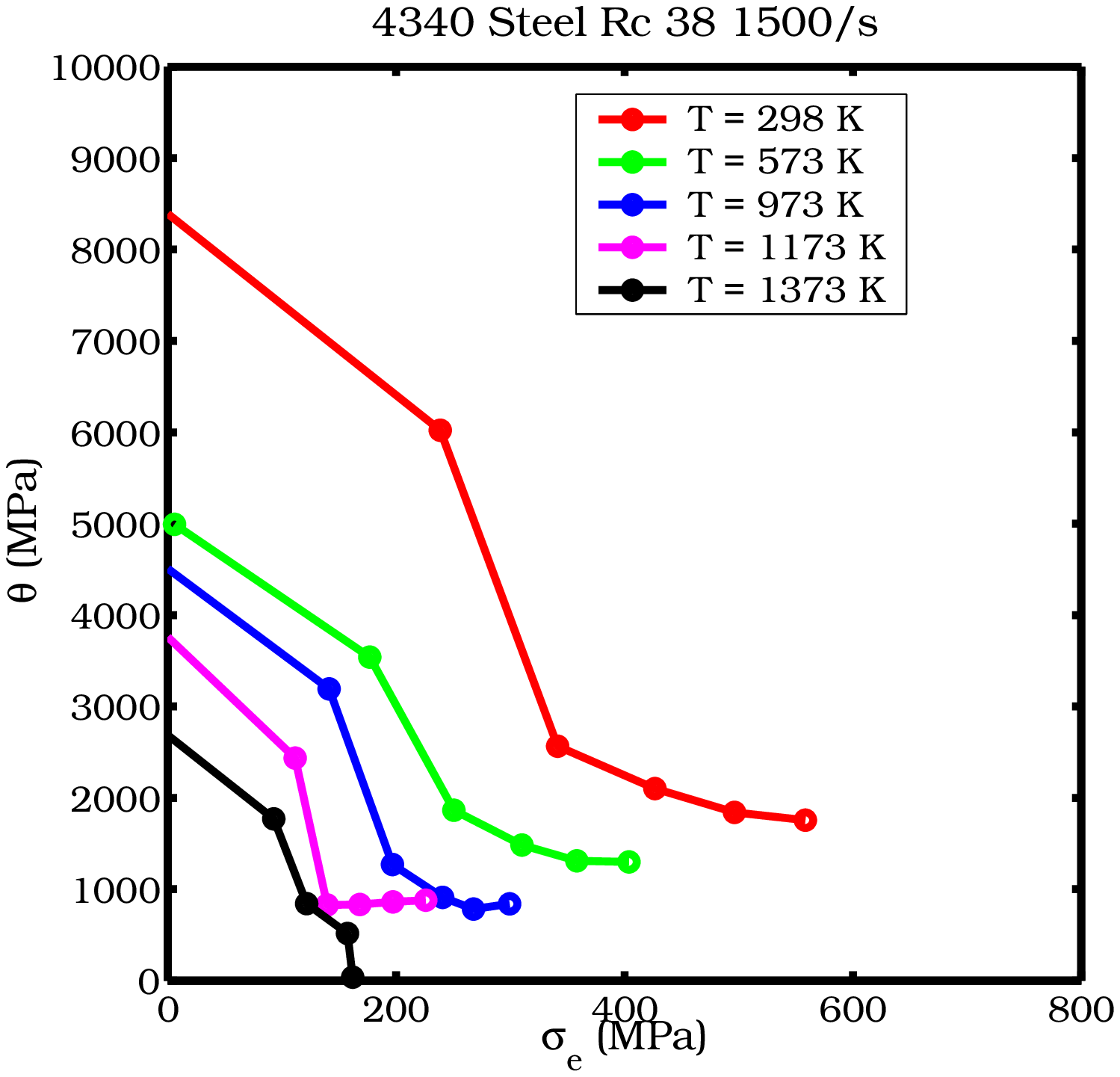}}\\
    (a) $\sigma_e$ vs. $\Ep$ ($R_c$ 38 1500 /s) \hspace{1in} 
    (b) $\theta$ vs. $\sigma_e$ ($R_c$ 38 1500 /s) \\
    \vspace{20pt}
    \scalebox{0.45}{\includegraphics{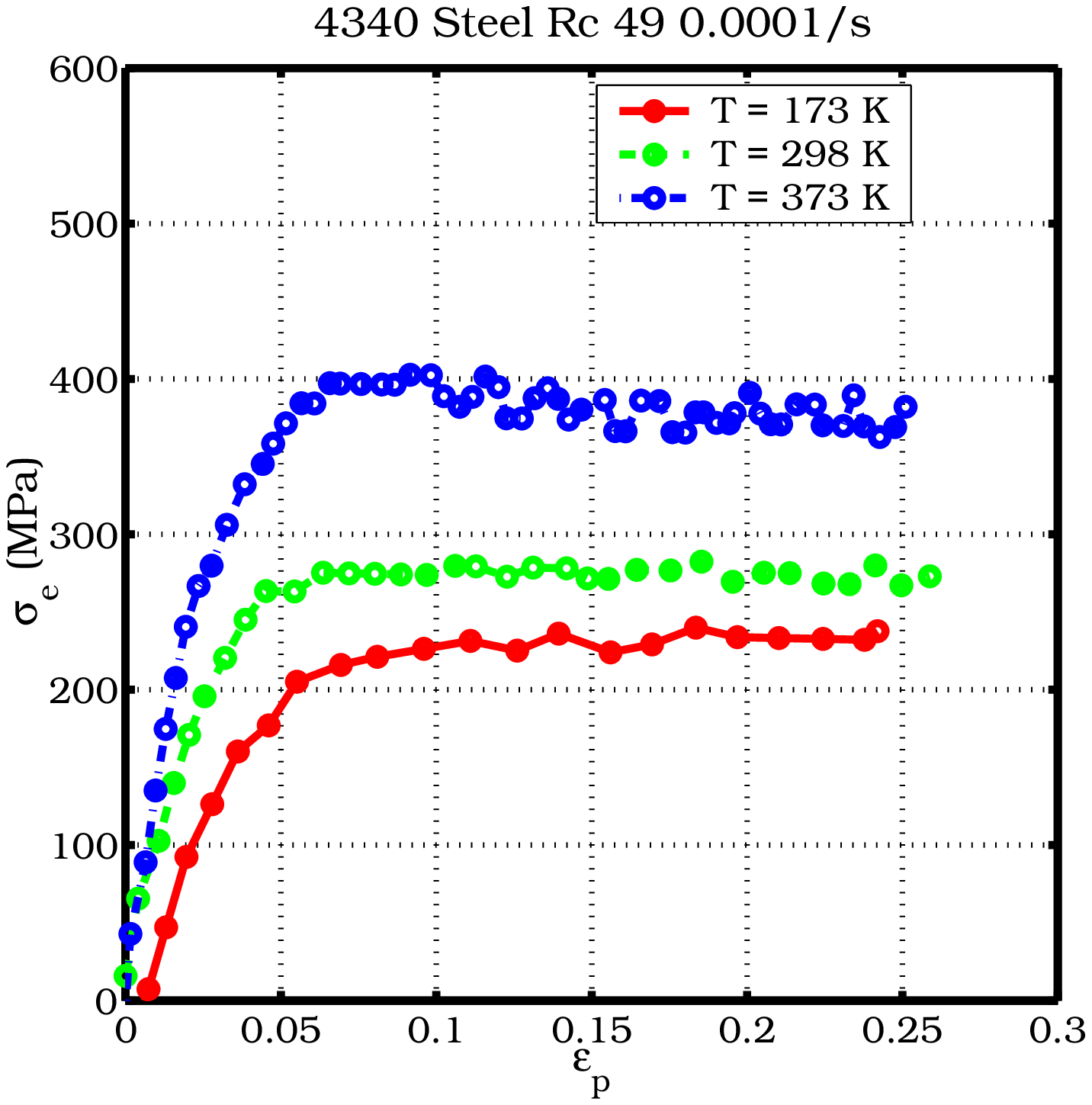}
                    \includegraphics{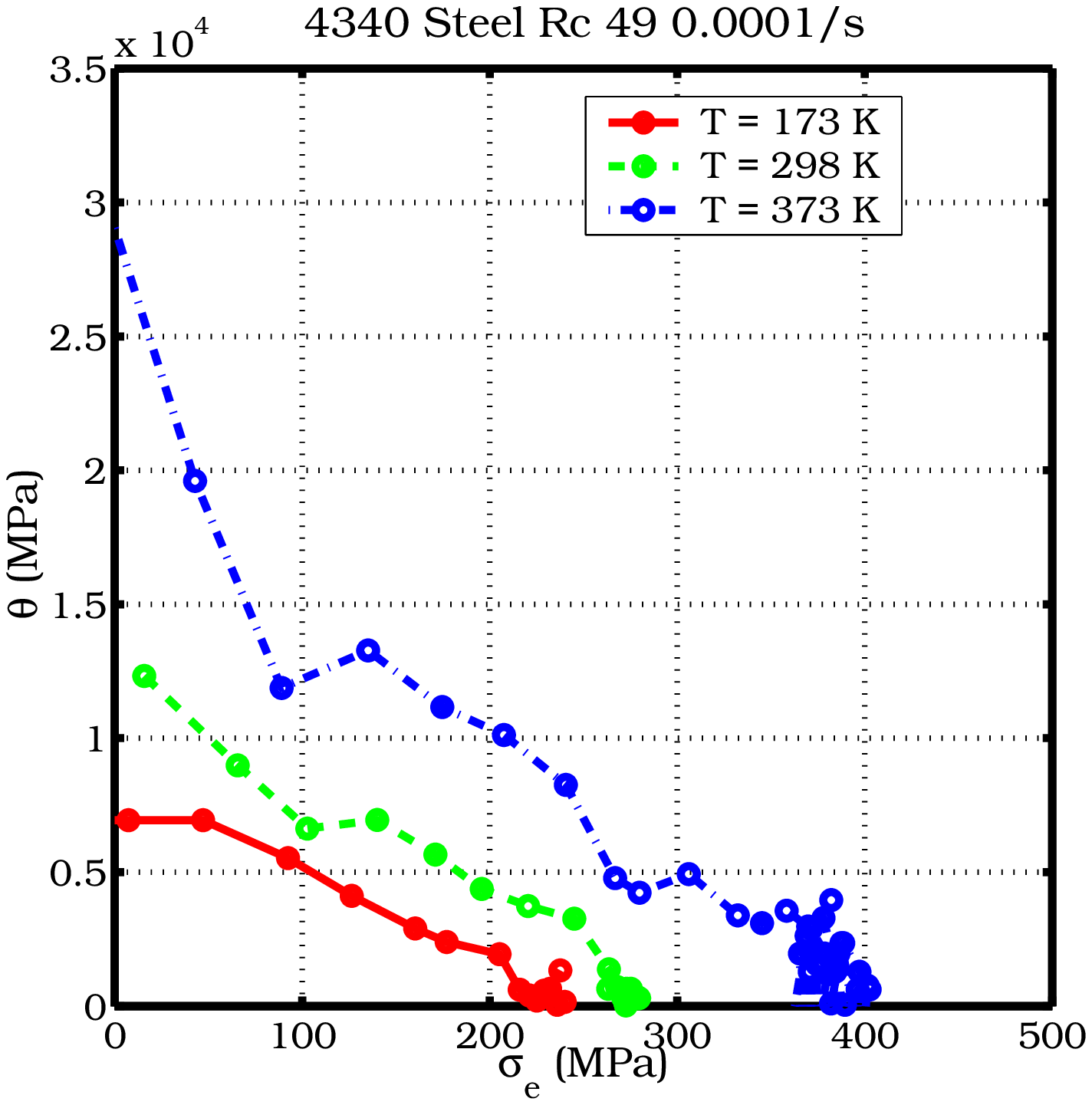}}\\
    (c) $\sigma_e$ vs. $\Ep$ ($R_c$ 49 0.0001 /s) \hspace{1in} 
    (d) $\theta$ vs. $\sigma_e$ ($R_c$ 49 0.0001 /s) \\
    \caption{Plots used to determine the saturation value ($\sigma_{es}$)
             of the structure evolution stress ($\sigma_e$).}
    \label{fig:SigmaevsEp}
  \end{figure}

  Representative plots of $\sigma_e$ versus the plastic strain are shown in 
  Figure~\ref{fig:SigmaevsEp}(a) and the corresponding $\theta$ versus
  $\sigma_e$ plots are shown in Figure~\ref{fig:SigmaevsEp}(b) (for the 
  $R_c$ 38 temper; strain rate of 1500 /s).  Similar plots for the 
  $R_c$ 49 temper for a strain rate of 0.0001 /s are shown in 
  Figures~\ref{fig:SigmaevsEp}(c) and (d).  The plotted value of the 
  tangent modulus ($\theta$) is the mean of the tangent moduli at each 
  value of $\sigma_e$ (except for the end points where a single value is
  used).  The saturation stress ($\sigma_{es}$) is the value at 
  which $\sigma_e$ becomes constant or $\theta$ is zero.  Note that
  errors in the fitting of $\sigma_i$ and $g_{0i}$ can cause the computed
  value of $\sigma_e$ to nonzero at zero plastic strain.
  
\begin{figure}[p]
    \centering
    \scalebox{0.45}{\includegraphics{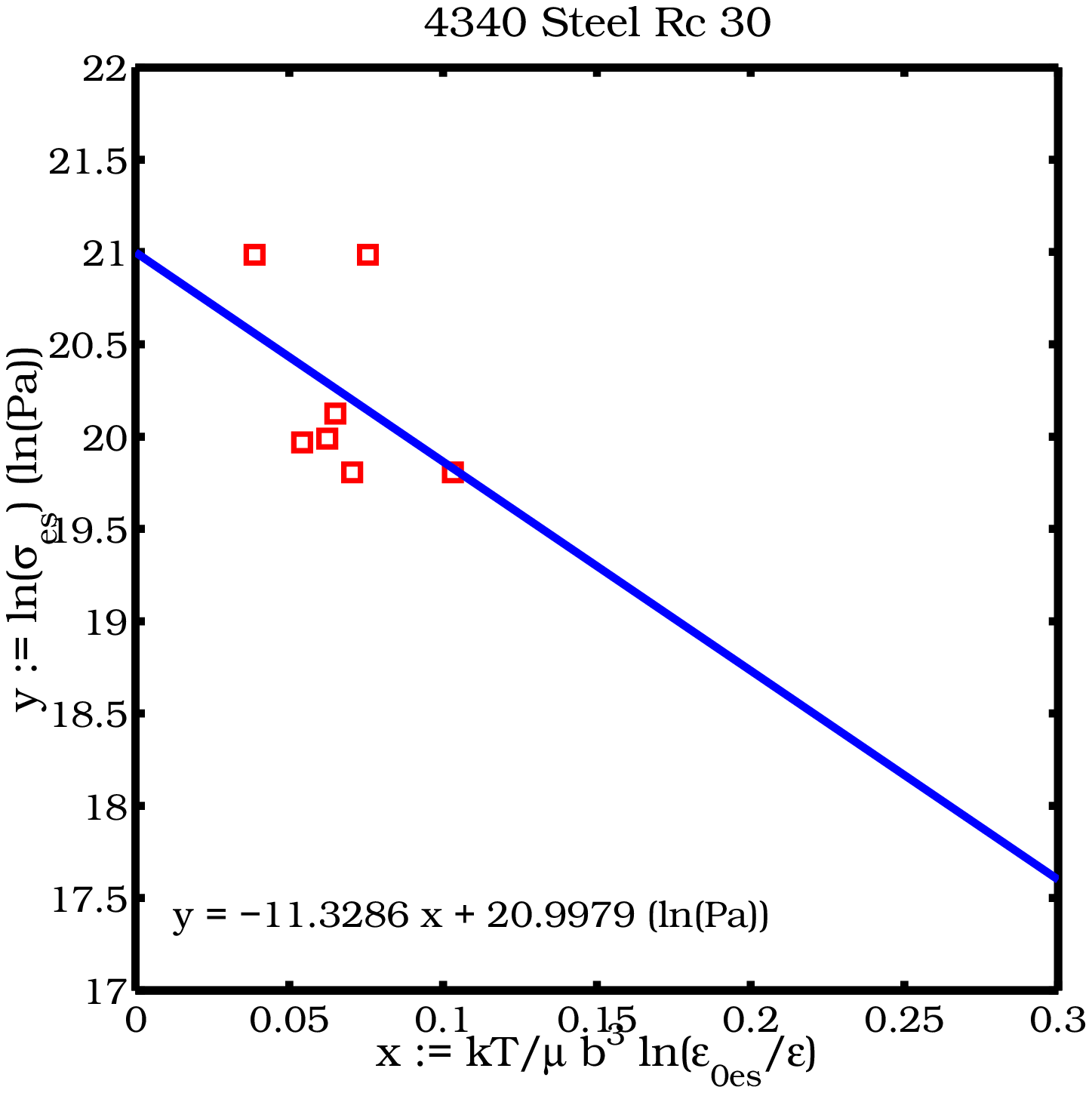}
                    \includegraphics{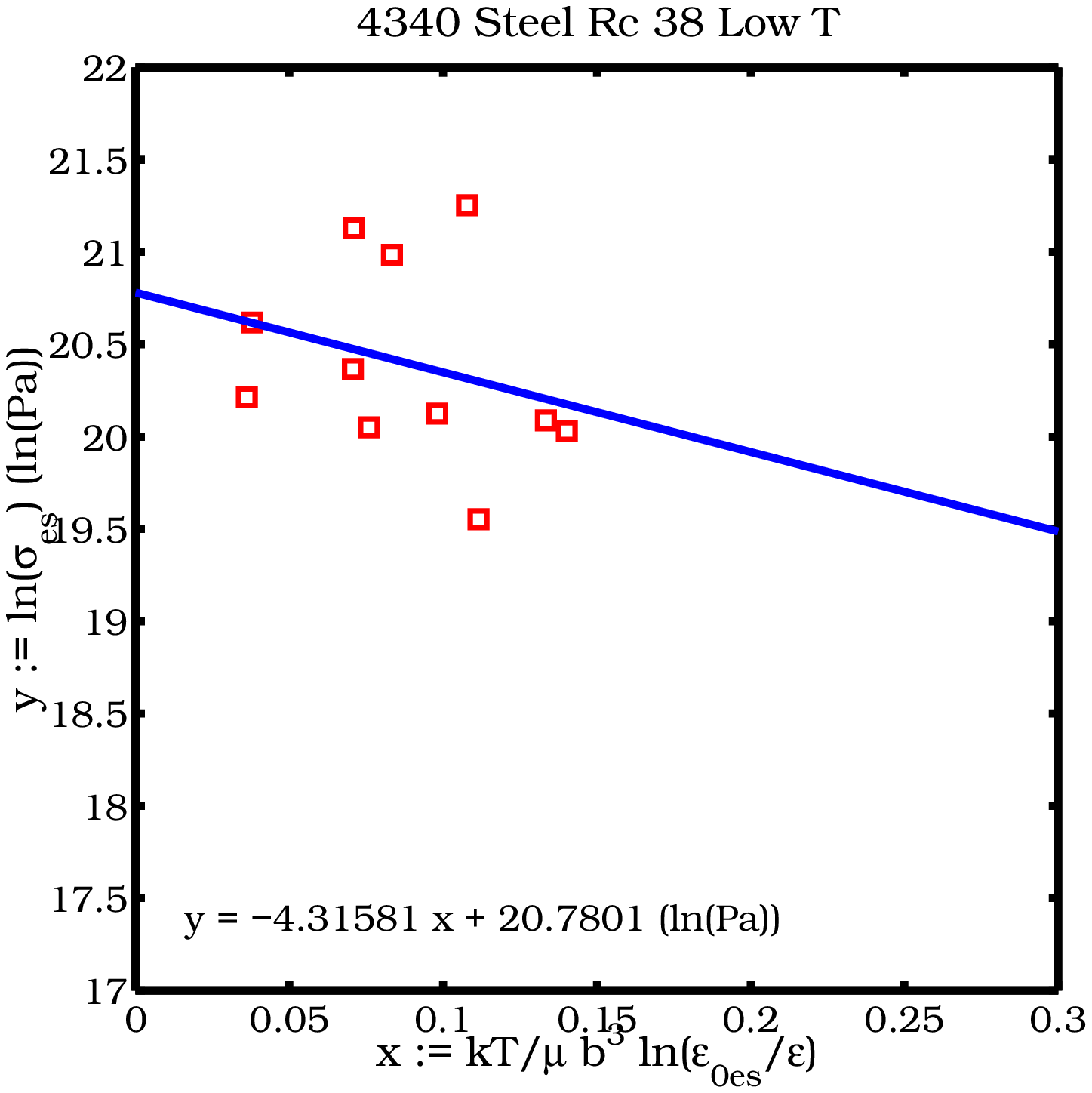}}\\
    (a) $R_c$ = 30 \hspace{2in} 
    (b) $R_c$ = 38 \\
    \vspace{20pt}
    \scalebox{0.45}{\includegraphics{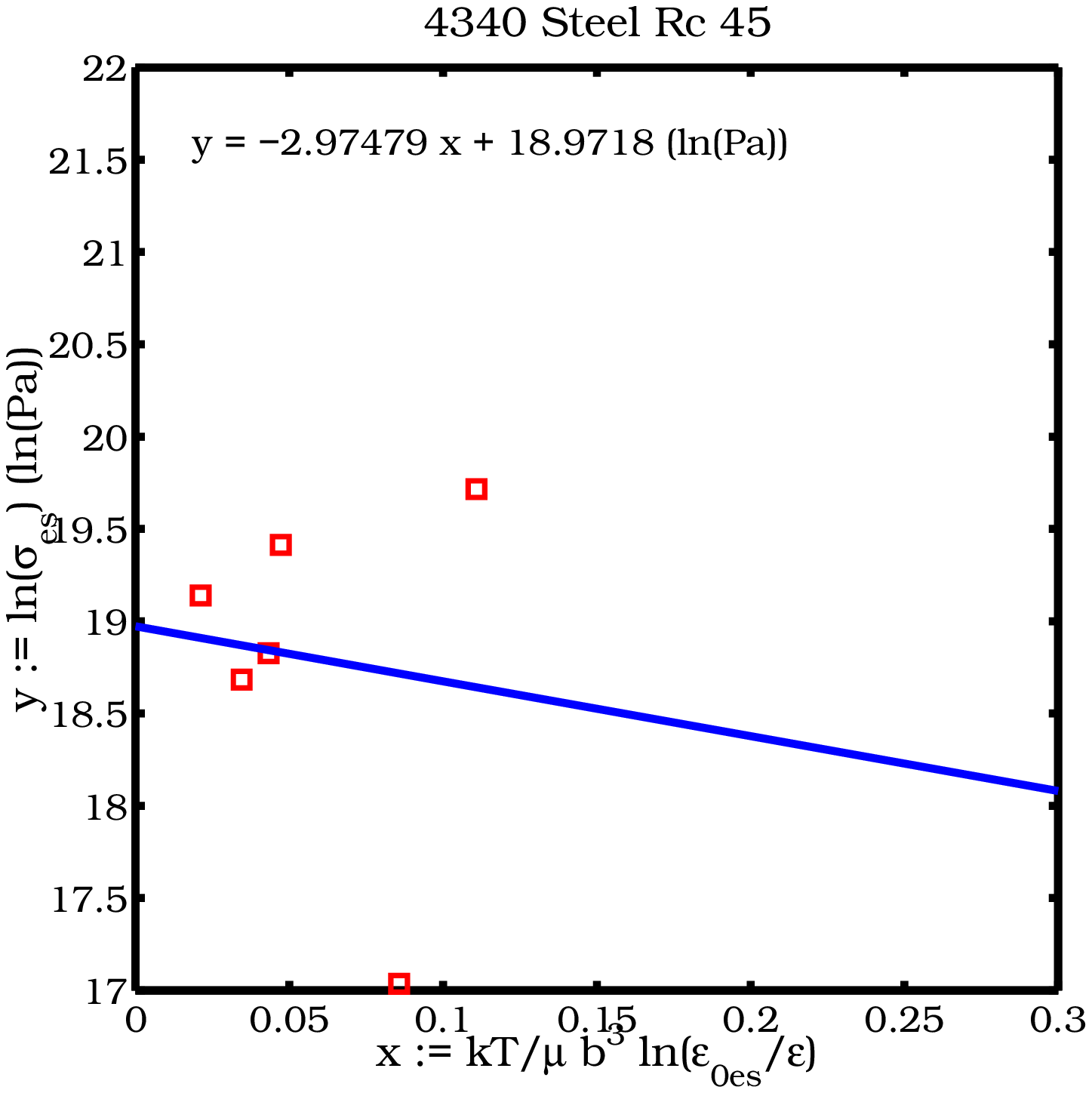}
                    \includegraphics{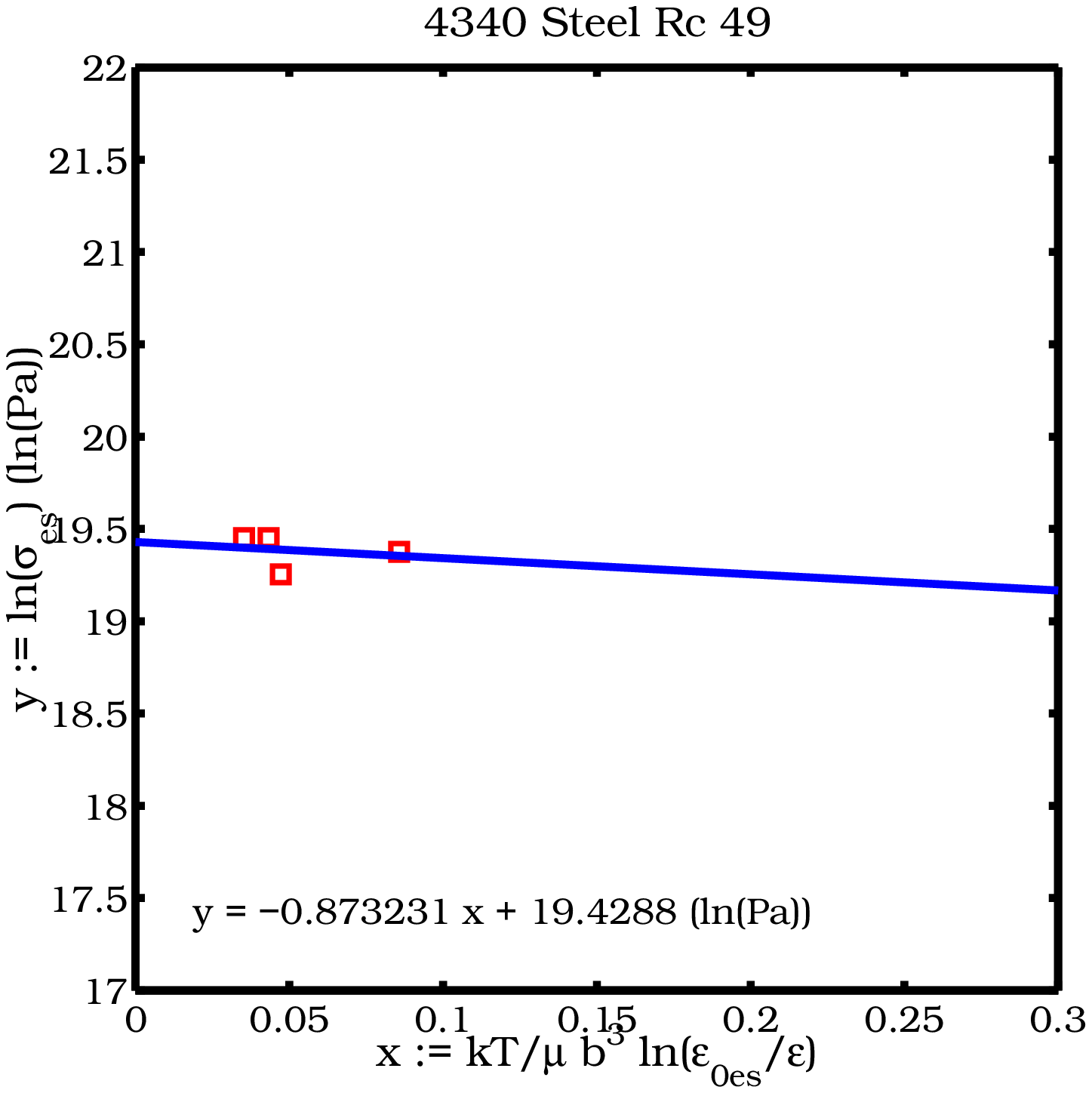}}\\
    (c) $R_c$ = 45 \hspace{2in} 
    (d) $R_c$ = 49 \\
    \caption{Fisher plots for the structure evolution dependent
             component of the MTS model for the $\alpha$ phase of
             various tempers of 4340 steel.}
    \label{fig:FisherSige}
  \end{figure}

  The raw data used to plot the Fisher plots for $\sigma_{0es}$ and 
  $g_{0es}$ are given in Appendix~\ref{app:FisherSige}.  These data
  are plotted in Figures~\ref{fig:FisherSige}(a), (b), (c), and (d).
  The straight line fit to the data for the $R_c$ 30 temper is shown 
  in Figure~\ref{fig:FisherSige}(a).   The fit to the $\alpha$ phase
  for $R_c$ 38 4340 steel is shown in Figure~\ref{fig:FisherSige}(b).
  Similar Fisher plots for $R_c$ 45 and $R_c$ 49 4340 steel are shown in
  Figures~\ref{fig:FisherSige}(c) and (d).  

  The correlation between the modified Arrhenius relation and the data 
  is quite poor.  Considering the fact that special care has been taken to 
  determine the value of $\sigma_{es}$, the poor fit appears to suggest that
  the strain dependent part of the mechanical threshold stress does not 
  follow an Arrhenius relation.  However, we do not have information on 
  the error in the experimental data and therefore cannot be confident about 
  such a conclusion.  We continue to assume, following ~\cite{Goto00} for 
  HY-100, that a modified Arrhenius type of temperature and strain rate 
  dependence is acceptable for the strain dependent part of the yield
  stress of 4340 steel.

\begin{figure}[t]
    \centering
    \scalebox{0.45}{\includegraphics{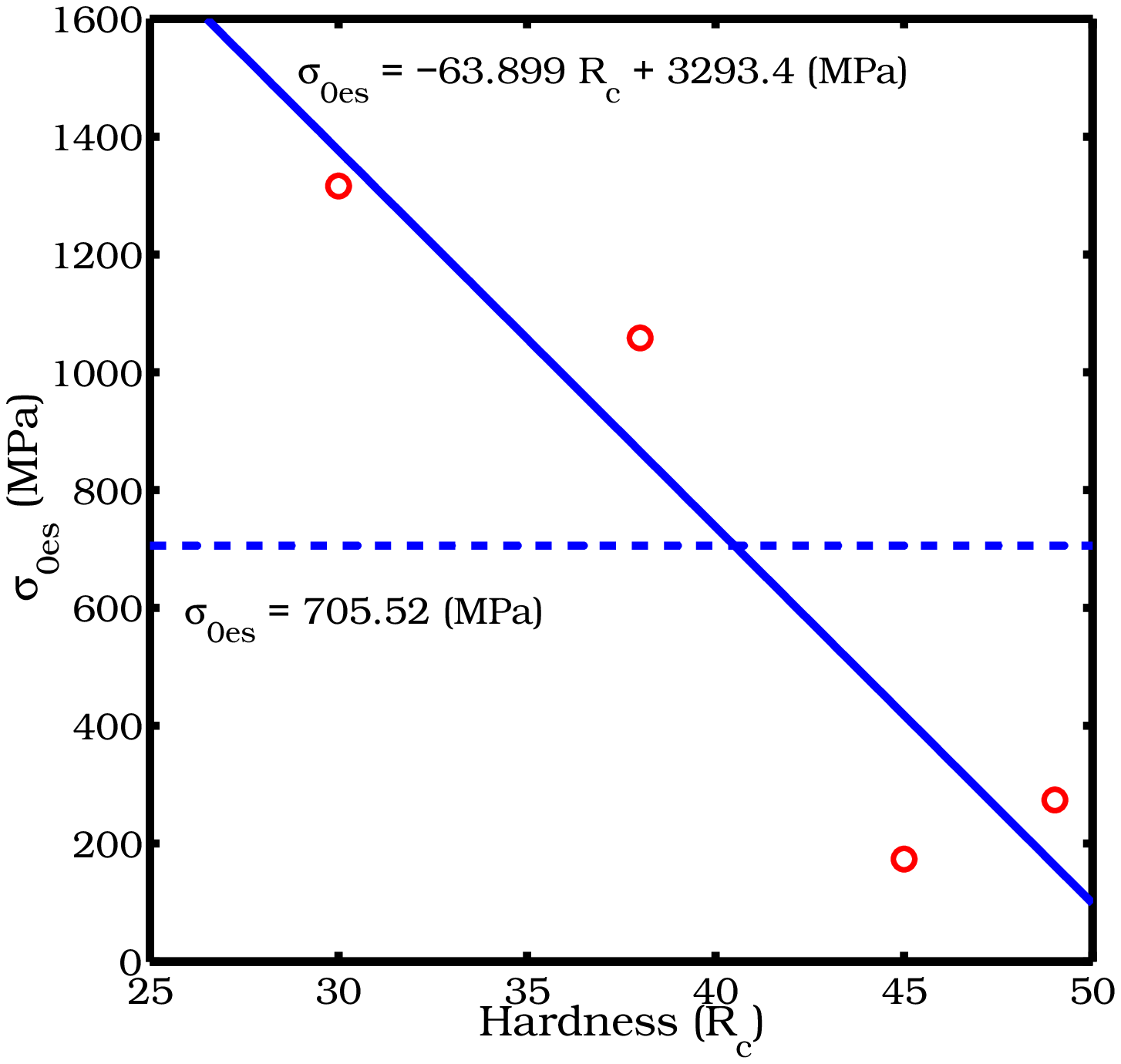}
                    \includegraphics{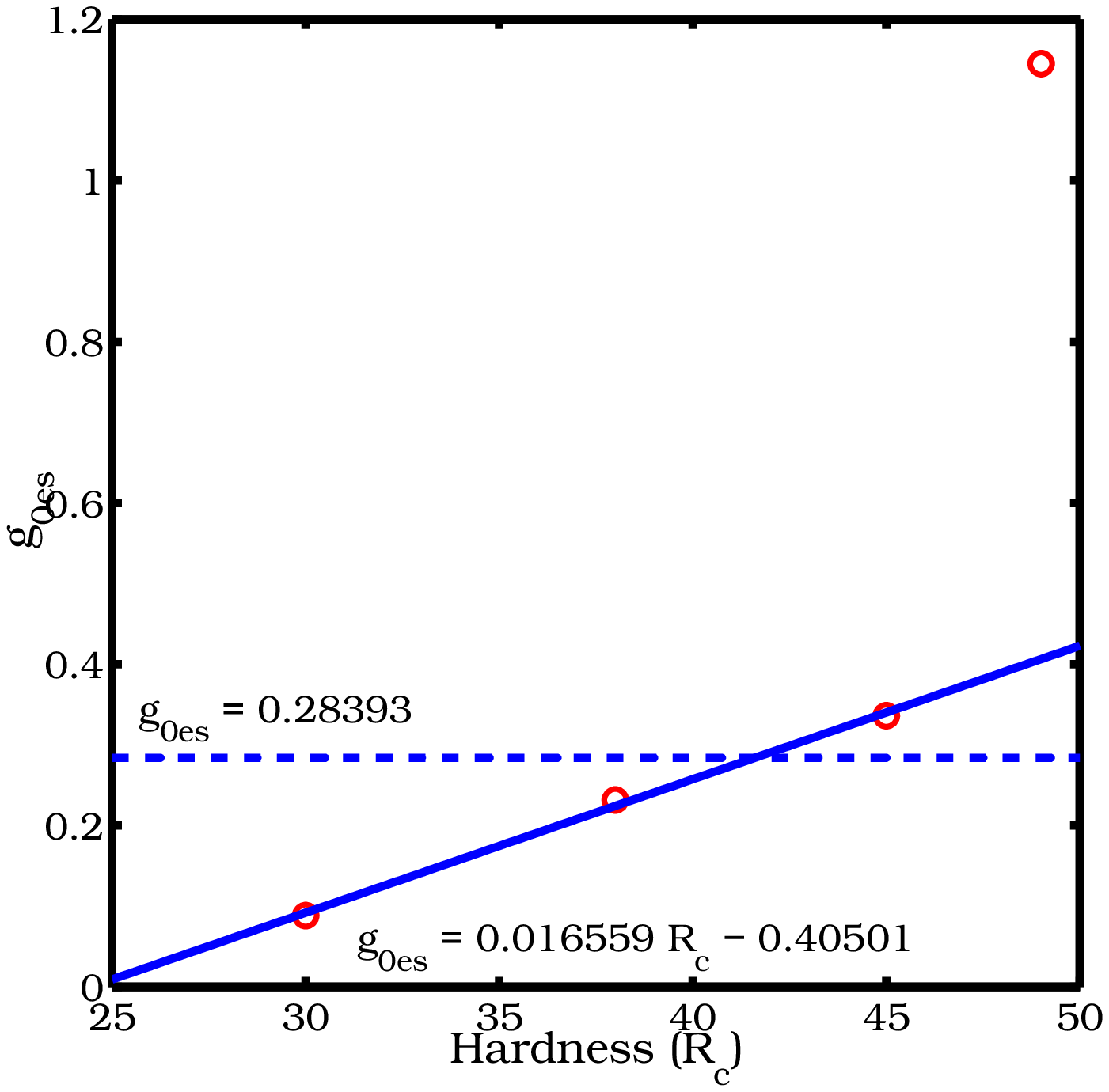}}\\
    (a) $\sigma_{0es} = - 63.9 R_c + 3293.4$ (MPa) \hspace{0.5in} 
    (b) $g_{0i} = 0.01656 R_c - 0.405$ \\
    \caption{Values of $\sigma_{0es}$ and $g_{0es}$ obtained from the 
             Fisher plots for various tempers of the $\alpha$ phase
             of 4340 steel.  The dashed lines show the median values
             of the parameters.}
    \label{fig:SigmaesGoes}
  \end{figure}

  Values of $\sigma_{0es}$ and $g_{0es}$ computed from the Fisher plots are 
  shown in Table~\ref{tab:SigmaesGoes}.  The value of the saturation stress
  decreases with increasing hardness while the normalized activation energy
  (at 0 K) increases with increasing hardness.  For intermediate tempers
  a median value of 0.284 is assumed for $g_{0es}$ and the mean value
  of 705.5 MPa is assumed for $\sigma_{0es}$.  Straight line fits
  to the data, as shown in Figures~\ref{fig:SigmaesGoes}(a) and (b), 
  could also be used to determine the values of $g_{0es}$ and $\sigma_{0es}$ 
  for intermediate tempers of 4340 steel. 

  Fits to the data for temperatures greater than 1040 K give us values
  of $\sigma_{0es}$ and $g_{0es}$ for the $\gamma$ phase of 4340 steel. 
  The values of these parameters at such high temperatures are
  $g_{0es} = $ 0.294 and $\sigma_{0es} = $ 478.36 MPa.
  \begin{table}[t]
    \centering
    \caption{Values of $\sigma_{0es}$ and $g_{0es}$ for four tempers of 
             4340 steel.}
    \begin{tabular}{ccc}
       \hline
       Hardness ($R_c$) & $\sigma_{0es}$ (MPa) & $g_{0es}$ \\
       \hline
       30 & 1316.1 & 0.088 \\
       38 & 1058.4 & 0.232 \\
       45 &  173.5 & 0.336\\
       49 &  274.9 & 1.245 \\
       \hline
    \end{tabular}
    \label{tab:SigmaesGoes}
  \end{table}

\subsection{Determination of hardening rate $\theta$}
  The modified Voce rule for the hardening rate ($\theta$)
  (equation (\ref{eq:theta})) is purely empirical.  To determine the 
  temperature and strain rate dependence of $\theta$, we plot the variation 
  of $\theta$ versus the normalized structure evolution stress assuming 
  hyperbolic tangent dependence of the rate of hardening on the 
  mechanical threshold stress.  We assume that $\alpha = 3$.

\begin{figure}[p]
    \centering
    \scalebox{0.45}{\includegraphics{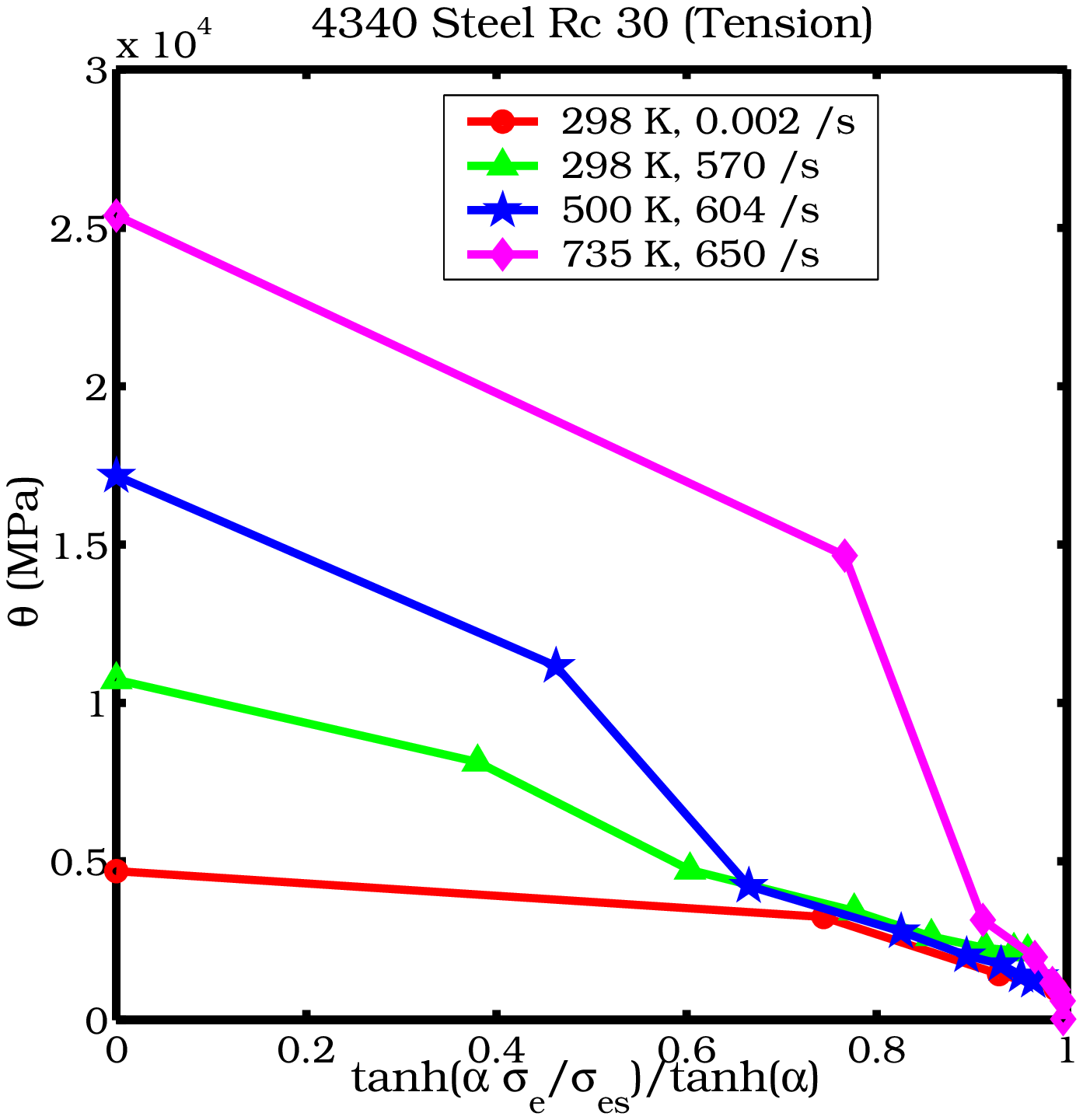}
                    \includegraphics{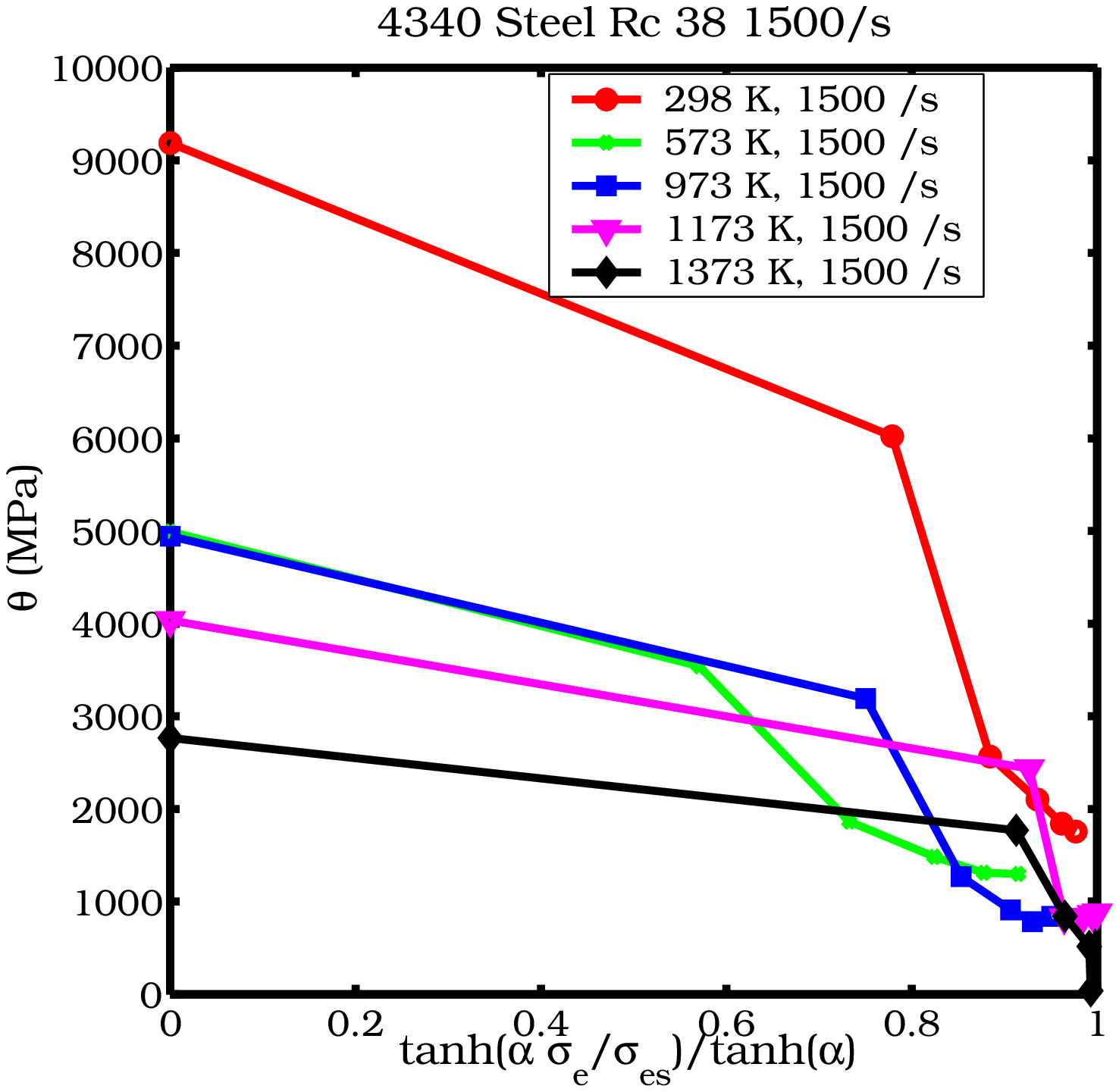}}\\
    (a) $R_c$ 30, Tension \hspace{1in} 
    (b) $R_c$ 38, 1500/s \\
    \vspace{12pt}
    \scalebox{0.45}{\includegraphics{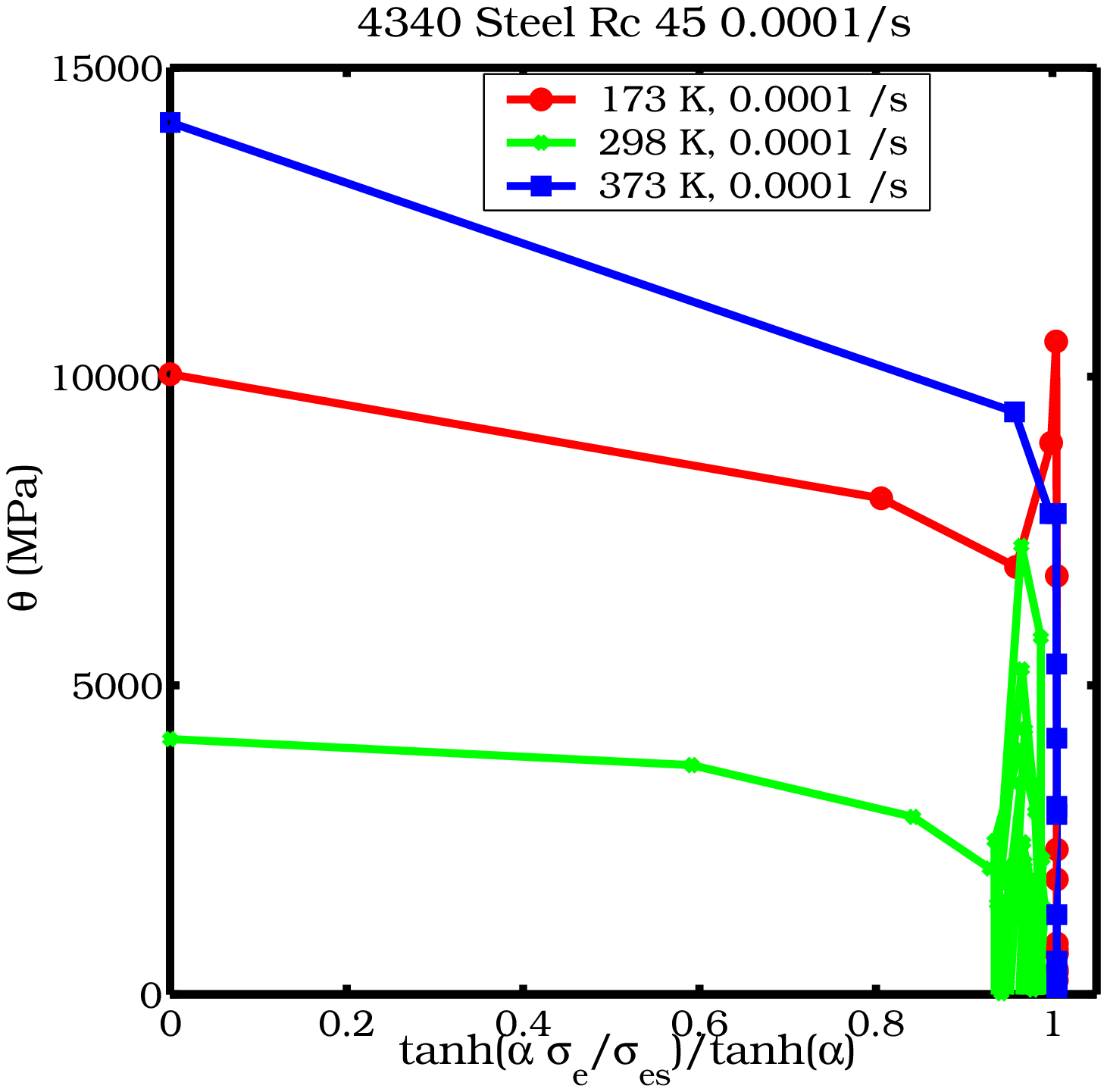}
                    \includegraphics{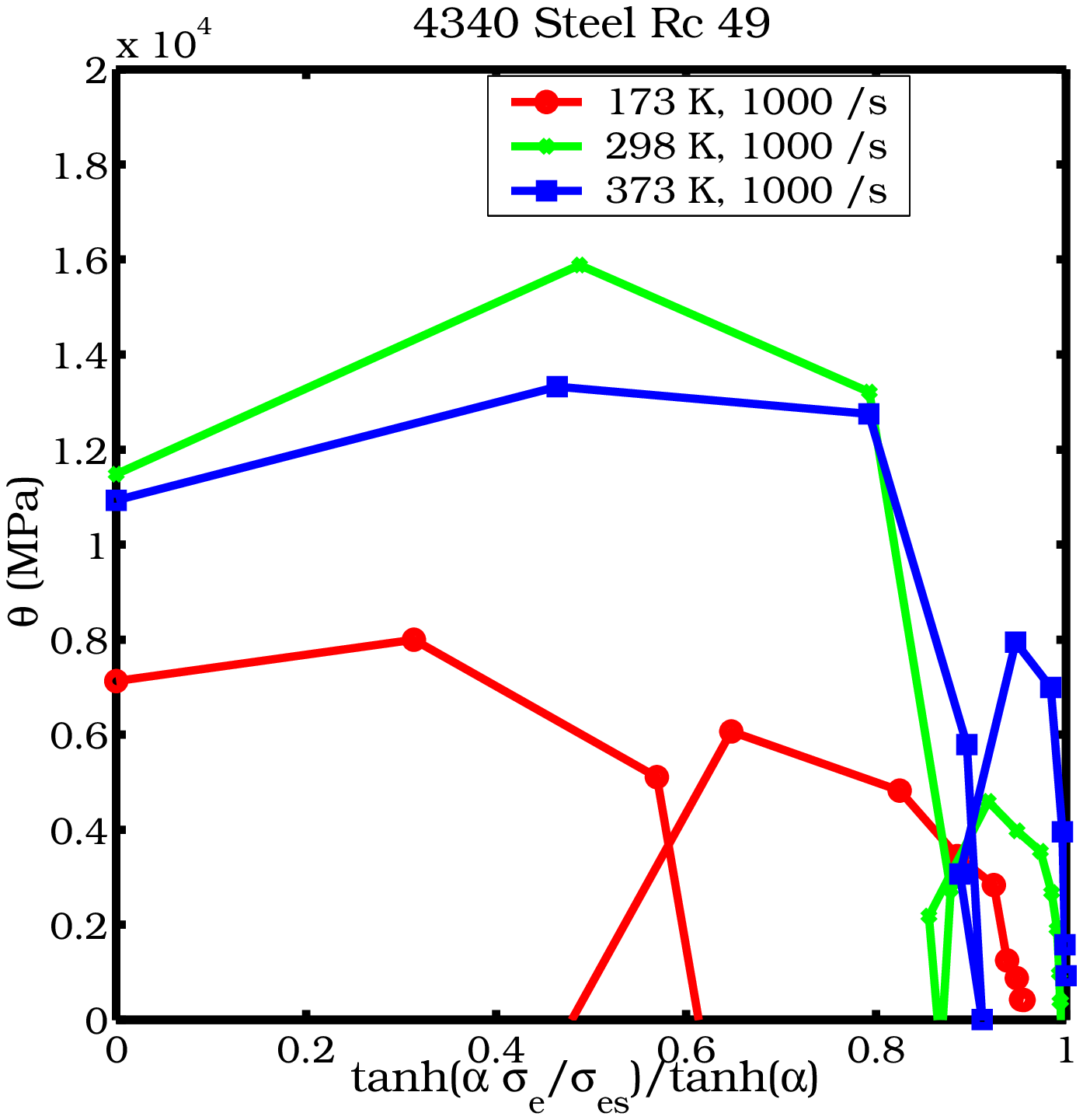}}\\
    (c) $R_c$ 45, 0.0001/s \hspace{1in} 
    (d) $R_c$ 49, 1000/s \\
    \caption{Plots used to determine $\theta_0$ as a function of temperature
             and strain rate. $\alpha$ = 3.}
    \label{fig:thetavsFx}
  \end{figure}

  Figures~\ref{fig:thetavsFx}(a), (b), (c), and (d) show some representative 
  plots of the variation 
  of $\theta$ with $F := \tanh(\alpha\sigma_e/\sigma_{es})/\tanh(\alpha)$.
  As the plots show, the value of $\theta_1$ (the value of $\theta$
  at $F = 1$) can be assumed to be zero for most of the data.  

  It is observed from Figure~\ref{fig:thetavsFx}(a) that there
  is a strong strain rate dependence of $\theta$ that appears to override
  the expected decrease with increase in temperature for the $R_c$ 30
  temper of 4340 steel.  It can also been seen that $\theta$ is almost
  constant at 298 K and 0.002/s strain rate reflecting linear 
  hardening.  However, the hyperbolic tangent rule appears to be a good
  approximation at higher temperatures and strain rates.  
  
  The plot for $R_c$ 38 4340 steel (Figure~\ref{fig:thetavsFx}(b)) shows
  a strong temperature dependence of $\theta$ with the hardening rate
  decreasing with increasing temperature.  The same behavior is observed
  for all high strain rate data.  However, for the data at a strain 
  rate of 0.0002/s, there is an increase in $\theta$ with increasing
  temperature.  Figures~\ref{fig:thetavsFx}(c) and (d) also show an increase
  in $\theta$ with temperature.  These reflect an anomaly in the 
  constitutive behavior of 4340 steel for relatively low temperatures 
  (below 400 K) (\citet{Tanimura86}) that cannot be modeled 
  continuously using an Arrhenius law and needs to be characterized in
  more detail.

\begin{figure}[p]
    \centering
    \scalebox{0.40}{\includegraphics{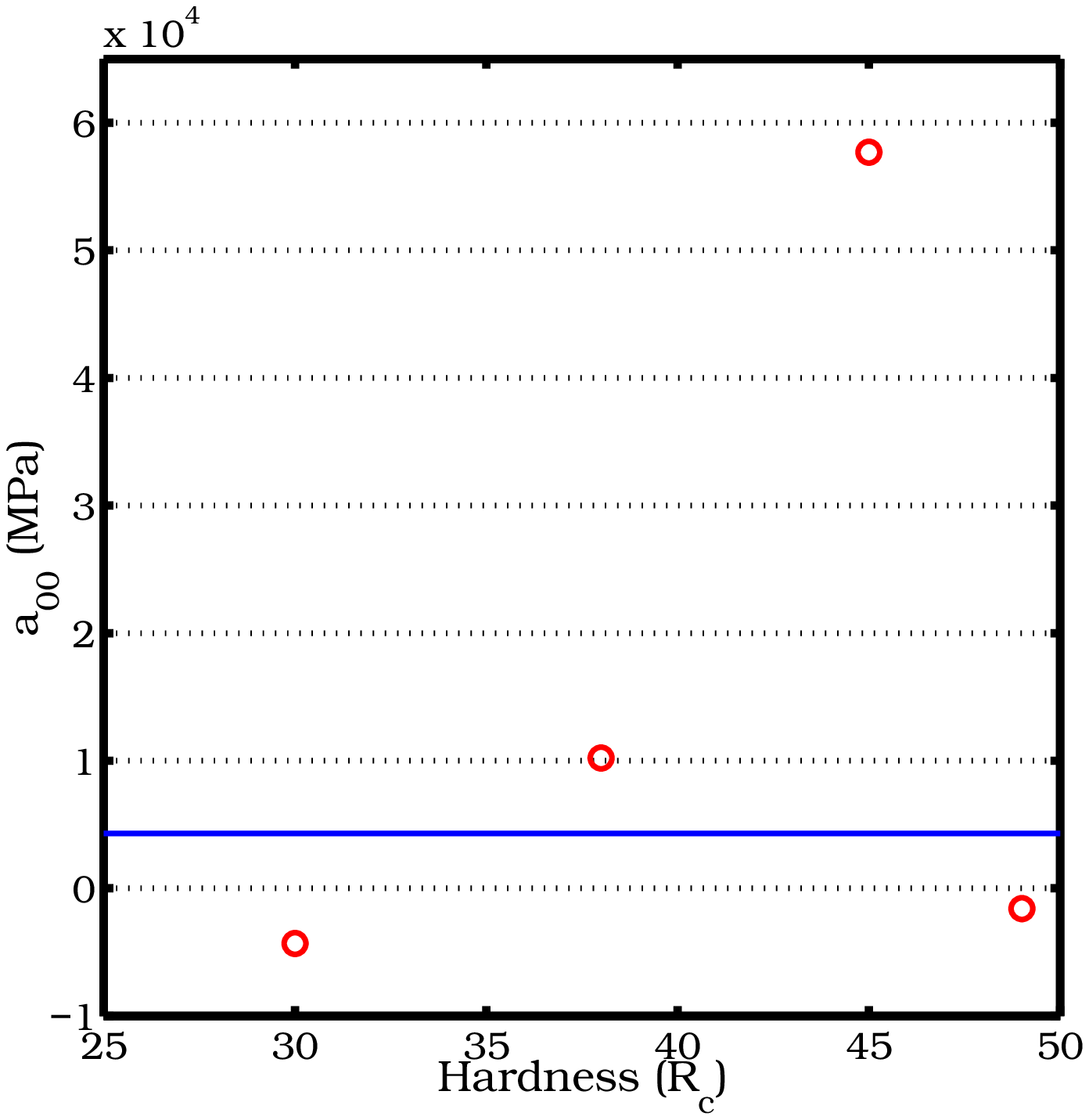}
                    \includegraphics{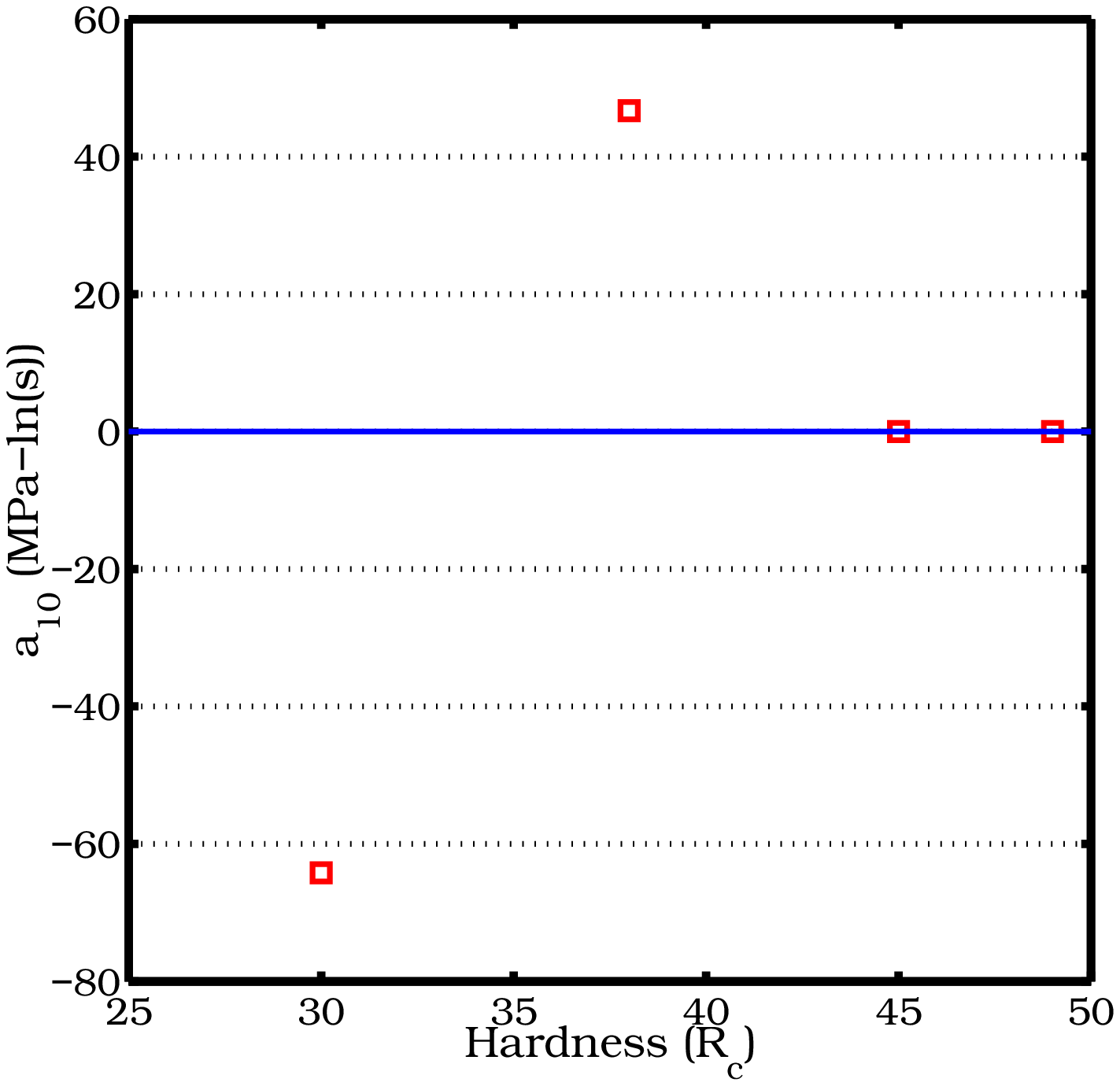}}\\
    (a) $a_{00}$ vs. $R_c$ \hspace{1in}
    (b) $a_{10}$ vs. $R_c$ \\
    \vspace{12pt}
    \scalebox{0.40}{\includegraphics{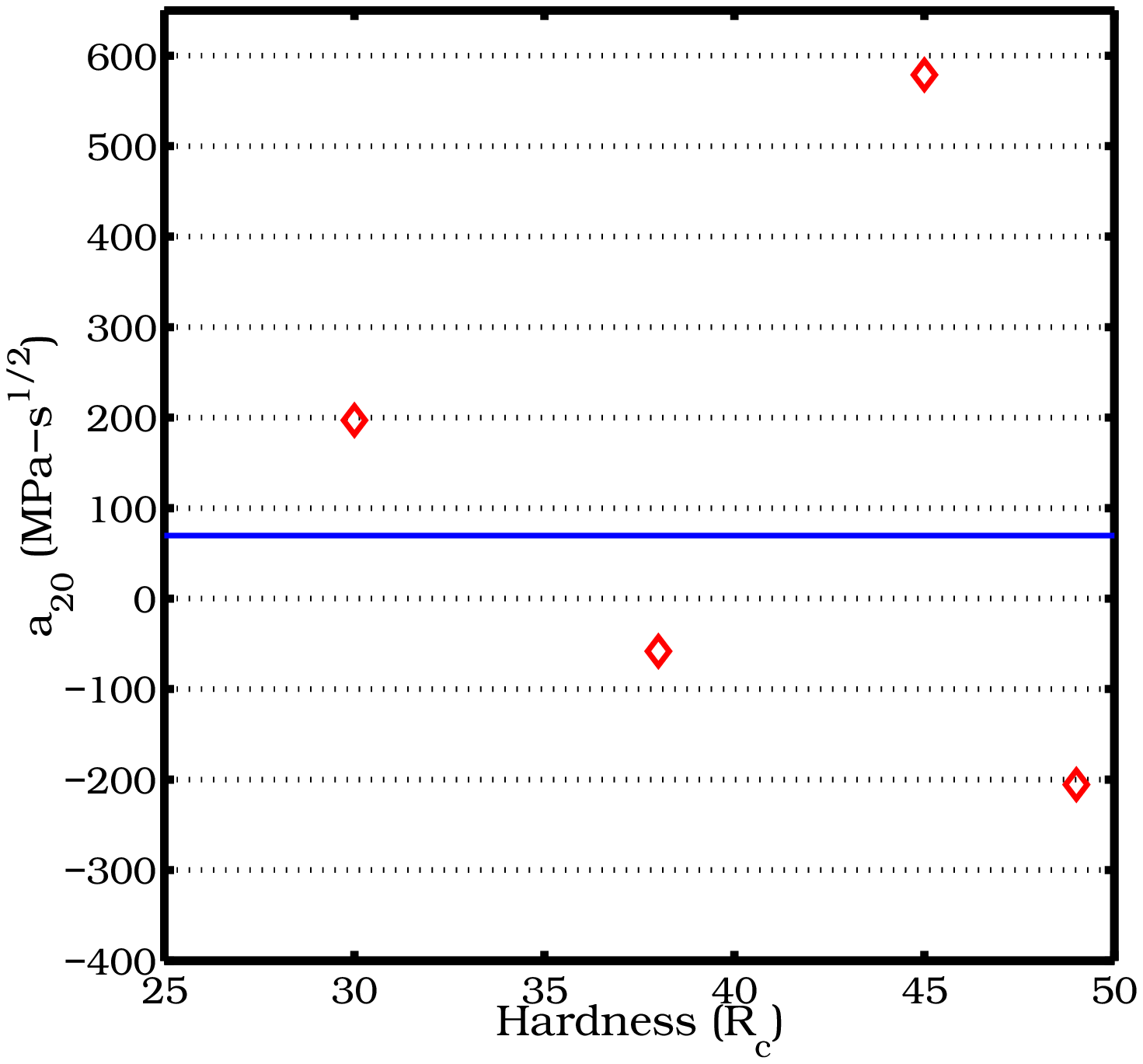}
                    \includegraphics{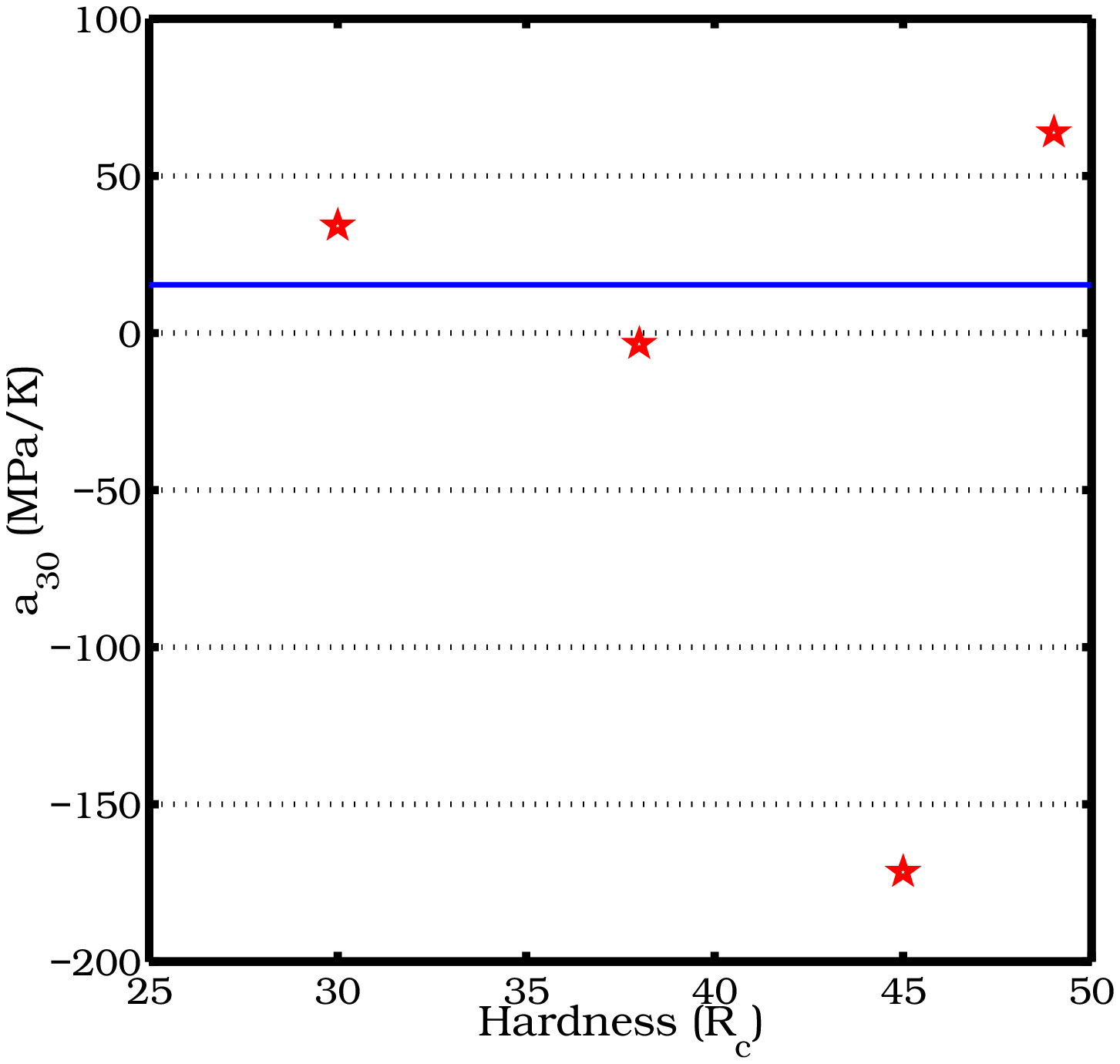}}\\
    (c) $a_{20}$ vs. $R_c$ \hspace{1in}
    (d) $a_{30}$ vs. $R_c$ \\
    \caption{Variation of the constants fit to the hardening rate 
             equation for various tempers of 4340 steel.}
    \label{fig:avsRc}
  \end{figure}

  Fits to the experimental data of the form shown in equation 
  (\ref{eq:theta_0}) have been attempted.  The resulting values of 
  $a_{00}$, $a_{01}$, $a_{02}$, and $a_{03}$ are plotted as functions
  of $R_c$ in Figures~\ref{fig:avsRc}(a), (b), (c), and (d), respectively. 
  The points show the values of the constants for individual tempers
  while the solid line shows the median value.
  These figures show that the constants vary considerably and also change
  sign for different tempers.  On average, the strain rate dependence is
  small for all the tempers but significant.  The $R_c$ 30 and $R_c$ 45
  data points in Figure~\ref{fig:avsRc}(d) reflect an increase in
  hardening rate with temperature that is nonphysical at high temperatures.
  
\begin{figure}[t]
    \centering
    \scalebox{0.55}{\includegraphics{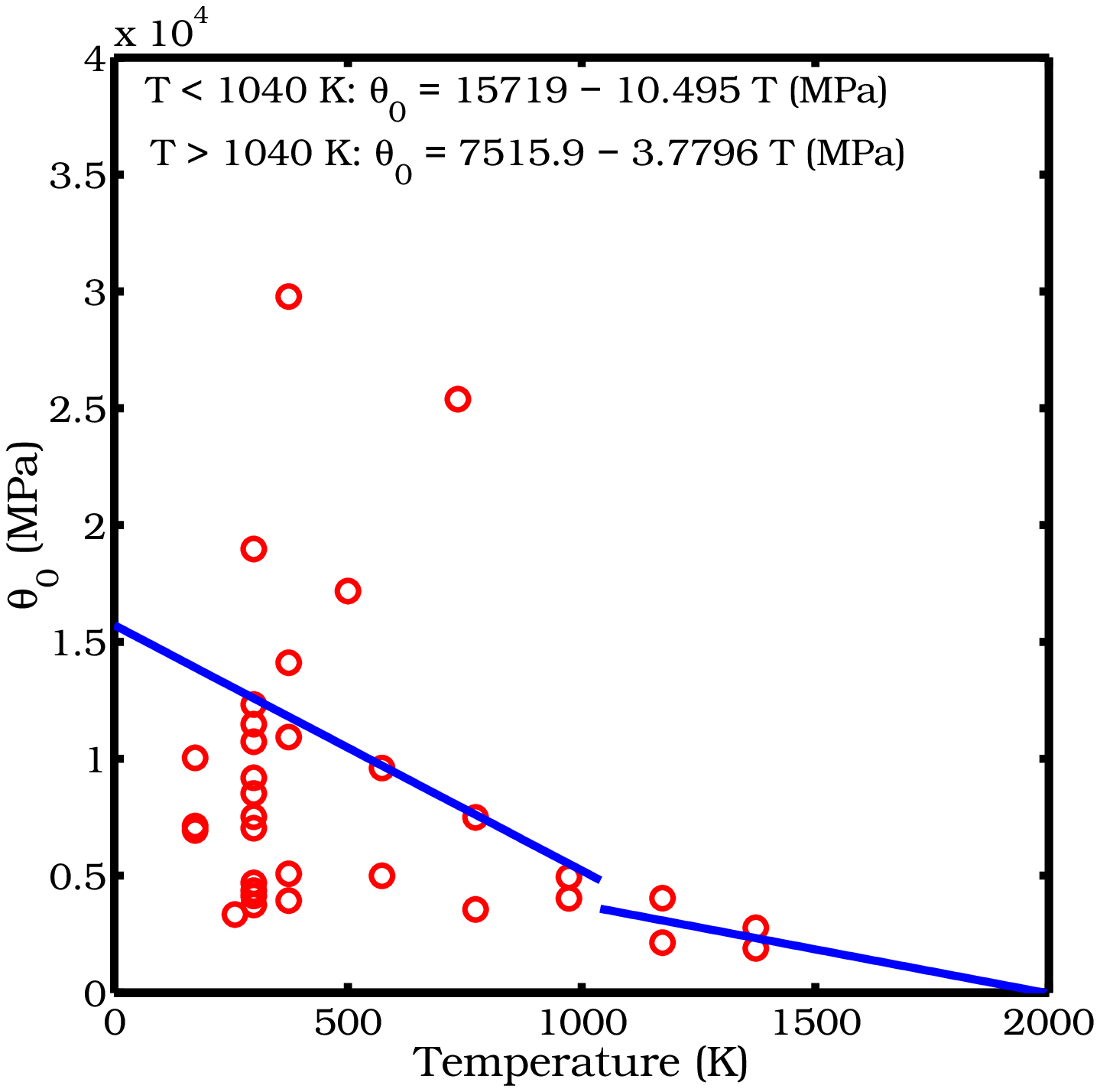}}
    \caption{Variation of $\theta_0$ with temperature.}
    \label{fig:theta0vsT}
  \end{figure}

  Instead of using these fits to the experimental data, we have decided
  to ignore the strain rate dependence of the hardening rate and fit 
  a curve to all the data taking only temperature dependence into 
  account (as shown in Figure~\ref{fig:theta0vsT}).  Distinctions have
  not been made between various tempers of 4340 steel to determine this
  average hardening rate.  However, we do divide the data into two parts
  based on the $\alpha$-$\gamma$ phase transition temperature.
  
   The resulting equations for $\theta_0$ as functions of temperature are
   \begin{equation} 
     \theta_0 = 
     \begin{cases}
        15719 - 10.495~T ~\text{(MPa)} & \text{for}~ T < 1040 K \\
        7516 - 3.7796~T ~\text{(MPa)} & \text{for}~ T > 1040 K
     \end{cases}
   \end{equation} 
   This completes the determination of the parameters for the MTS model.
   
\section{Comparison of MTS model predictions and experimental data} 
         \label{sec:MTSComp}
  The performance of the MTS model for 4340 steel is compared to experimental
  data in this section.  In the figures that follow, the MTS predictions are
  shown as dotted lines while the experimental data are shown as solid 
  lines with symbols indicting the conditions of the test.  Isothermal
  conditions have been assumed for strain rates less than 500/s and 
  adiabatic heating is assumed to occurs at higher strain rates.

\begin{figure}[p]
    \centering
    \scalebox{0.50}{\includegraphics{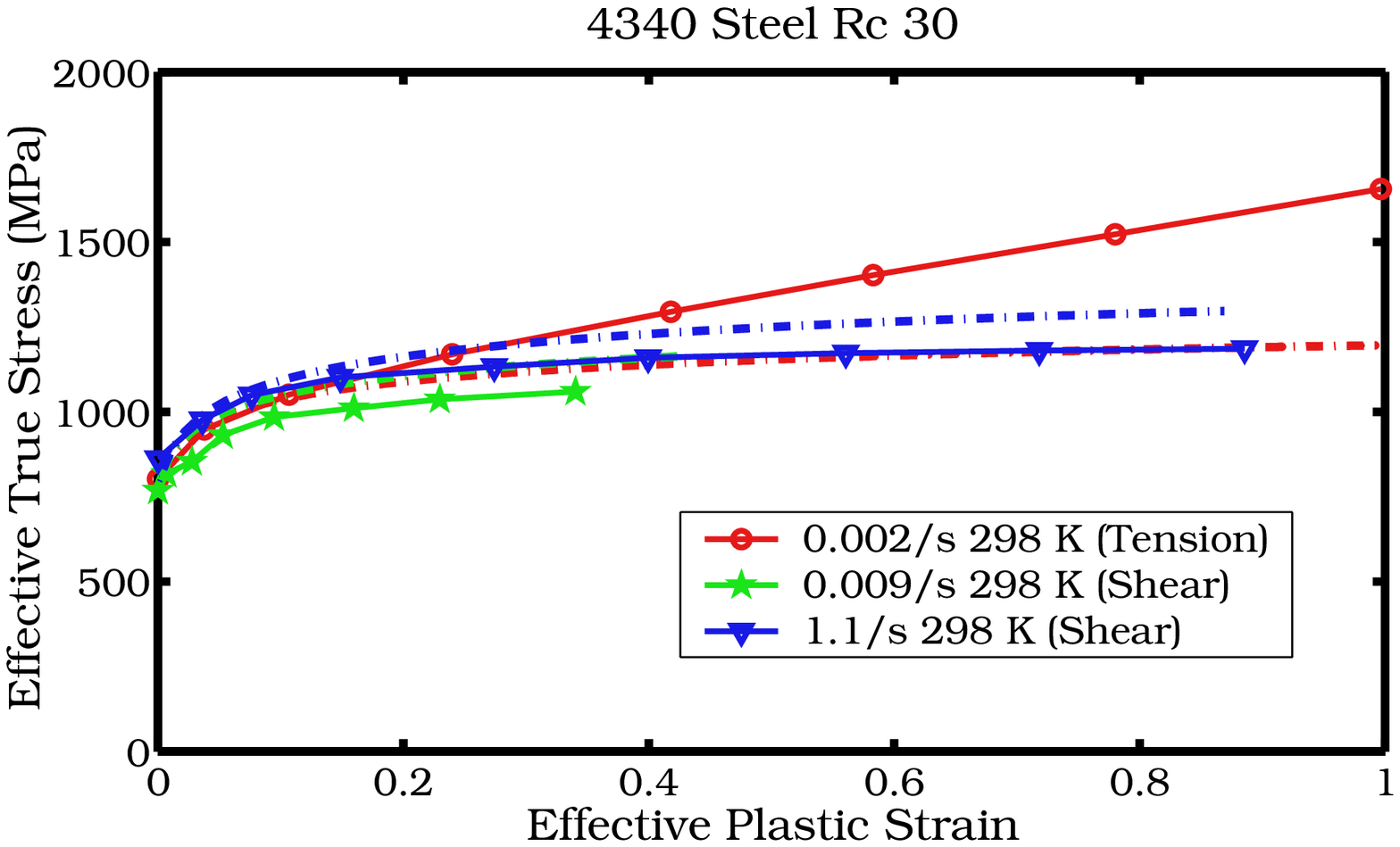}}\\
    (a) Low strain rates. \\
    \vspace{12pt}
    \scalebox{0.50}{\includegraphics{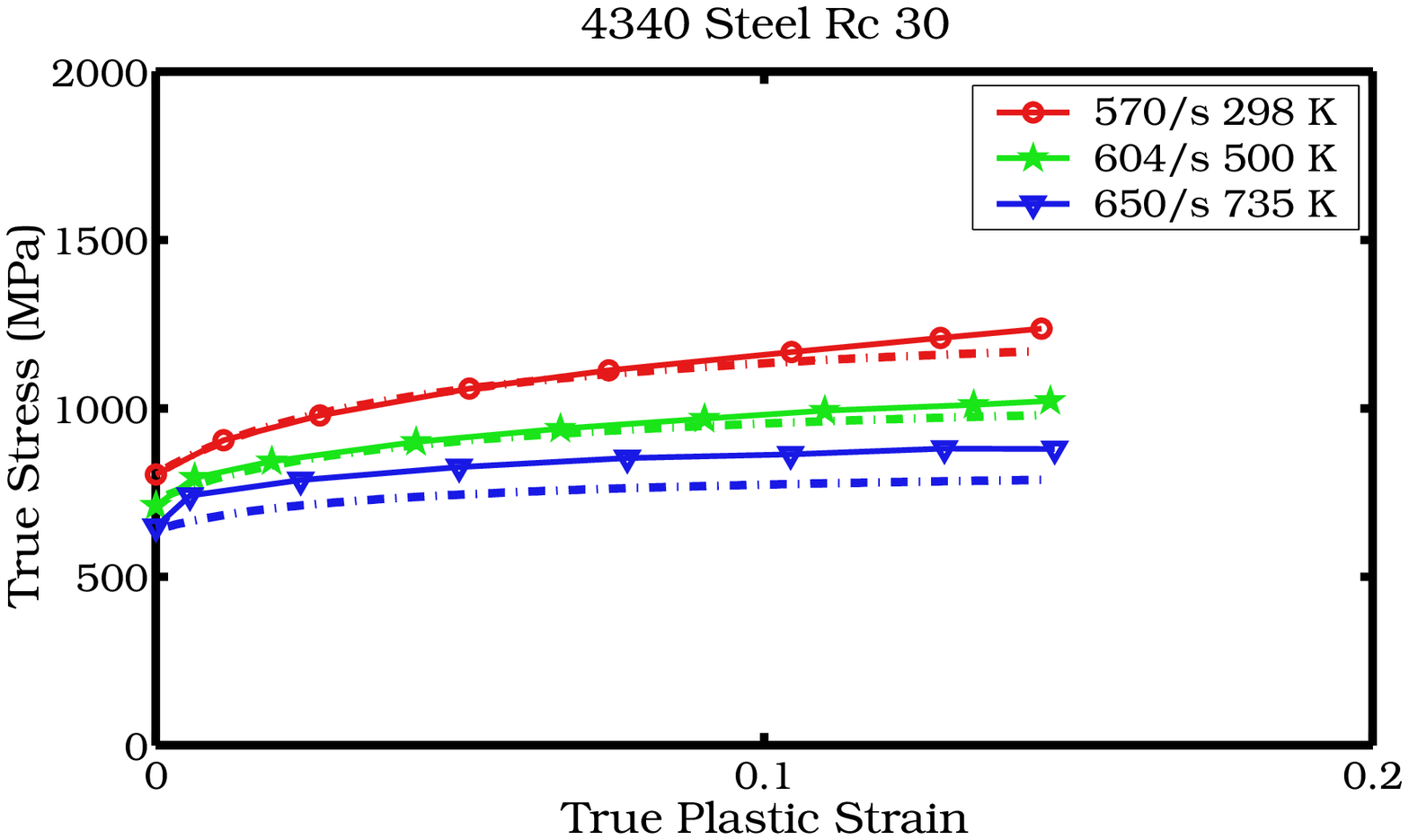}}\\
    (b) High strain rates. 
    \caption{Comparison of MTS prediction with experimental data
             from \citet{Johnson85} for the $R_c$ 30 temper of 
             4340 steel.}
    \label{fig:validRc30}
  \end{figure}

  Figure~\ref{fig:validRc30}(a) shows the low strain rate experimental
  data and the corresponding MTS predictions for the $R_c$ 30 temper
  of 4340 steel.  Comparisons for moderately high strain rates and high 
  temperatures for the $R_c$ 30 temper are shown in 
  Figure~\ref{fig:validRc30}(b).  The model matches the experimental
  curves quite well for low strain rates (keeping in mind the difference
  between the stress-strain curves in tension and in shear).  The high
  strain rate curves are also accurately reproduced though there is 
  some error in the initial hardening modulus for the 650 /s and 735 K
  case.  This error can be eliminated if the effect of strain rate is
  included in the expression for $\theta_0$.  The maximum modeling 
  error for this temper varies between 5\% to 10\%.

\begin{figure}[p]
    \centering
    \scalebox{0.50}{\includegraphics{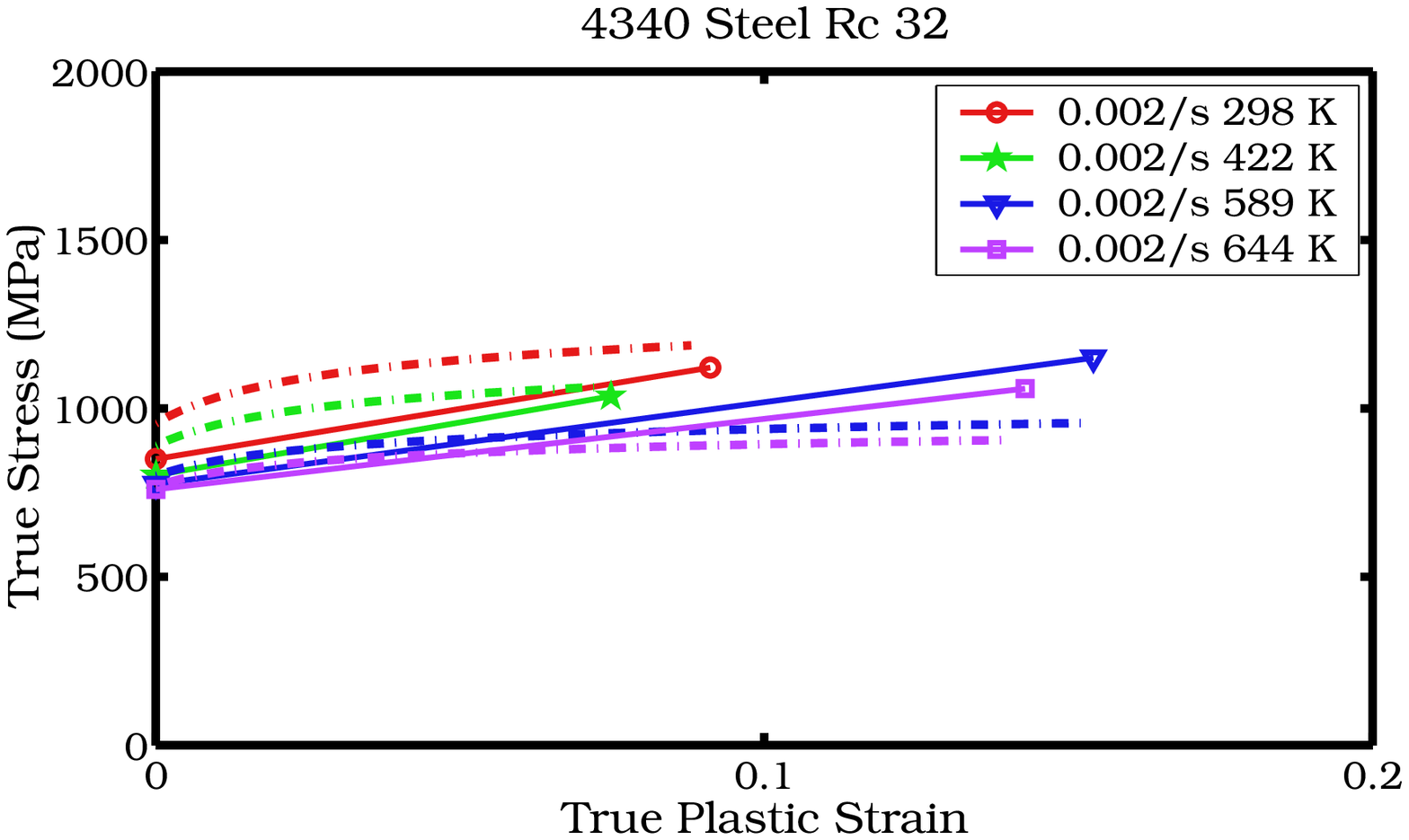}}
    \caption{Comparison of MTS prediction with experimental data
             from \citet{ASMH96} for the $R_c$ 32 temper of 
             4340 steel.}
    \label{fig:validRc32}
  \end{figure}

  We have not used the $R_c$ 32 experimental data to fit the MTS model
  parameters.  As a check of the appropriateness of the relation between
  the parameters and the $R_c$ hardness number, we have plotted the
  MTS predictions versus the experimental data for this temper in 
  Figure~\ref{fig:validRc32}.  Our model predicts a stronger temperature
  dependence for this temper than the experimental data.  However, the
  initial high temperature yield stress is reproduced quite accurately
  while the ultimate tensile stress is reproduced well for the lower
  temperatures.

\begin{figure}[t]
    \centering
    \scalebox{0.50}{\includegraphics{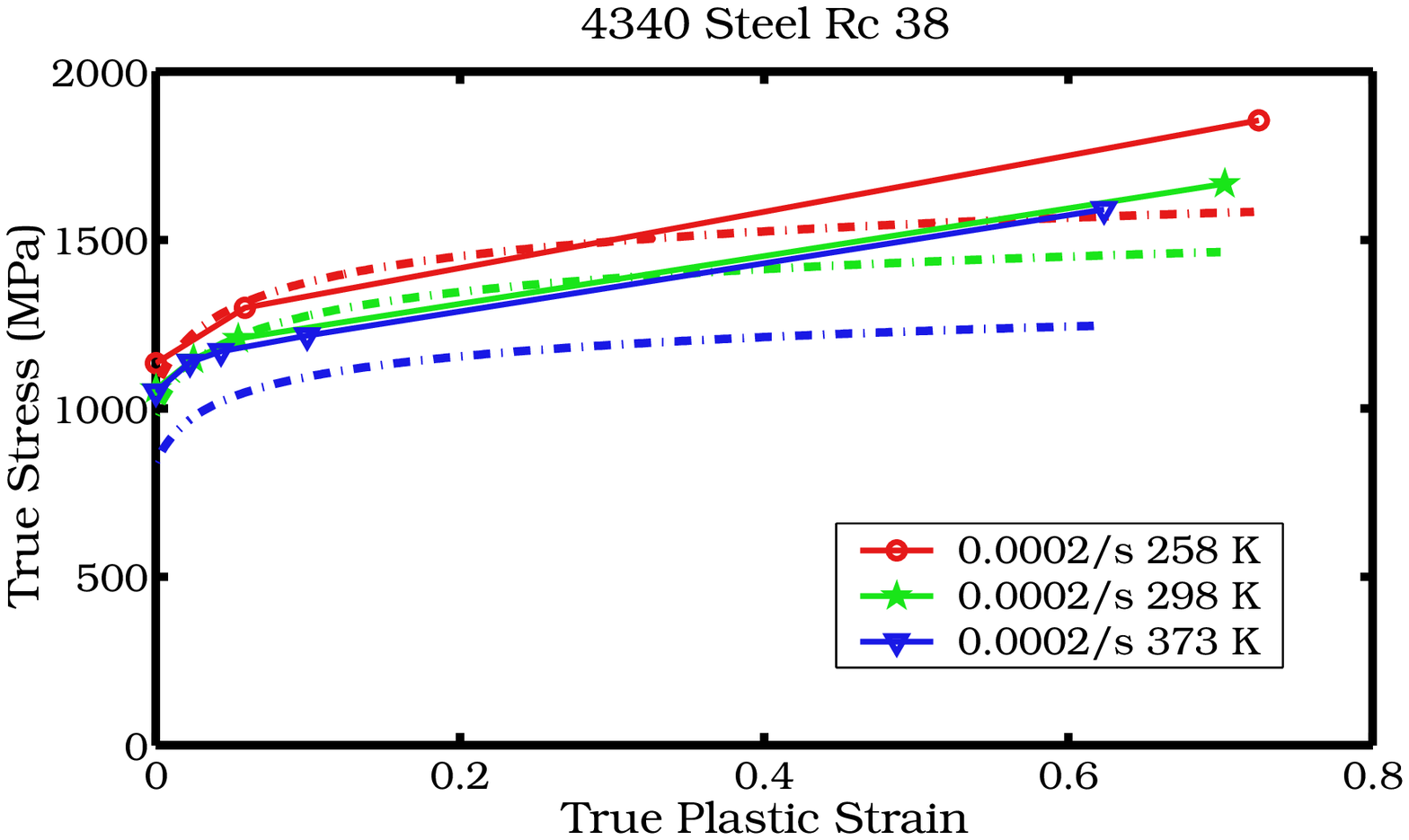}}
    \caption{Comparison of MTS prediction with experimental data
             from \citet{Larson61} for the $R_c$ 38 temper of 
             4340 steel at 0.0002/s strain rate.}
    \label{fig:validRc38Lo}
  \end{figure}

\begin{figure}[p]
    \centering
    \scalebox{0.50}{\includegraphics{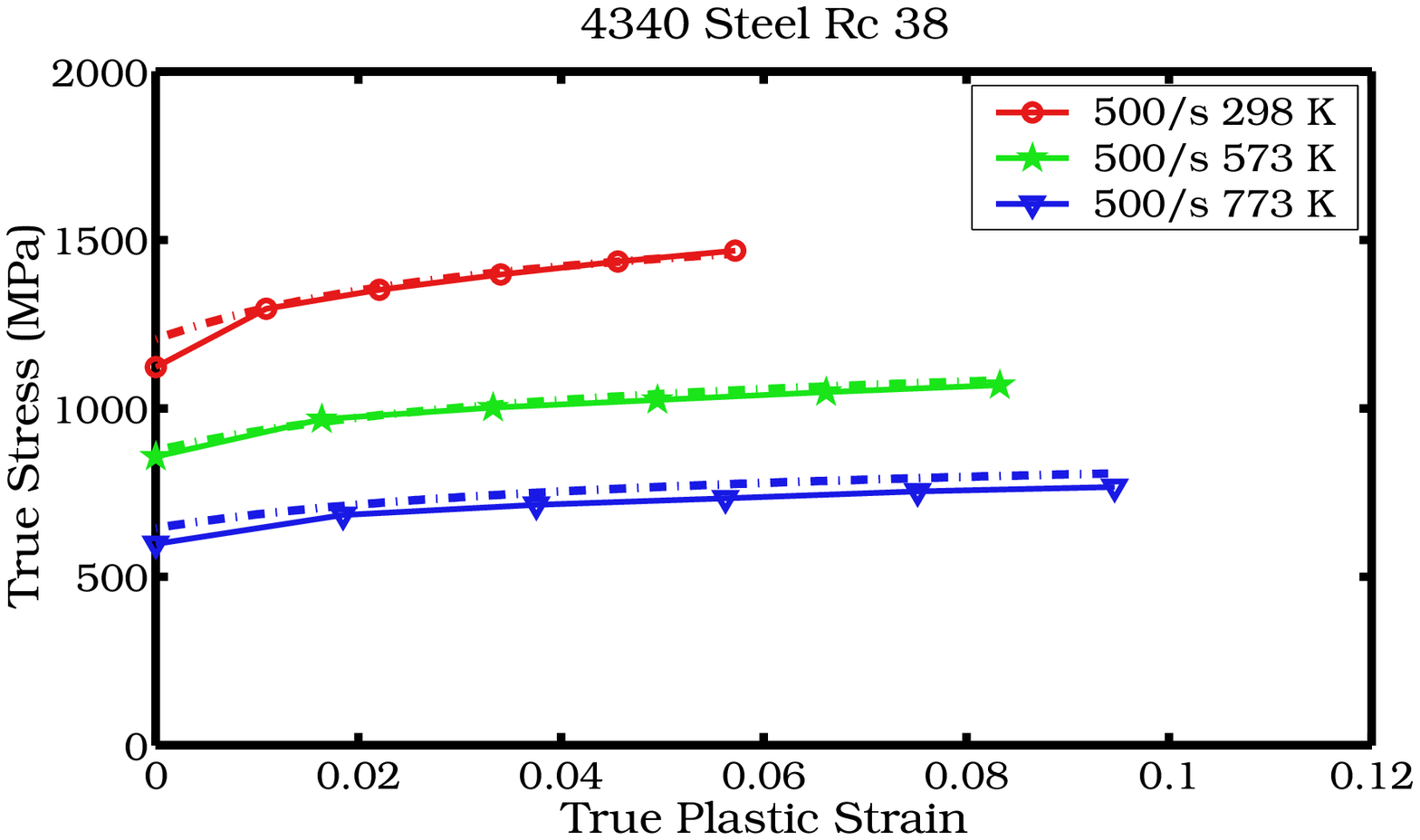}} \\
    (a) Strain Rate = 500 /s \\
    \vspace{12pt}
    \scalebox{0.50}{\includegraphics{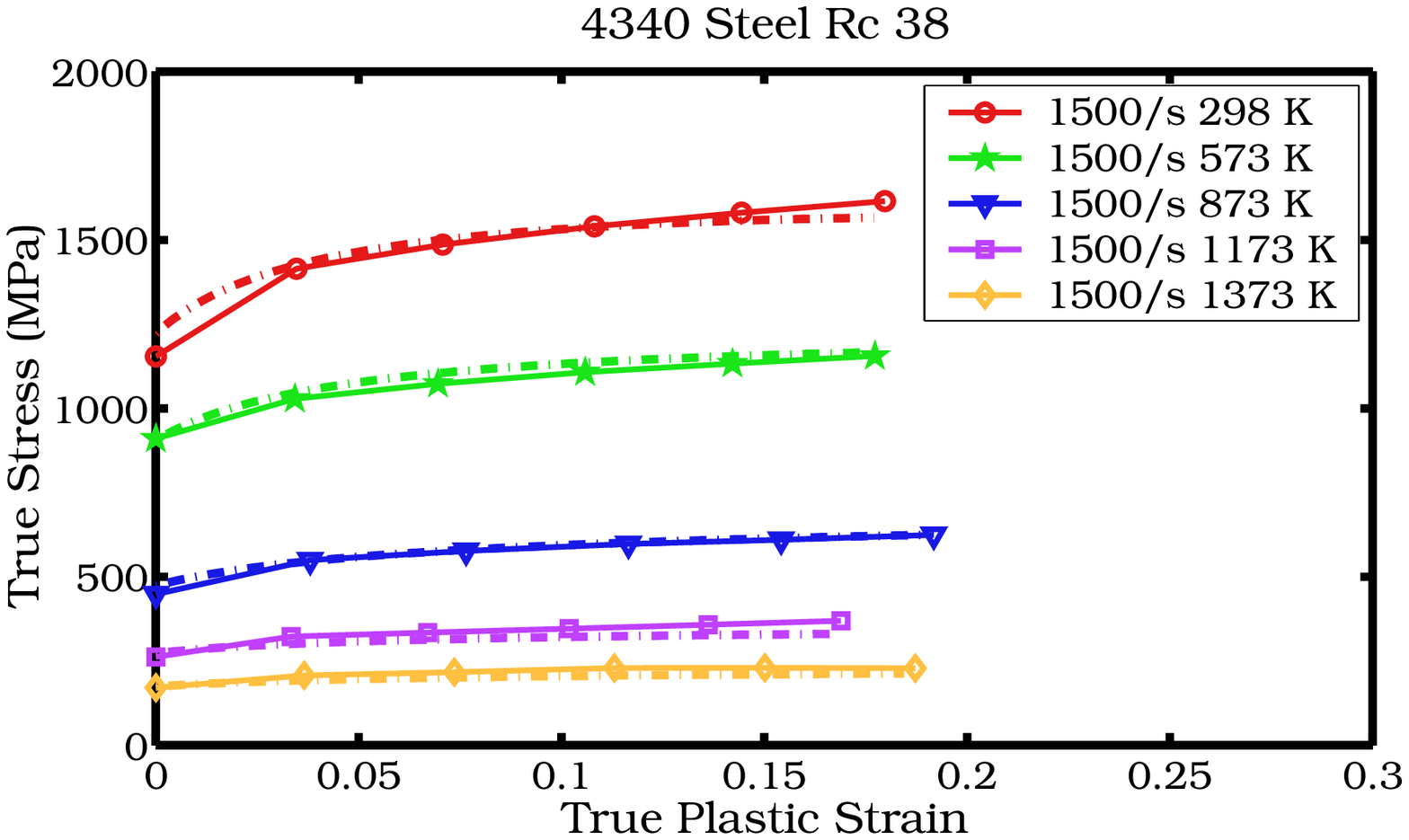}} \\
    (b) Strain Rate = 1500 /s \\
    \vspace{12pt}
    \scalebox{0.50}{\includegraphics{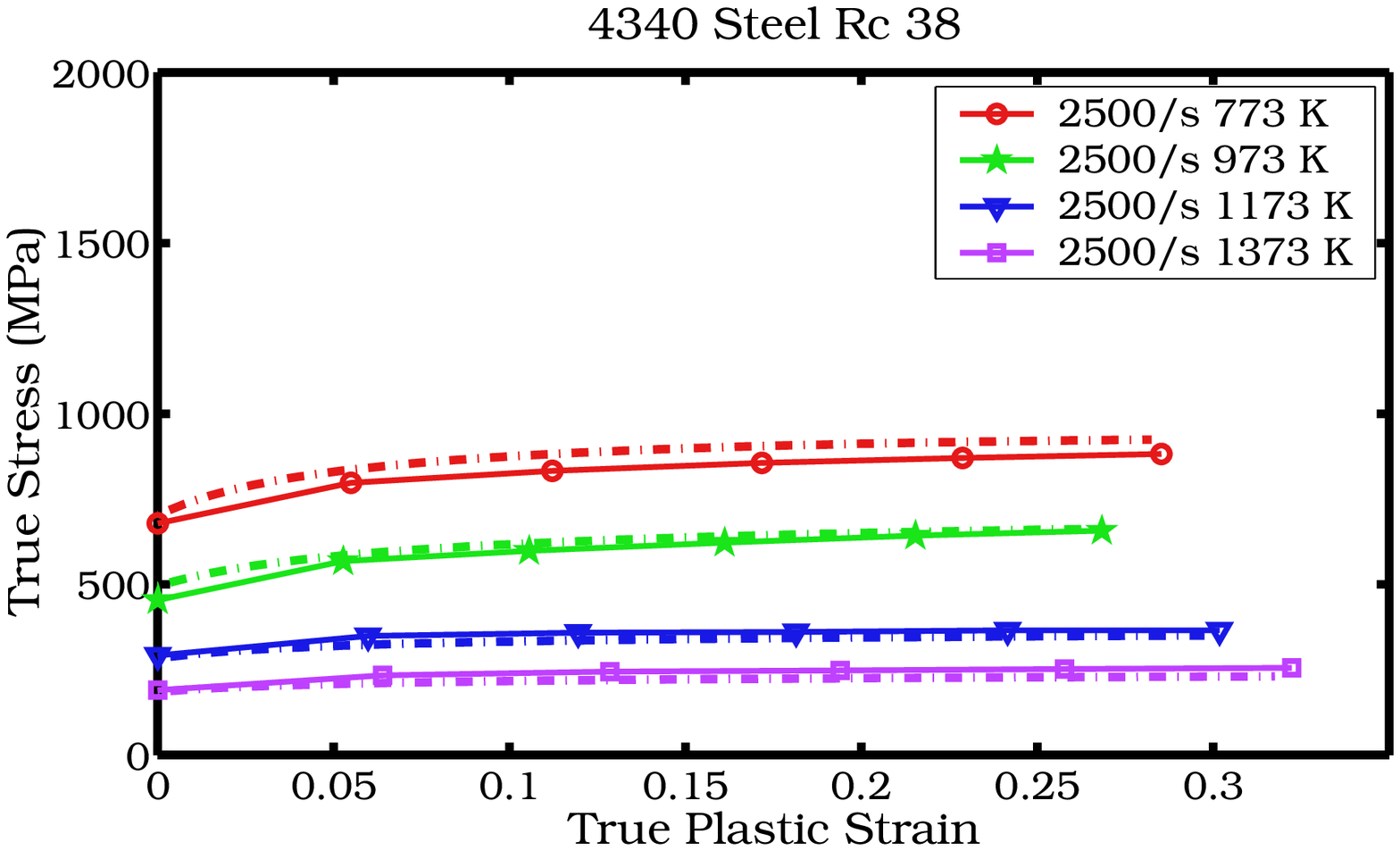}} \\
    (c) Strain Rate = 2500 /s 
    \caption{Comparison of MTS prediction with experimental data
             from \citet{Lee97} for the $R_c$ 38 temper of 
             4340 steel at high strain rates.}
    \label{fig:validRc38Hi}
  \end{figure}

  The low strain rate stress-strain curves for $R_c$ 38 4340 steel
  are shown in Figure~\ref{fig:validRc38Lo}.  High strain rate 
  stress-strain curves for the $R_c$ 38 temper are shown in 
  Figures~\ref{fig:validRc38Hi}(a), (b), and (c).  The saturation
  stress predicted at low strain rates is around 20\% smaller than
  the observed values at large strains.  The anomaly at 373 K is 
  not modeled accurately by the MTS parameters used.  On the other
  hand, the high strain rate data are reproduced quite accurately
  by the MTS model with a modeling error of around 5\% for all 
  temperatures.
  
\begin{figure}[t]
    \centering
    \scalebox{0.50}{\includegraphics{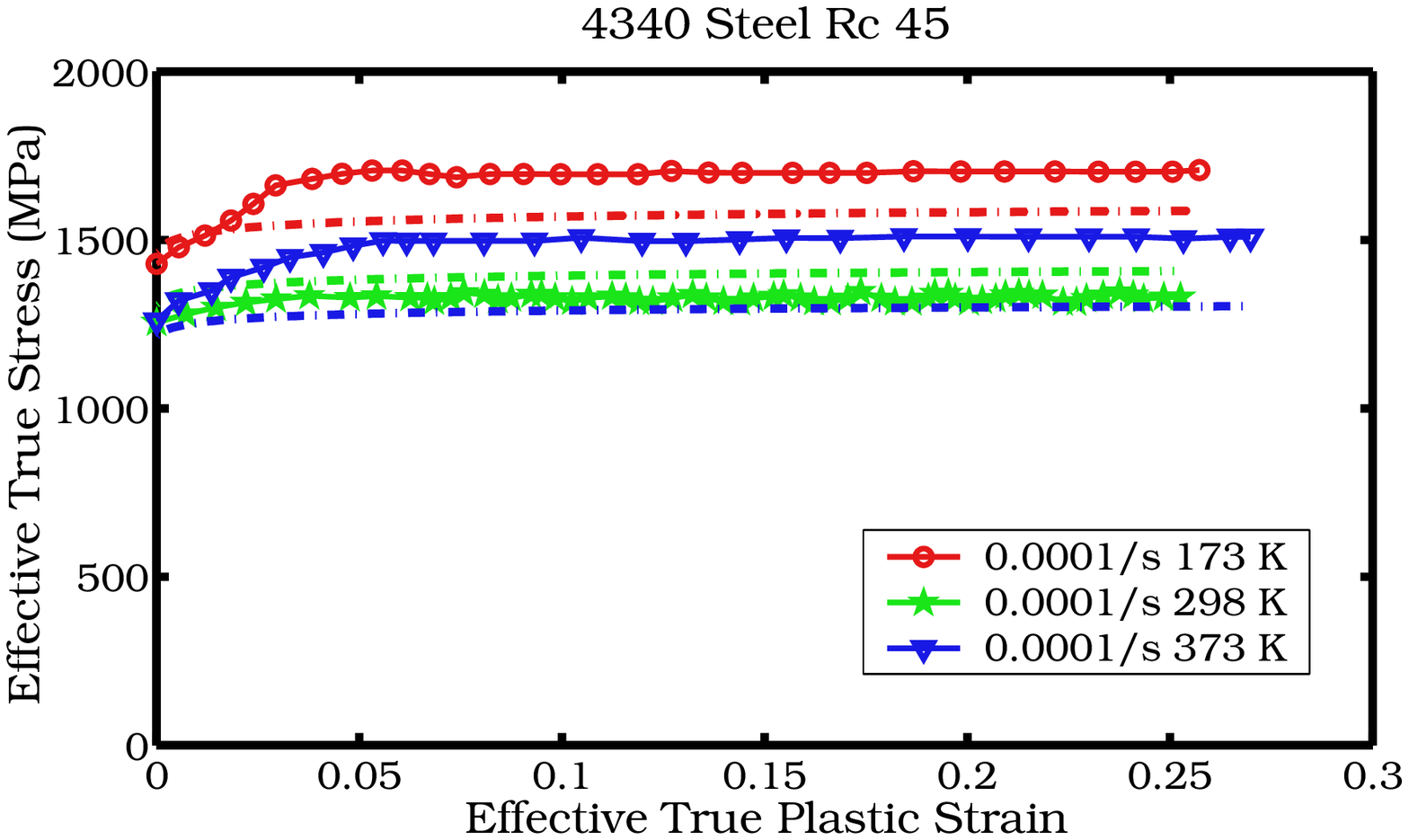}} \\
    (a) Strain Rate = 0.0001 /s \\
    \vspace{12pt}
    \scalebox{0.50}{\includegraphics{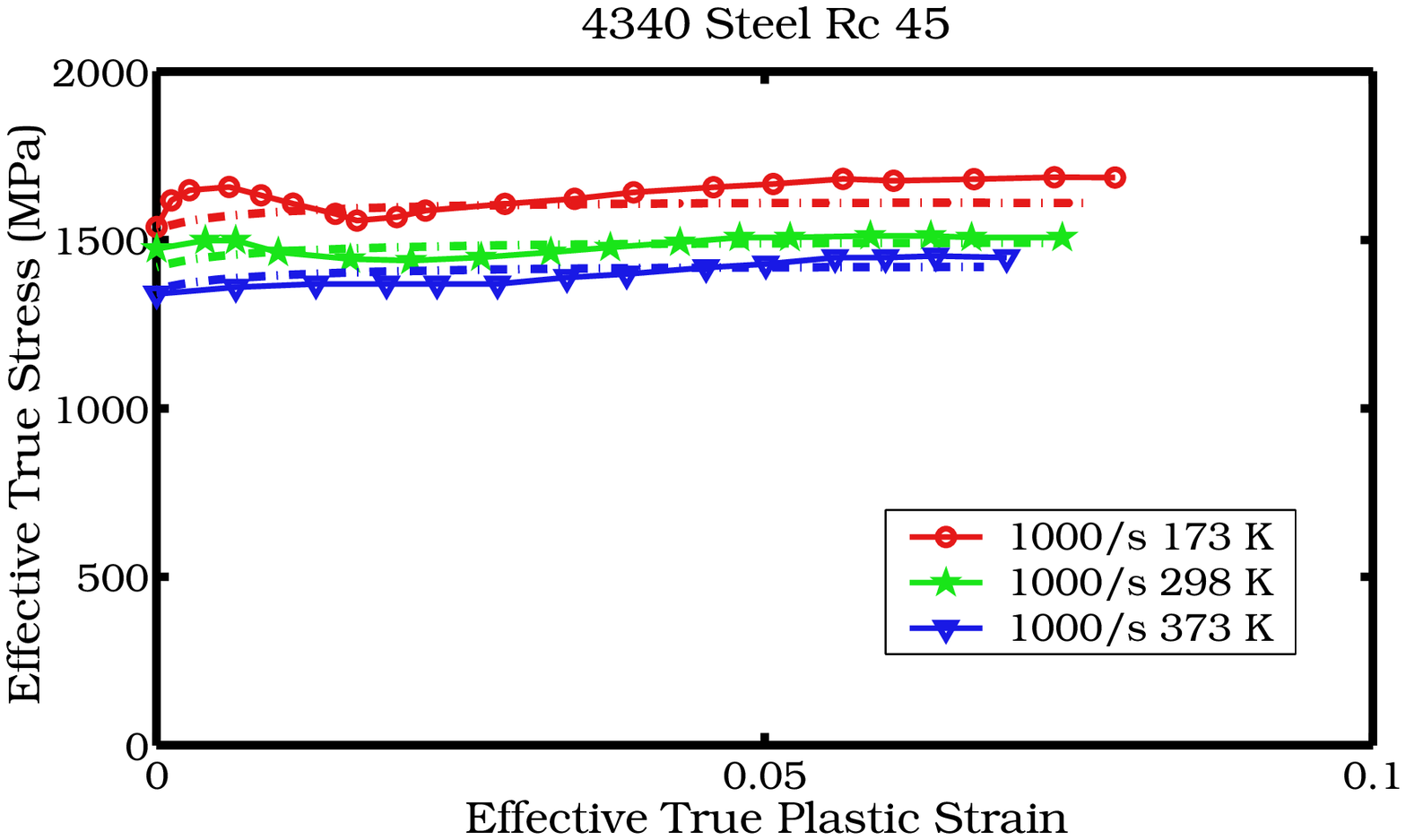}} \\
    (b) Strain Rate = 1000 /s 
    \caption{Comparison of MTS prediction with experimental data
             from \citet{Chi89} for the $R_c$ 45 temper of 
             4340 steel.}
    \label{fig:validRc45}
  \end{figure}

  Experimental data for the $R_c$ 45 temper are compared with MTS 
  predictions in Figures~\ref{fig:validRc45} (a) and (b).  The MTS
  model underpredicts the low strain rate yield stress and initial
  hardening modulus by around 15\% for both the 173 K and 373 K data.
  The prediction is within 10\% for the 298 K data.  The anomaly 
  at 373 K is clearly visible for the low strain rate plots shown 
  in Figure~\ref{fig:validRc45}(a).  The high strain rate data are 
  reproduced quite accurately for all three temperatures and the error
  is less than 10\%.
  
\begin{figure}[t]
    \centering
    \scalebox{0.50}{\includegraphics{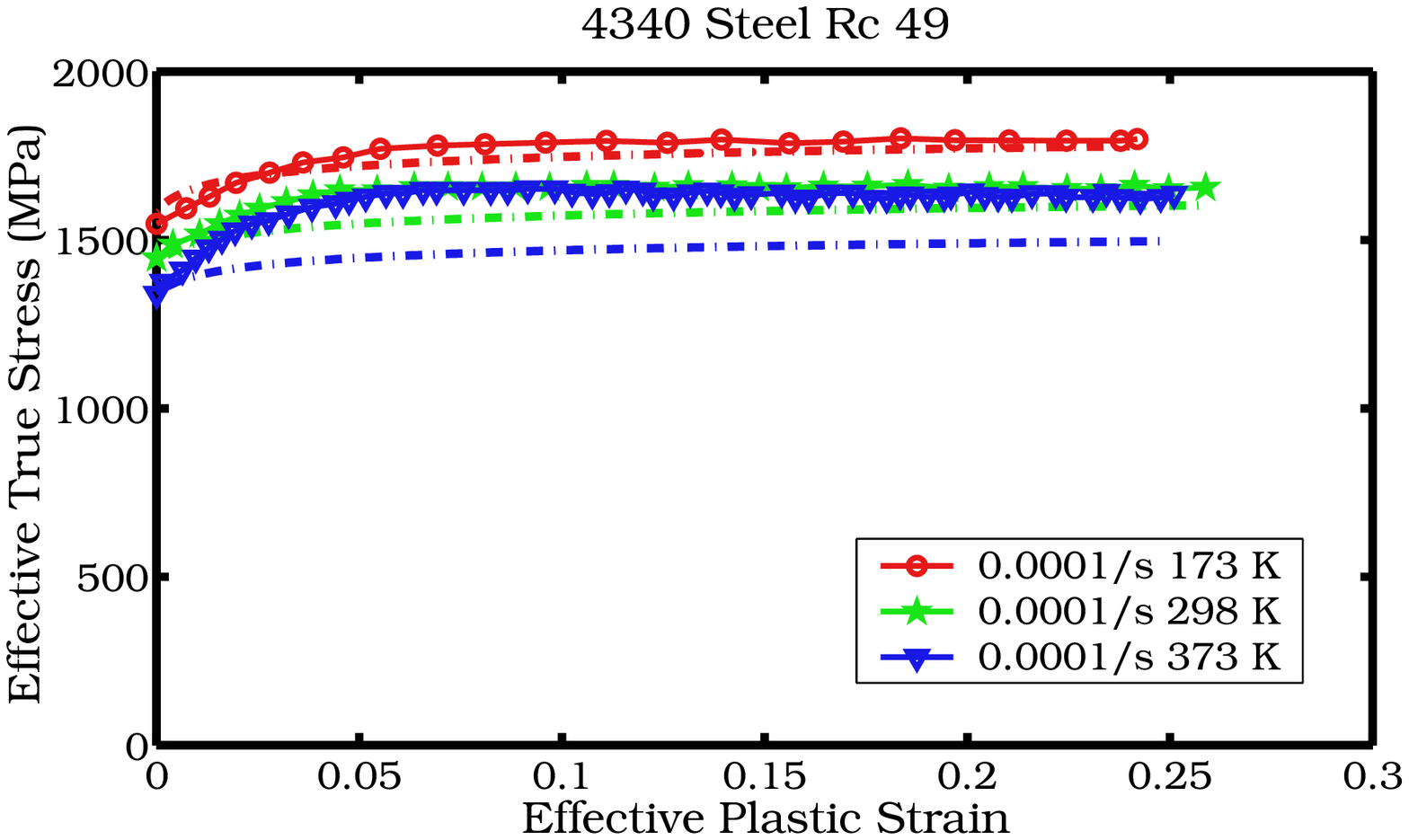}} \\
    (a) Strain Rate = 0.0001 /s \\
    \vspace{12pt}
    \scalebox{0.50}{\includegraphics{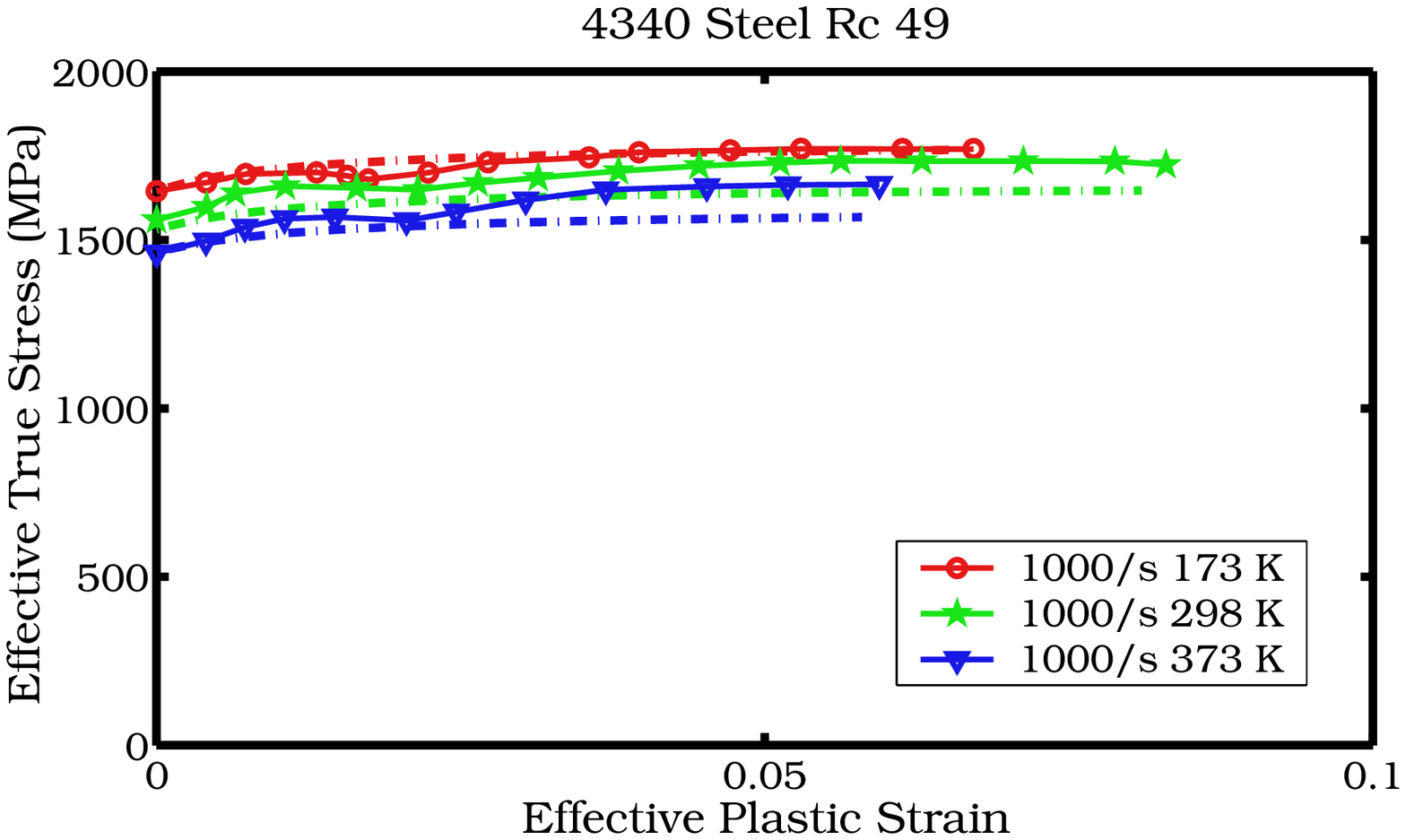}} \\
    (b) Strain Rate = 1000 /s 
    \caption{Comparison of MTS prediction with experimental data
             from \citet{Chi89} for the $R_c$ 49 temper of 
             4340 steel.}
    \label{fig:validRc49}
  \end{figure}

  Comparisons for the $R_c$ 49 temper are shown in 
  Figures~\ref{fig:validRc49} (a) and (b).  The model predicts the 
  experimental data quite accurately for 173 K and 298 K at a strain
  rate of 0.0001/s.  However, the anomalous behavior at 373K is not 
  predicted and a modeling error of around 15\% is observed for this
  temperature.  For the high strain rate cases shown in 
  Figure~\ref{fig:validRc49}(b), the initial hardening modulus is
  under-predicted and saturation is predicted at a lower stress than
  observed.  In this case, the modeling error is around 10\%.

   The comparisons of the MTS model predictions with experimental data
   shows that the predictions are all within an error of 20\% for the
   range of data examined.  If we assume that the standard deviation of
   the experimental data is around 5\% (\citet{Hanson05}) then the 
   maximum modeling error is around 15\% with around a 5\% mean.  This
   error is quite acceptable for numerical simulations, provided the
   simulations are conducted within the range of conditions used to
   fit the data.

\section{MTS model predictions over an extended range of conditions}
  In this section, we compare the yield stresses predicted for a $R_c$ 40
  temper of 4340 steel by the MTS model with those predicted by the 
  Johnson-Cook (JC) model.  A large range of strain rates and temperatures
  is explored.  In the plots shown below, the yield stress ($\sigma_y$) is
  the Cauchy stress, the plastic strain ($\Ep$) is the true plastic
  strain, the temperatures ($T$) are the initial temperatures and
  the strain rates $\Epdot{}$ are the nominal strain rates.  The effect
  of pressure on the density and melting temperature has been ignored in the
  MTS calculations presented in this section.  The Johnson-Cook model and
  relevant parameters are discussed in Appendix~\ref{app:JC}.

  \subsection{Yield stress versus plastic strain}
\begin{figure}[p]
    \centering
    \scalebox{0.40}{\includegraphics{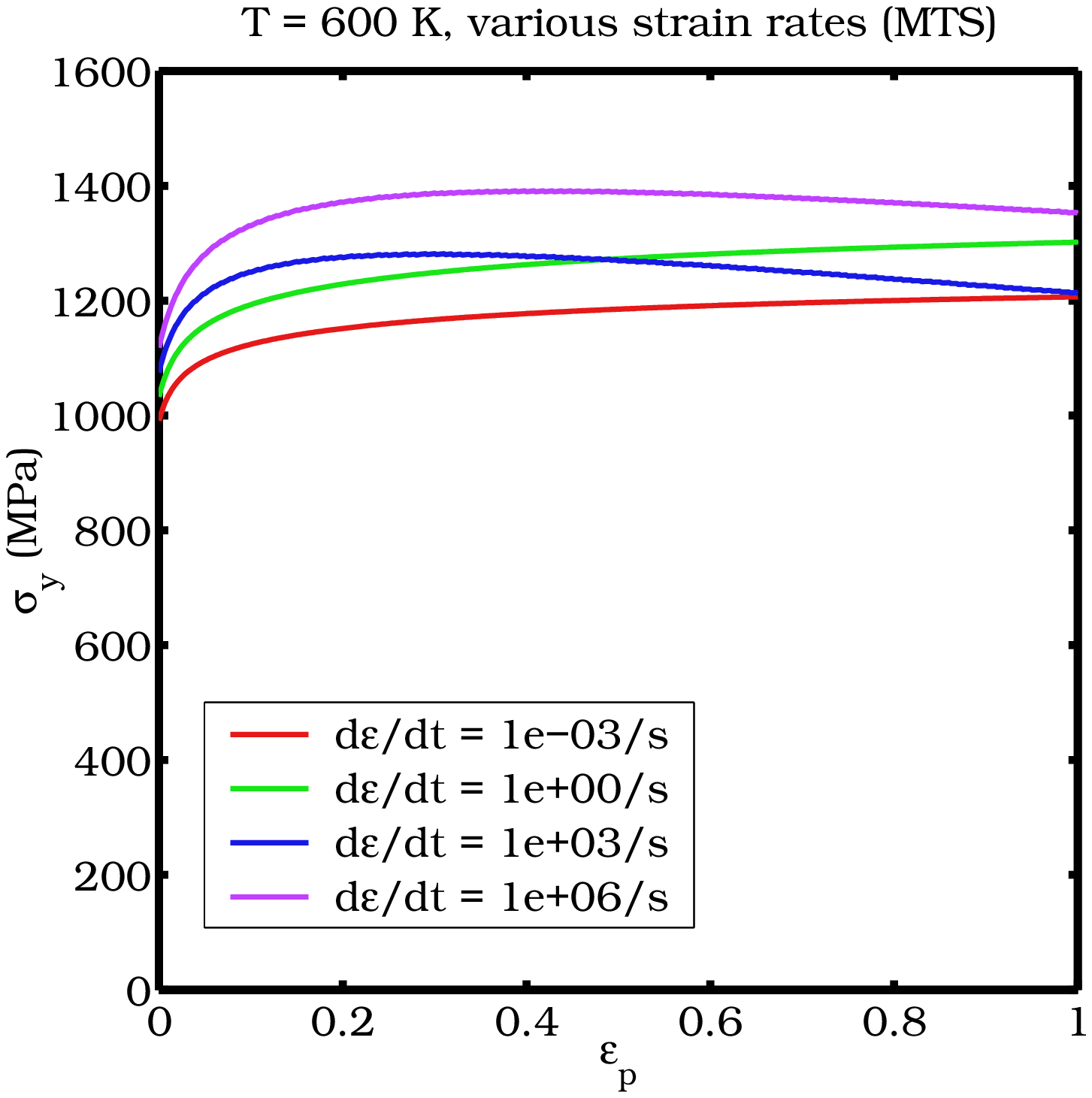}
                    \includegraphics{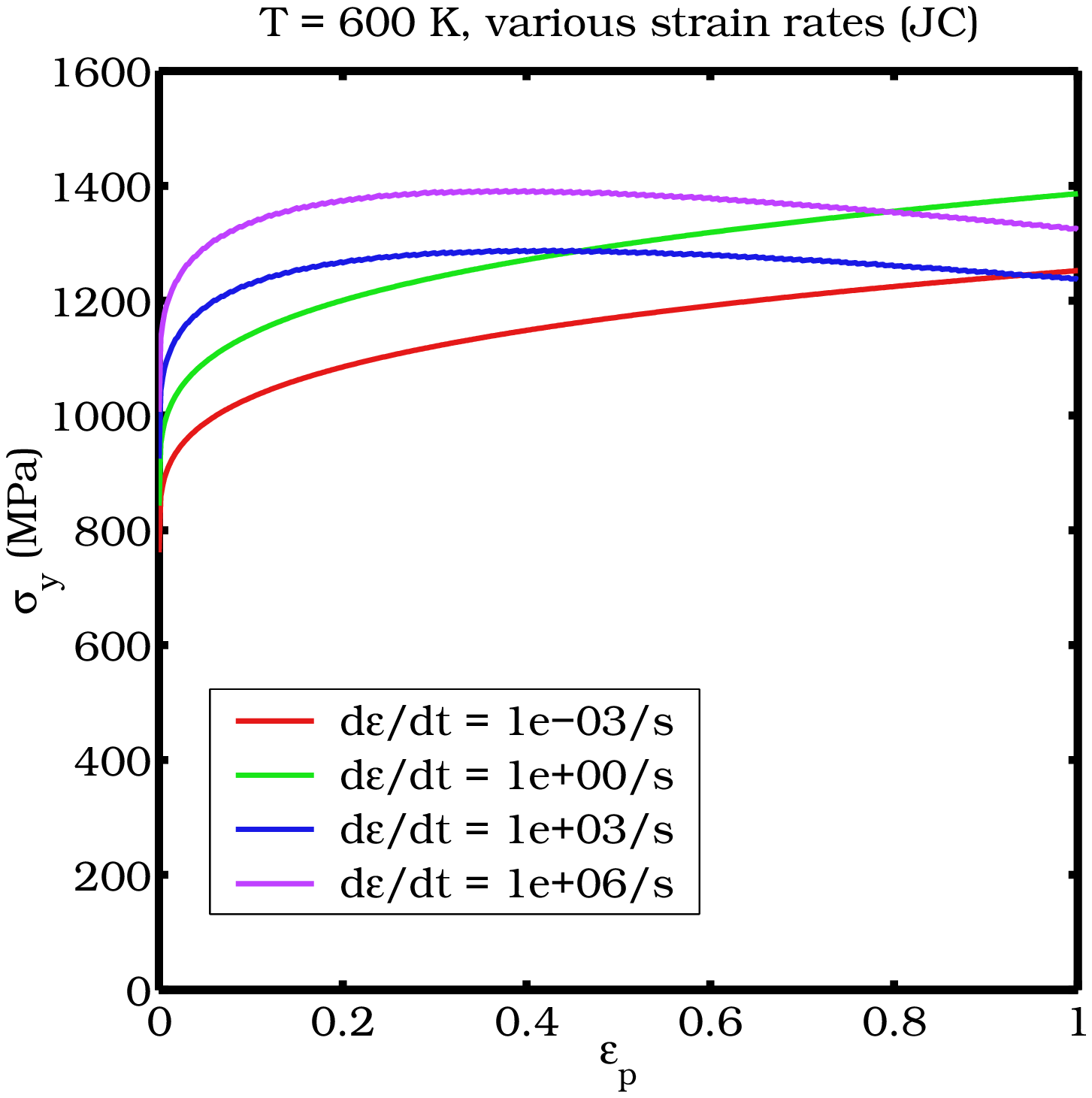}} \\
    (a) MTS Prediction. \hspace{1in} (b) JC Prediction. 
    \caption{Comparison of MTS and JC predictions of yield stress versus
             plastic strain at various strain rates for $T_0$ = 600 K.}
    \label{fig:sigy_ep_epdot}
  \end{figure}

  Figures~\ref{fig:sigy_ep_epdot}(a) and (b) show the yield stress-plastic
  strain curves predicted by the MTS and JC models, respectively.  The
  initial temperature is 600 K and adiabatic heating is assumed for strain
  rates above 500 /s.  The strain rate dependence of the yield stress is
  less pronounced for the MTS model than for the JC model.  The hardening
  rate is higher at low strain rates for the JC model.  The expected 
  rapid increase in the yield stress at strain rates above
  1000 /s (\citet{Nicholas81}) is not predicted by either model.  This
  error is probably due to the limited high rate data used to determine
  the MTS model parameters.

\begin{figure}[p]
    \centering
    \scalebox{0.40}{\includegraphics{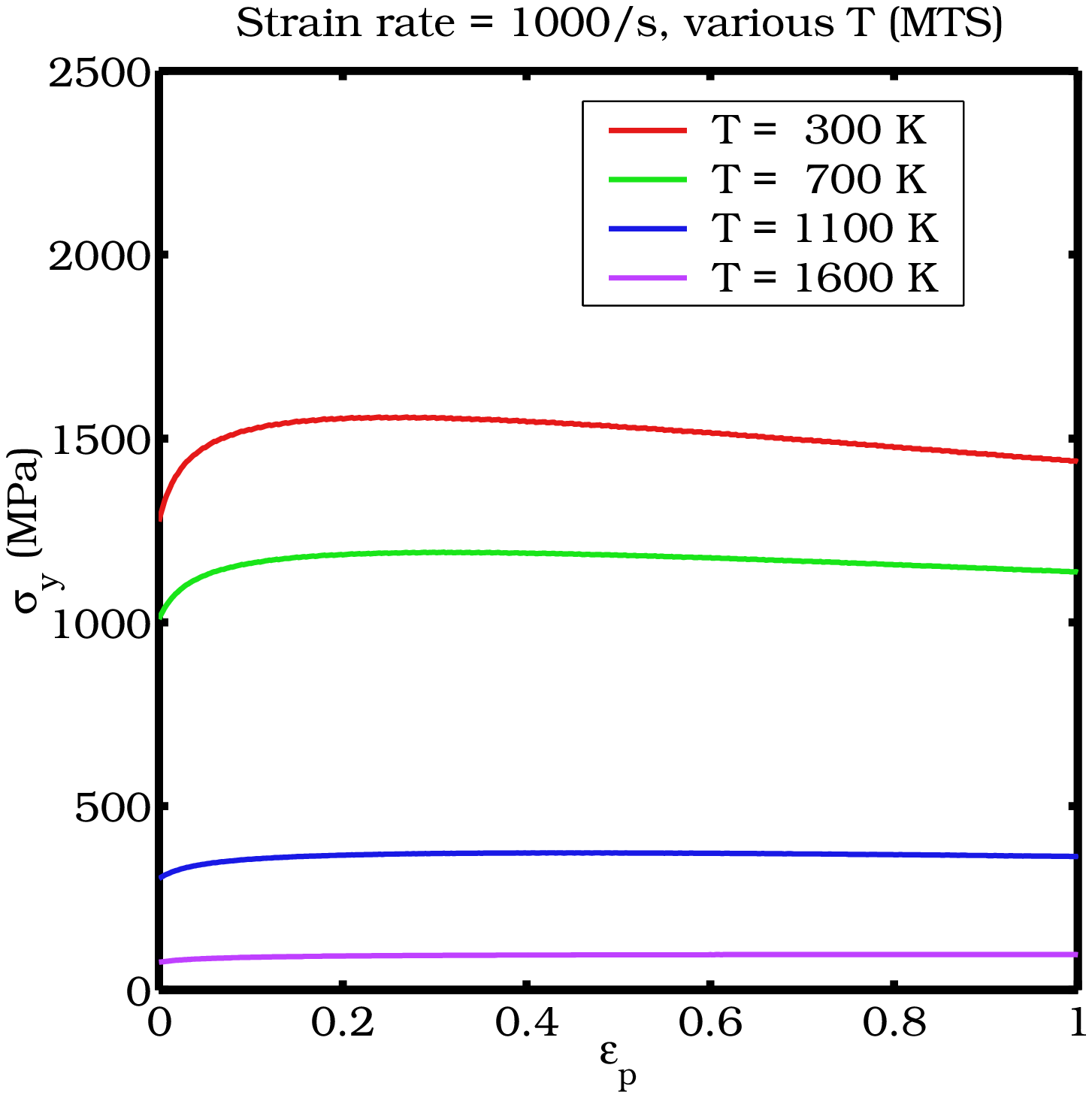}
                    \includegraphics{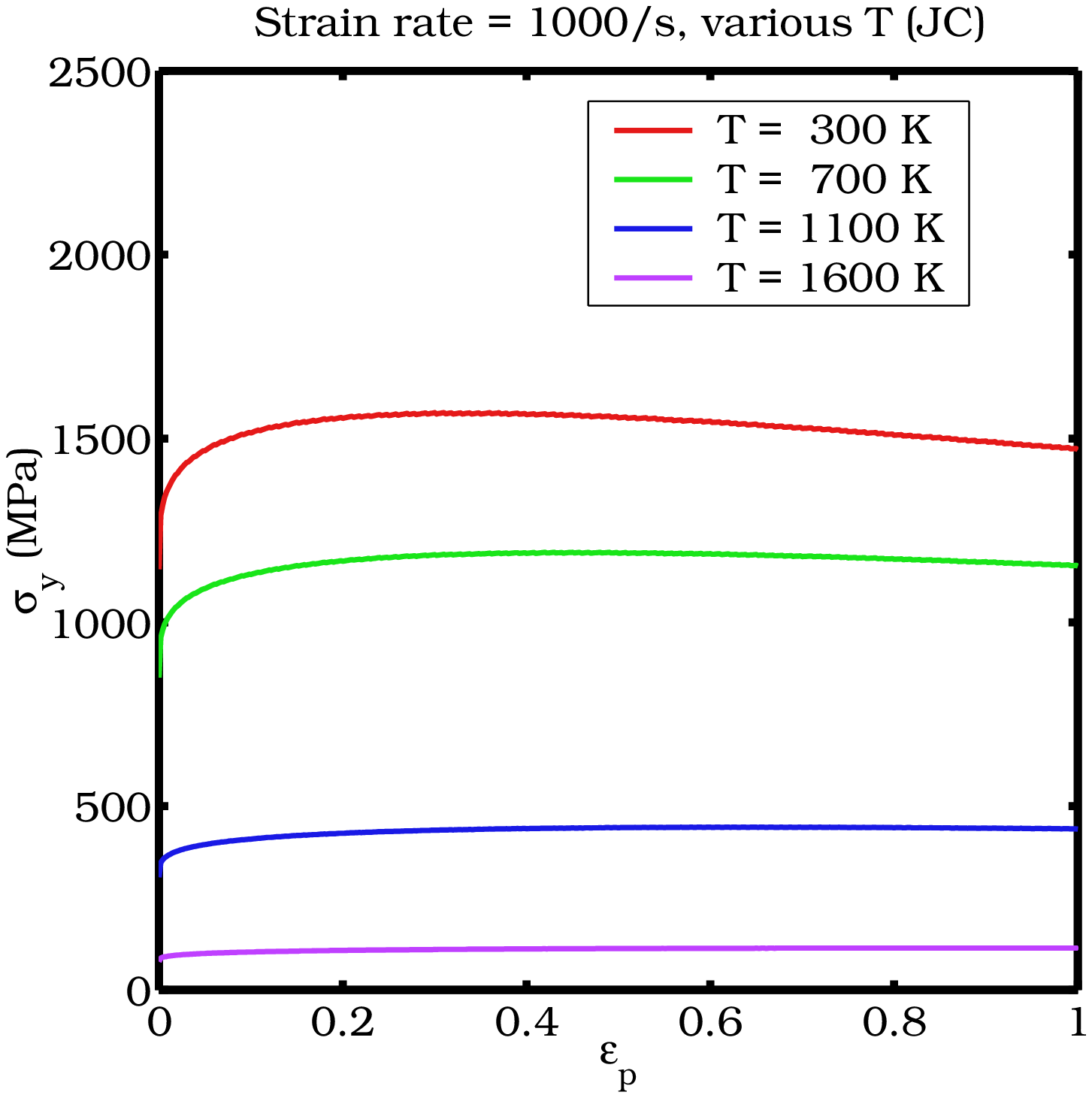}} \\
    (a) MTS Prediction. \hspace{1in} (b) JC Prediction. 
    \caption{Comparison of MTS and JC predictions of yield stress versus
             plastic strain at various strain rates for $\Epdot{}$ = 1000 /s.}
    \label{fig:sigy_ep_T}
  \end{figure}

  The temperature dependence of the yield stress for a strain rate of
  1000 /s is shown in Figures~\ref{fig:sigy_ep_T}(a) and (b).
  Both models predict similar stress-strain responses as a function 
  of temperature.  However, the initial yield stress is higher for the
  MTS model and the initial hardening rate is lower that that predicted
  by the JC model for initial temperatures of 300K and 700 K.  For the
  high temperature data, the MTS model predicts lower yield stresses.

  \subsection{Yield stress versus strain rate}
\begin{figure}[p]
    \centering
    \scalebox{0.40}{\includegraphics{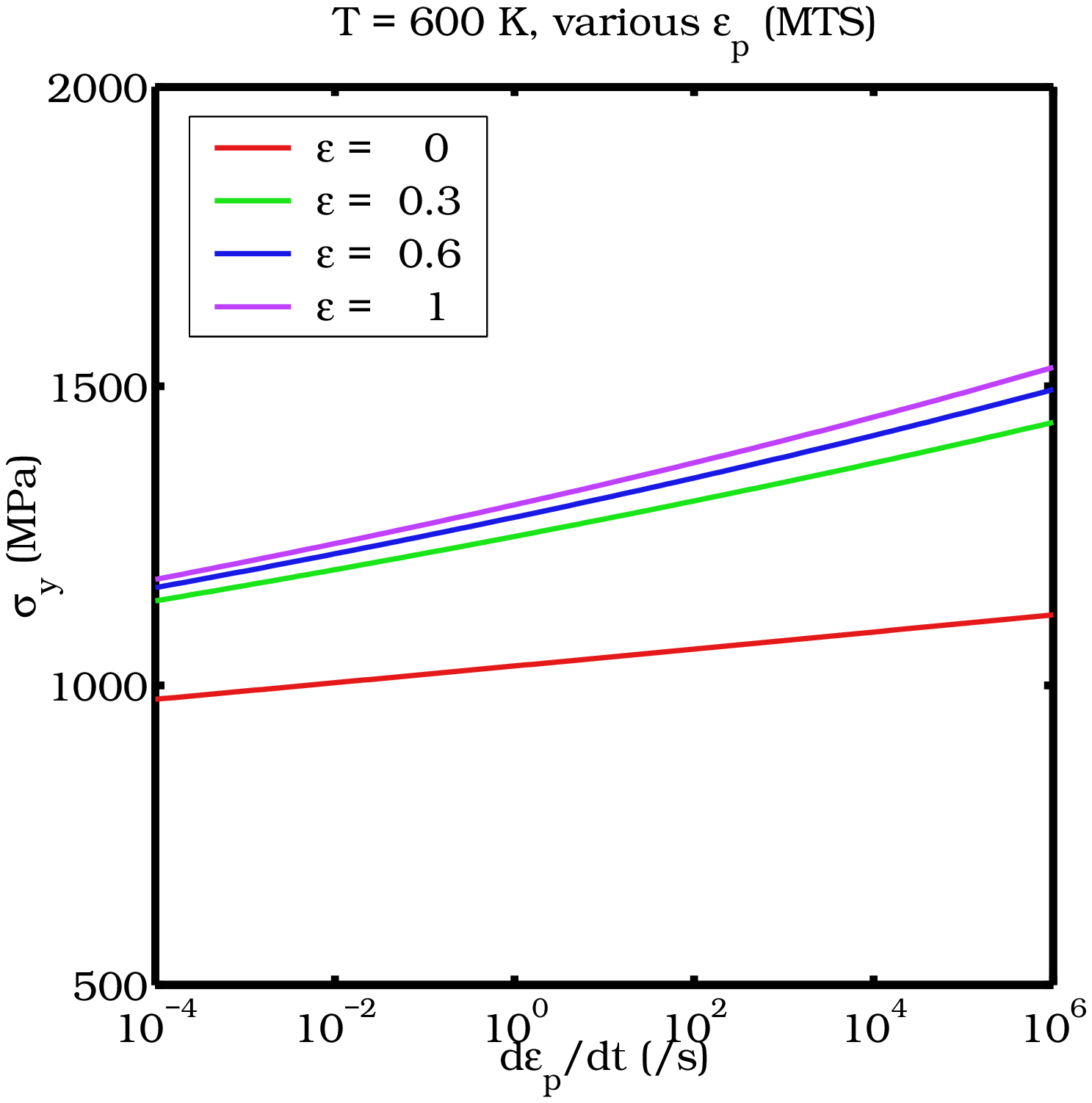}
                    \includegraphics{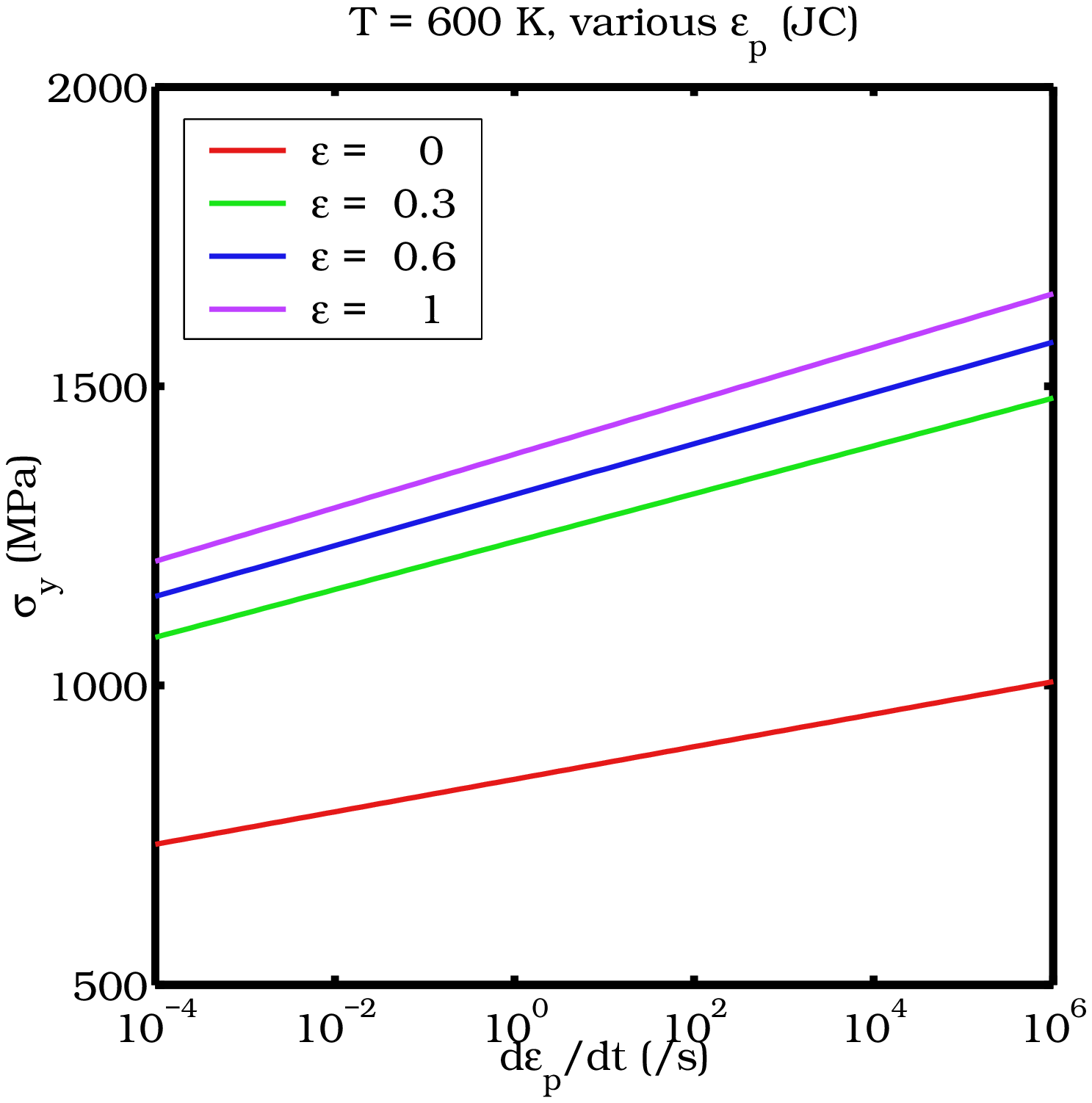}} \\
    (a) MTS Prediction. \hspace{1in} (b) JC Prediction. 
    \caption{Comparison of MTS and JC predictions of yield stress versus
             strain rate at various plastic strains for $T_0$ = 600 K.}
    \label{fig:sigy_epdot_ep}
  \end{figure}

  The strain rate dependence of the yield stress (at a temperature of 
  600 K) predicted by the MTS and JC models is shown in 
  Figures~\ref{fig:sigy_epdot_ep}(a) and (b), respectively.  The JC model
  shows a higher amount of strain hardening than the MTS model.   The
  strain rate hardening of the MTS model appears to be closer to 
  experimental observations (\citet{Nicholas81}) than the JC model.

\begin{figure}[p]
    \centering
    \scalebox{0.40}{\includegraphics{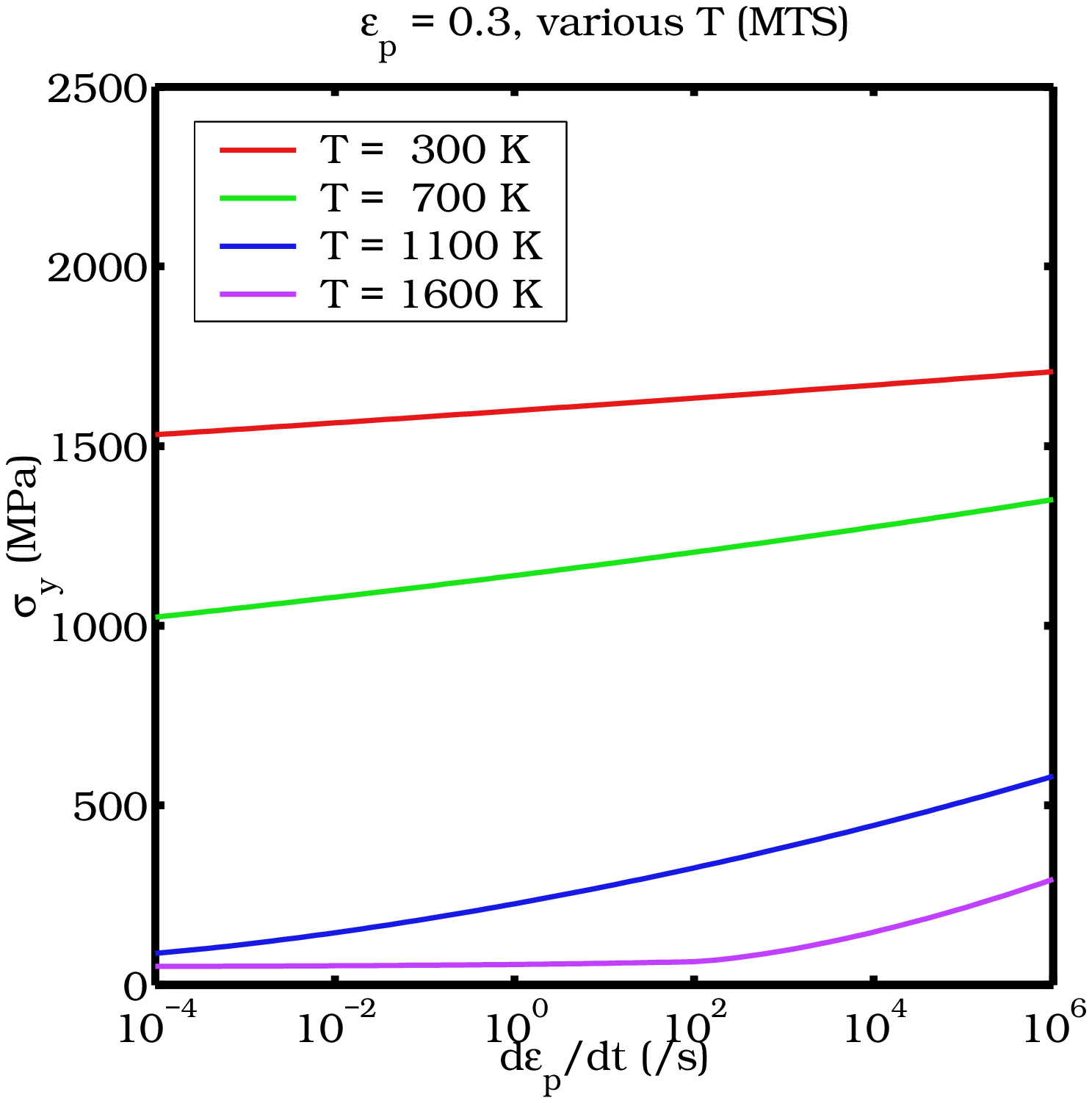}
                    \includegraphics{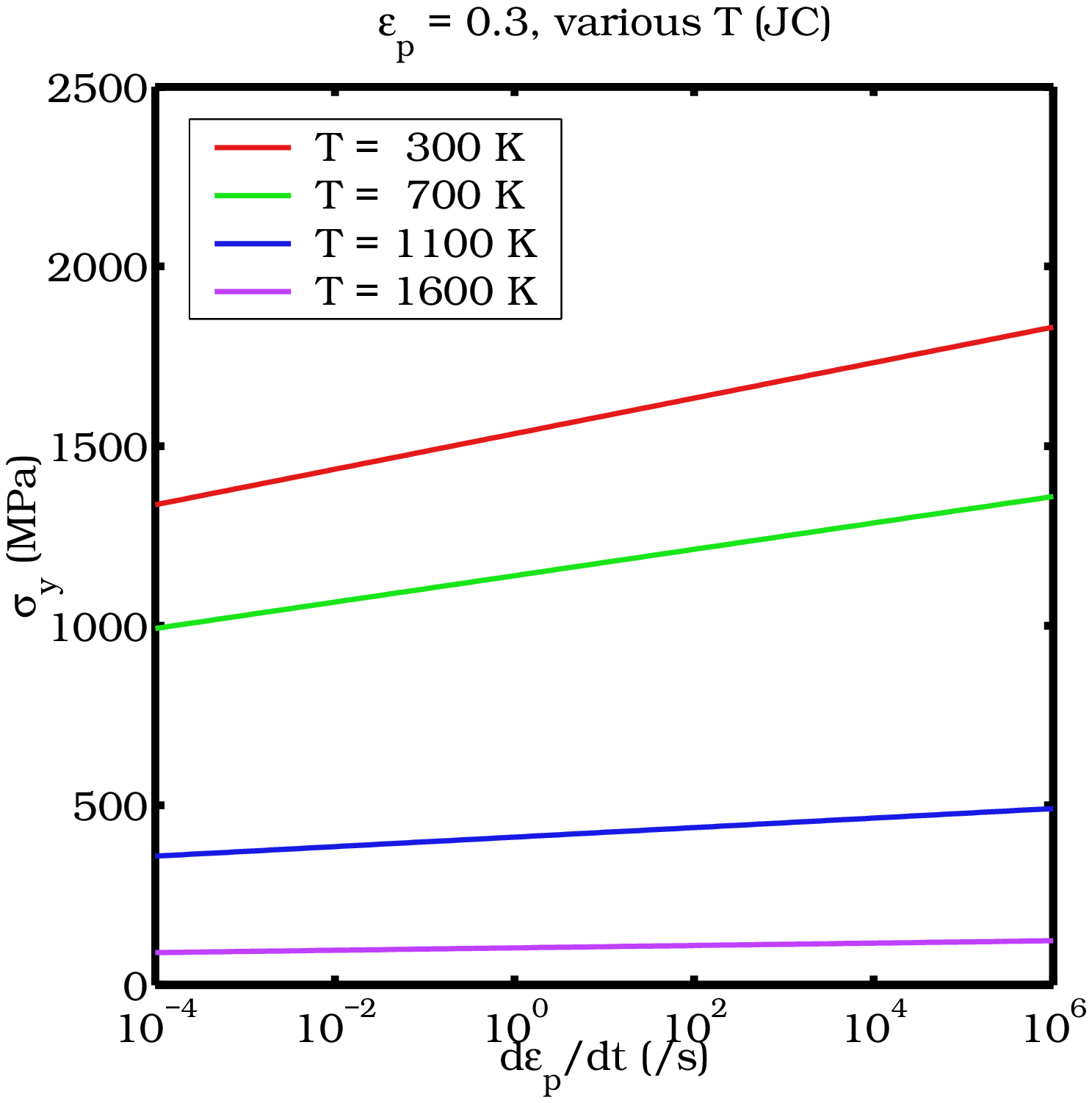}} \\
    (a) MTS Prediction. \hspace{1in} (b) JC Prediction. 
    \caption{Comparison of MTS and JC predictions of yield stress versus
             strain rate at various temperatures for $\Ep$ = 0.3 .}
    \label{fig:sigy_epdot_T}
  \end{figure}

  The temperature and strain rate dependence of the yield stress at a 
  plastic strain of 0.3 is shown in Figures~\ref{fig:sigy_epdot_T}(a) and
  (b).  Above the phase transition temperature, the MTS model predicts 
  more strain rate hardening than the JC model.  However, at 700 K, both 
  models predict quite similar yield stresses.  At room temperature, the
  JC model predicts a higher rate of strain rate hardening than the MTS model
  and is qualitatively closer to experimental observations.

  \subsection{Yield stress versus temperature}
\begin{figure}[p]
    \centering
    \scalebox{0.40}{\includegraphics{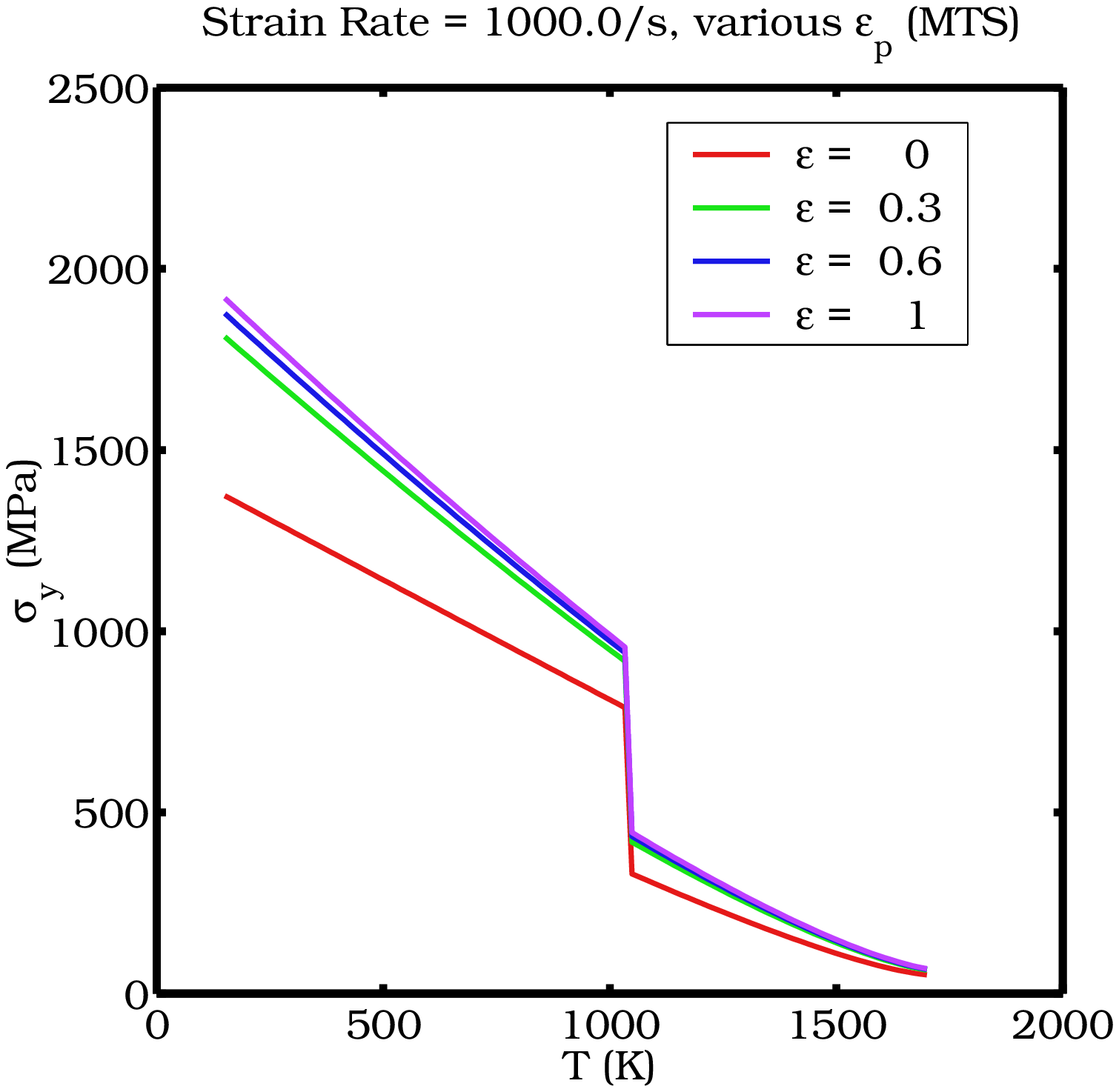}
                    \includegraphics{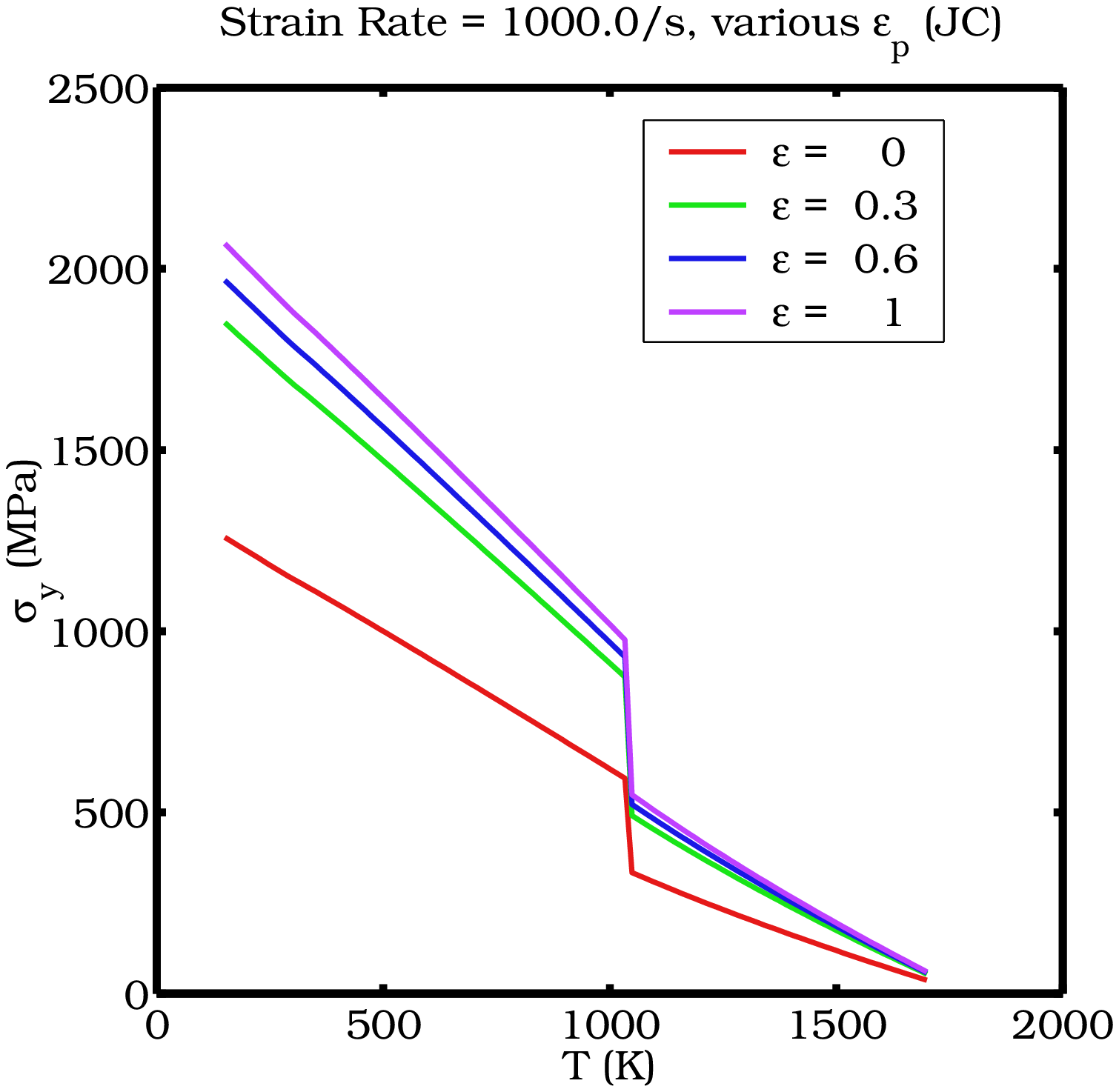}} \\
    (a) MTS Prediction. \hspace{1in} (b) JC Prediction. 
    \caption{Comparison of MTS and JC predictions of yield stress versus
             temperature at various plastic strains for $\Epdot{}$ = 1000/s.}
    \label{fig:sigy_T_ep}
  \end{figure}

  The temperature dependence of the yield stress for various plastic
  strains (at a strain rate of 1000 /s) is shown in 
  Figures~\ref{fig:sigy_T_ep}(a) and (b).  The sharp change in the 
  value of the yield stress at the phase transition temperature may
  be problematic for Newton methods used in the determination of the
  plastic strain rate.  We suggest that at temperatures close to the
  phase transition temperature, the high temperature parameters should
  be used in numerical computations.  The figures show that
  both the models predict similar rates of temperature dependence of 
  the yield stress.

\begin{figure}[p]
    \centering
    \scalebox{0.40}{\includegraphics{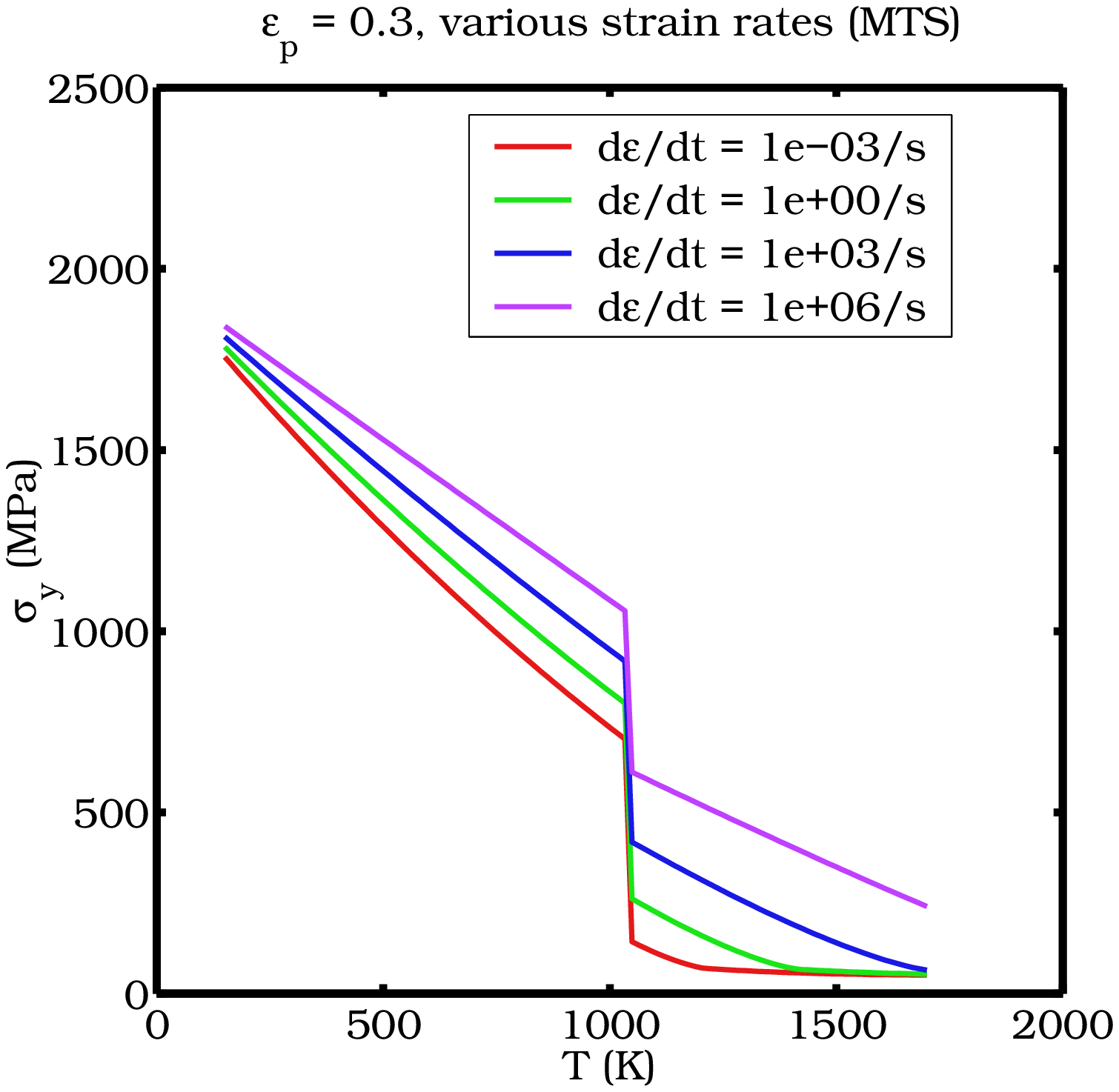}
                    \includegraphics{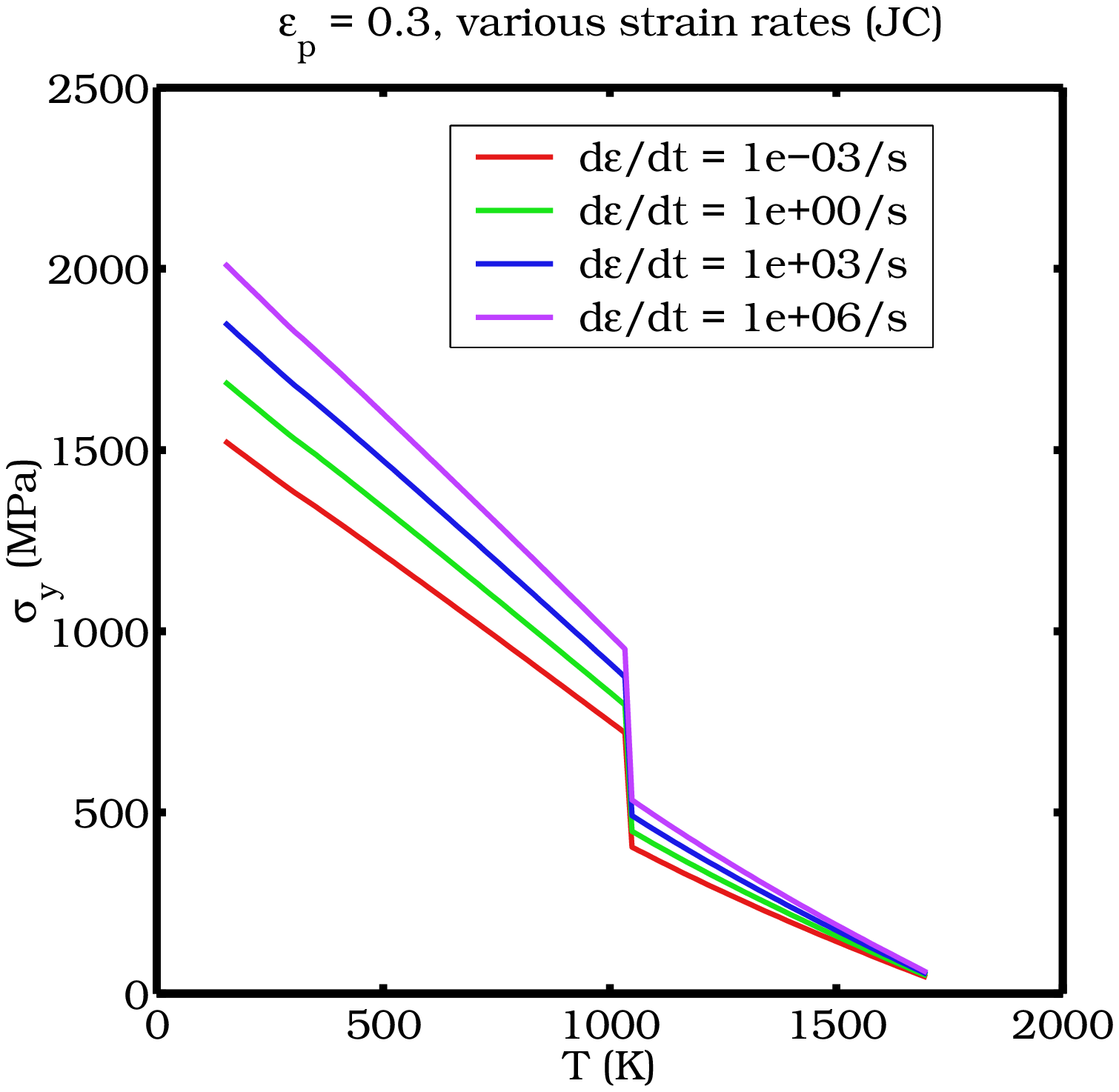}} \\
    (a) MTS Prediction. \hspace{1in} (b) JC Prediction. 
    \caption{Comparison of MTS and JC predictions of yield stress versus
             temperature at various strain rates for $\Ep$ = 0.3 .}
    \label{fig:sigy_T_epdot}
  \end{figure}

  The temperature dependence of the yield stress for various strain rates
  (at a plastic strain of 0.3) is shown in 
  Figures~\ref{fig:sigy_T_epdot}(a) and (b).  In this case, the MTS model
  predicts at smaller strain rate effect at low temperatures than the JC 
  model.  The strain rate dependence of the yield stress increases with 
  temperature for the MTS model while it decreases with temperature
  for the JC model.  The JC model appears to predict a more realistic 
  behavior because the thermal activation energy for dislocation motion
  is quite low at high temperatures.  However, the MTS model fits high
  temperature/high strain rate experimental data better than the JC model
  and we might be observing the correct behavior in the MTS model.  

  \subsection{Taylor impact tests}
  For further confirmation of the effectiveness of the MTS model, we have
  simulated three-dimensional Taylor impact tests using the Uintah code
  (\citet{Banerjee05}).  Details of the code, the algorithm used, 
  and the validation process have been discussed elsewhere 
  (\citet{Banerjee05,Banerjee05a}).  

\begin{figure}[p]
    \centering
    \scalebox{0.45}{\includegraphics{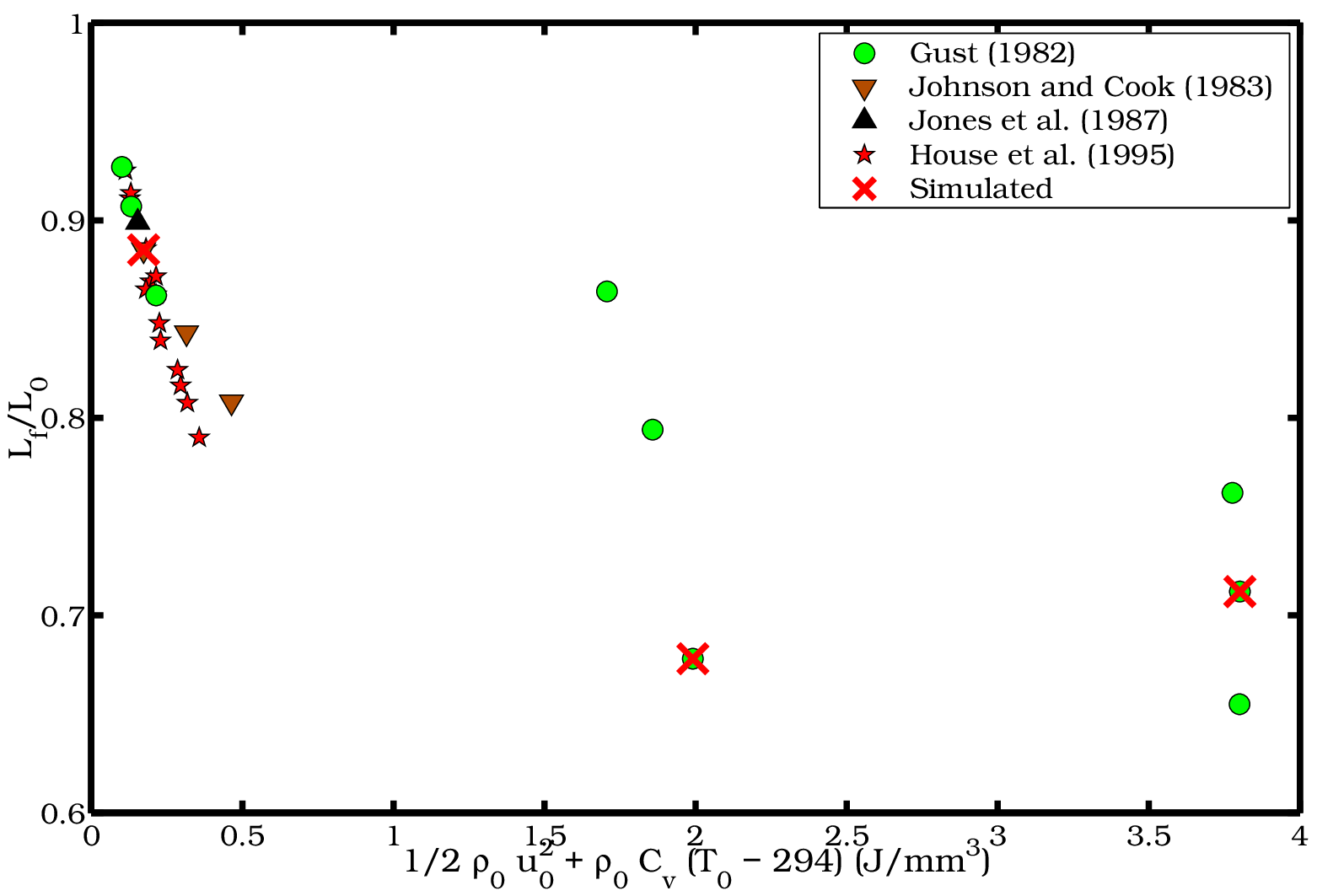}} 
    \caption{The ratio of the final length to the initial length of 
             Taylor impact specimens as a function of initial energy
             density.  The experimental data are from \citet{Gust82},
             \citet{Johnson83}, \citet{Jones87}, and \citet{House95}.
             The tests that we have simulated are marked with crosses.}
    \label{fig:StLfExpt}
  \end{figure}

  It is well known that the final length of a Taylor impact cylinder 
  scales with the initial velocity.  Figure~\ref{fig:StLfExpt} shows 
  some experimental data on the final length of cylindrical Taylor impact 
  specimens as a function of initial velocity.  We are interested in 
  temperatures higher than room temperature.  For clarity, we have separated 
  the high temperature tests from the room temperature tests by adding an 
  initial internal energy component to the initial kinetic energy density.
  We have simulated three Taylor tests at three energy levels (marked with
  crosses on the plot).

  The four cases that we have simulated have the following initial
  conditions:
  \begin{enumerate}
    \item {\bf Case 1:} $R_c$ = 30; $L_0$ = 25.4 mm; $D_0$ = 7.62 mm; 
          $U_0$ = 208 m/s; $T_0$ = 298 K; Source \citet{Johnson83}.
    \item {\bf Case 2:} $R_c$ = 40; $L_0$ = 30.0 mm; $D_0$ = 6.0 mm; 
          $U_0$ = 312 m/s; $T_0$ = 725 K; Source \citet{Gust82}.
    \item {\bf Case 3:} $R_c$ = 40; $L_0$ = 30.0 mm; $D_0$ = 6.0 mm; 
          $U_0$ = 160 m/s; $T_0$ = 1285 K; Source \citet{Gust82}.
    \item {\bf Case 4:} $R_c$ = 40; $L_0$ = 30.0 mm; $D_0$ = 6.0 mm; 
          $U_0$ = 612 m/s; $T_0$ = 725 K; 
  \end{enumerate}

  The MTS model parameters for the $R_c$ 30 temper of 4340 steel have 
  been given earlier.  The MTS parameters for the $R_c$ 40 temper 
  of 4340 steel can be calculated either using the linear fit for various 
  hardness levels (shown in Figure~\ref{fig:SigmaiGoialpha}) or by a 
  linear interpolation between the $R_c$ 38 and the $R_c$ 45 values.
  MTS model parameters at temperatures above 1040 K take the high temperature
  values discussed earlier.  The initial yield stress in the 
  Johnson-Cook model is obtained from the $R_c$-$\sigma_0$ relation 
  given in Appendix~\ref{app:JC}.  

\begin{figure}[p]
    \centering
    \scalebox{0.45}{\includegraphics{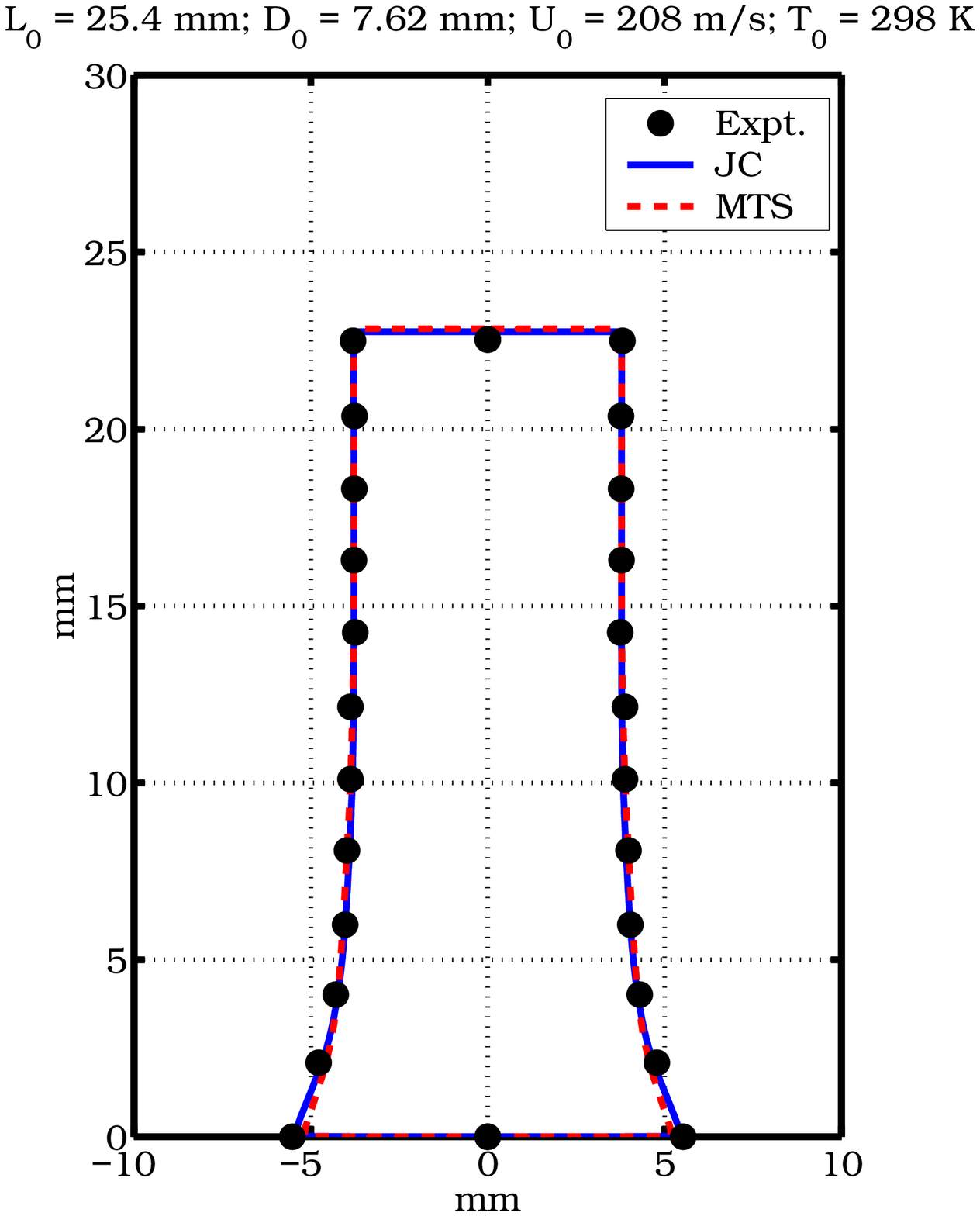} \hspace{12pt}
                    \includegraphics{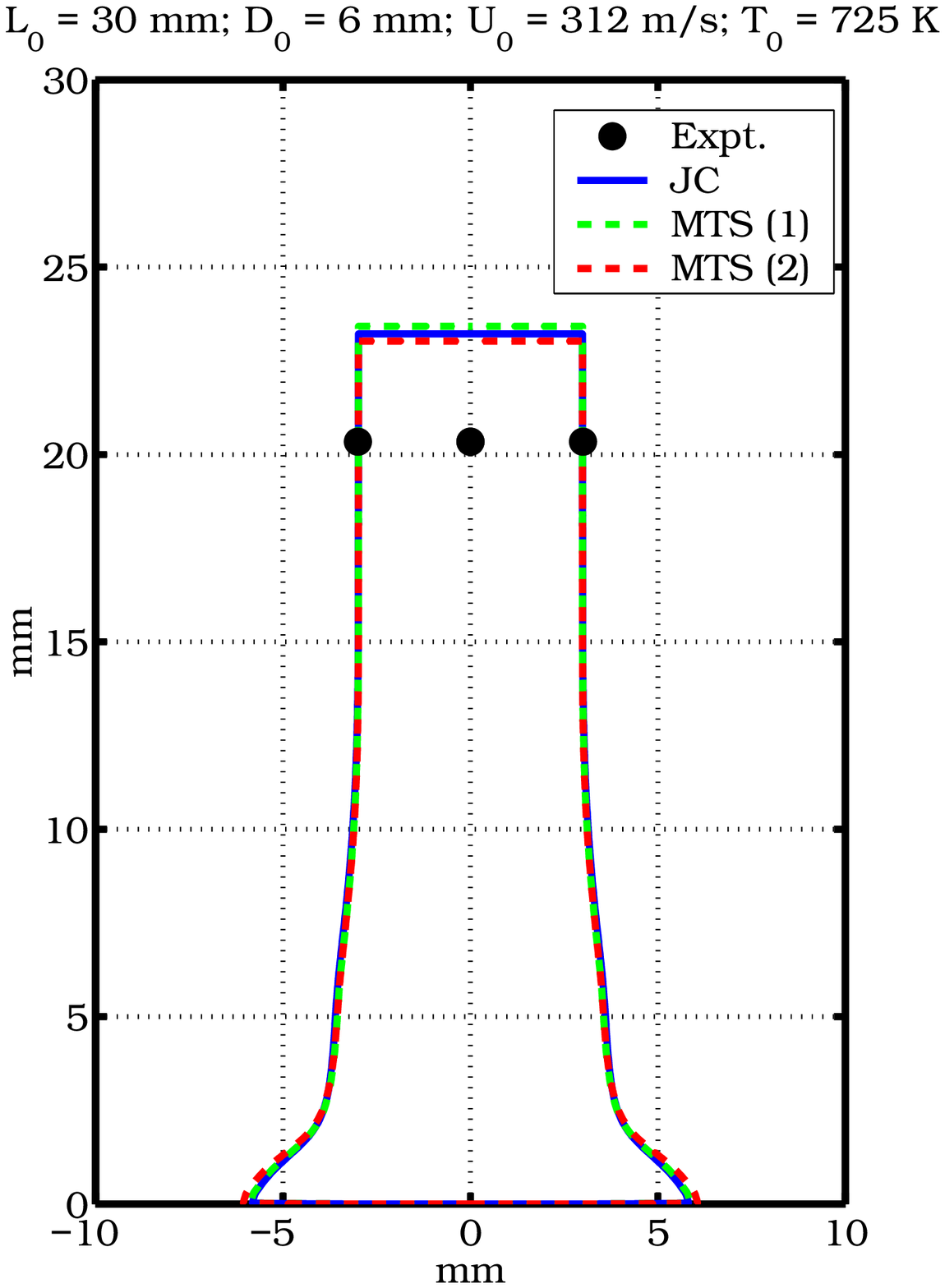}}\\
    (a) Case 1. \hspace{1.5in} (b) Case 2. \\
    \vspace{12pt}
    \scalebox{0.45}{\includegraphics{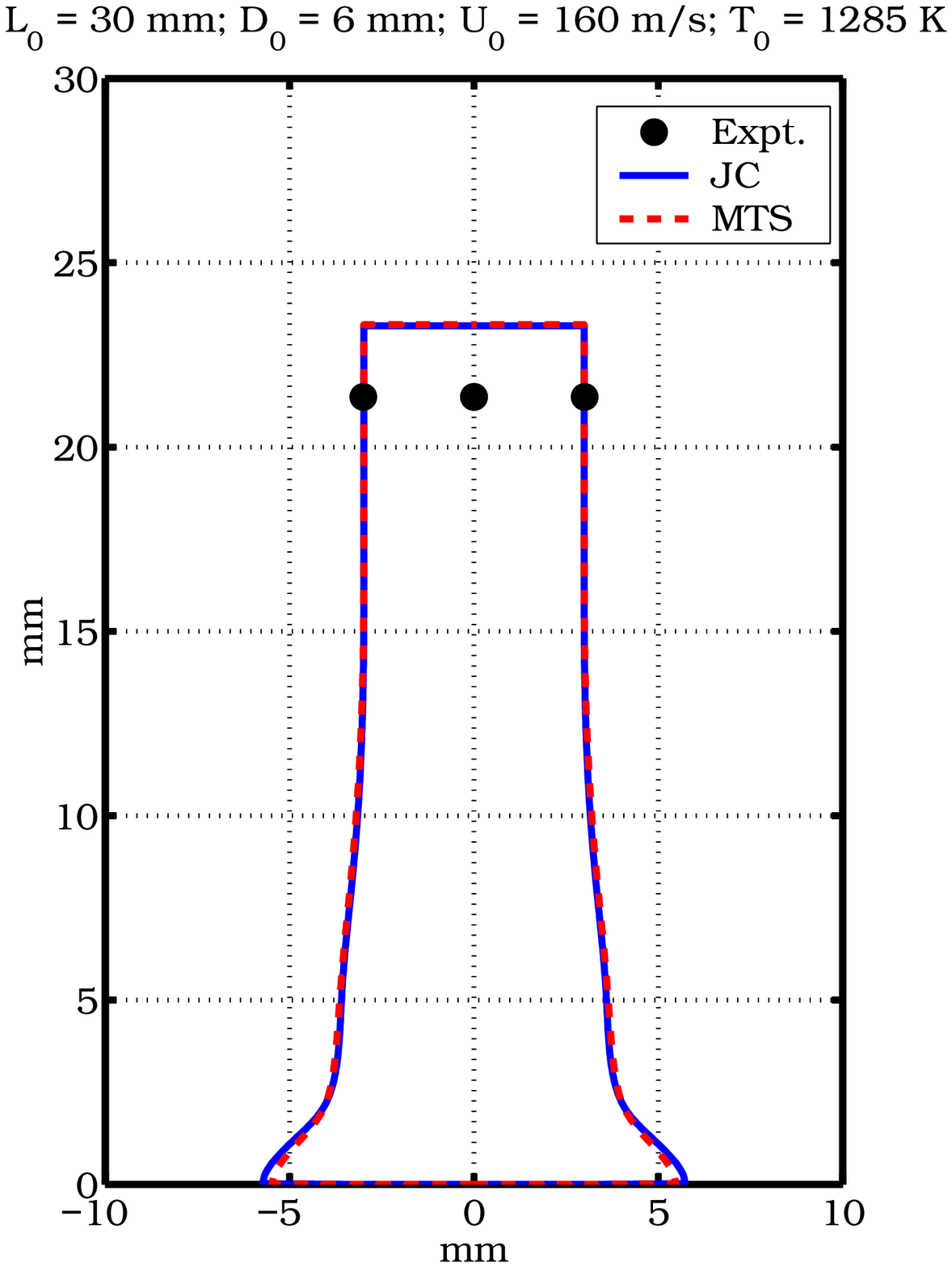} \hspace{12pt}
                    \includegraphics{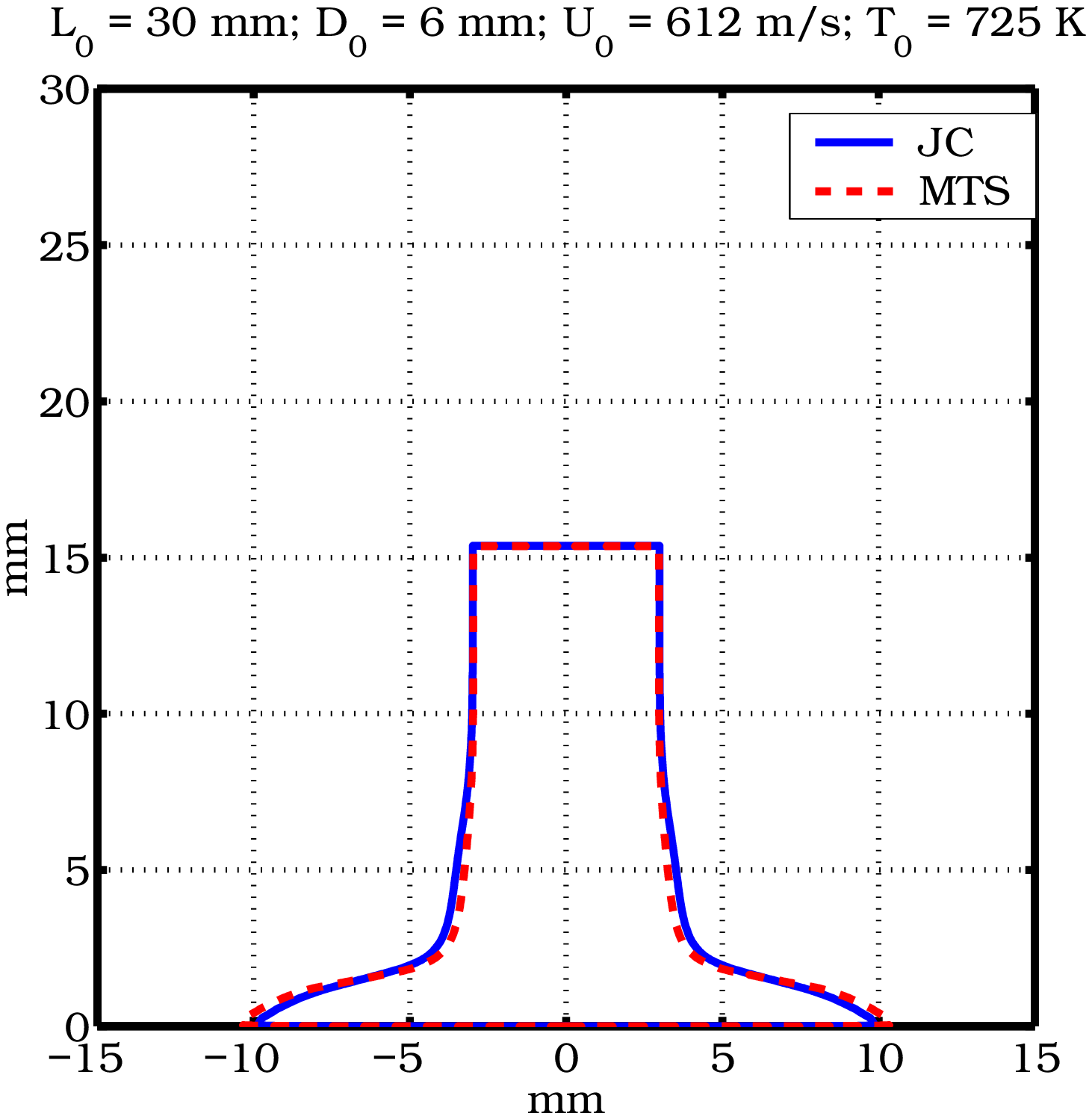}} \\
    (c) Case 3. \hspace{1.5in} (d) Case 4.
    \caption{Comparison of MTS and JC predictions of final Taylor specimen
             profiles with experimental results.}
    \label{fig:profileSt}
  \end{figure}

  The computed final profiles are compared with 
  the experimental data in Figures~\ref{fig:profileSt}(a), (b), (c), and (d). 

  For the room temperature
  test (Figure~\ref{fig:profileSt}(a)), the Johnson-Cook model accurately
  predicts the final length, the mushroom diameter, and the overall profile.
  The MTS model underestimates the mushroom diameter by 0.25 mm.  This
  difference is within experimental variation (see \citet{House95}). 

  The simulations at 725 K (Figure~\ref{fig:profileSt}(b)) overestimate
  the final length of the specimen.  The legend shows two MTS predictions
  for this case - MTS (1) and MTS (2).  MTS (1) uses parameters $\sigma_i$
  and $g_{0i}$ that have been obtained using the fits shown in
  Figure~\ref{fig:SigmaiGoialpha}.  MTS (2) used parameters obtained by
  linear interpolation between the $R_c$ 38 and $R_c$ 45 values.  The 
  MTS (2) simulation predicts a final length that is slightly less than
  that predicted by the MTS (1) and Johnson-Cook models.  The mushroom
  diameter is also slightly larger for the MTS (2) simulation.  

  The final length of the specimen for Case 2 is not predicted accurately 
  by either model.  We have confirmed that this error is not due to 
  discretization (note that volumetric locking does not occur with
  the explicit Material Point Method used in the simulations).  
  Plots of energy and momentum have also shown that both quantities
  are conserved in these simulations.  The final mushroom diameter is 
  not provided by \cite{Gust82}.  However, the author mentions that no 
  fracture was observed in the specimen - discounting a smaller final 
  length due to fracture.  In the absence of more extensive high temperature 
  Taylor impact data it is unclear if the error is within experimental 
  variation or due to a fault with the models used.  

  The third case (Figure~\ref{fig:profileSt}(c)) was simulated at an initial 
  temperature of 1285 K (above the $\alpha$-$\gamma$ phase transition 
  temperature of iron).  The MTS and Johnson-Cook models predict almost 
  exactly the same behavior for this case.  The final length is 
  overestimated by both the models.  Notice that the final lengths
  shown in Figure~\ref{fig:StLfExpt} at or near this temperature and for 
  similar initial velocities vary from 0.65 to 0.75 of the initial length.
  The simulations predict a final length that is approximately 0.77 times
  the initial length - which is to the higher end of the range of
  expected final lengths.  The discrepancy may be because the models do
  not predict sufficient strain hardening at these high temperatures.
  
  In all three cases, the predictions from the MTS and the Johnson-Cook
  models are nearly identical.  To determine if any significant difference
  between the predictions of these models can be observed at higher strain
  rates, we simulated the geometry of Case 2 with a initial velocity of
  612 m/s.  The resulting profiles predicted by the MTS and the Johnson-Cook
  models are shown in Figure~\ref{fig:profileSt}(d).
  In this case, the MTS model predicts a slightly wider mushroom than the
  Johnson-Cook model.  The final predicted lengths are almost identical.
  Interestingly, the amount of strain hardening predicted by the MTS model
  is smaller than that predicted by the Johnson-Cook model (as can be
  observed from the secondary bulge in the cylinder above the mushroom).
  We conclude that the Johnson-Cook and MTS models presented in this
  paper predict almost identical elastic-plastic behavior in the range of 
  conditions explored.  Please note that quite different sets of data were
  used to determine the parameters of these models and hence the similarity
  of the results may indicate the underlying accuracy of the parameters.
  
\section{Remarks and Conclusions}\label{sec:conclude}
  We have determined parameters for the Mechanical Threshold Stress model
  and the Johnson-Cook model for
  various tempers of 4340 steel.  The predictions of the MTS model have
  been compared with experimental stress-strain data.  Yield stresses 
  predicted by the Johnson-Cook and the MTS model have been compared for
  a range of strain rates and temperatures.  Taylor impact tests have been
  simulated and the predicted profiles have been compared with experimental
  data.
  
  Some remarks and conclusions regarding this work are given below.
  \begin{enumerate}
    \item  The MTS and Johnson-Cook models predict similar stress-strain
           behaviors over a large range of strain rates and temperatures.
           Noting that the parameters for these models have been obtained 
           from different sets of experimental data, the similarity 
           of the results, especially in the Taylor test simulations, 
           is remarkable.  We suggest that this is an indication of the
           accuracy of the models and the simulations.  However, the 
           Taylor impact tests show that both models predict lower strains 
           at high temperatures than experiments suggest.  We are in the
           process of determining paramters for the Preston-Tonks-Wallace model
           (\citet{Preston03}) to check if the issue is model dependent.
    \item  The MTS model parameters are considerably easier to obtain than
           the Johnson-Cook parameters.  However, the MTS simulations of the
           Taylor impact tests take approximately 1.5 times longer than the
           Johnson-Cook simulations.  This is partly because the shear
           modulus and melting temperature models are not evaluated in
           the Johnson-Cook model simulations.  Also, the MTS model 
           involves more floating point operations than the Johnson-Cook model.
           The Johnson-Cook model is numerically more efficient than the 
           MTS model and is preferable for large numerical
           simulations involving 4340 steel.
    \item  The Nadal-LePoac shear modulus model and the 
           Burakovsky-Preston-Silbar melting temperature model involve 
           less data fitting and are the suggested models for elastic-plastic
           simulations over a large range of temperatures and strain rates.
           The specific heat model that we have presented leads to better
           predictions of the rate of temperature increase close to the
           $\alpha$-$\gamma$ phase transition of iron.  The shear modulus and
           melt temperature models are also valid in the range of strain rates
           of the order of 10$^8$ /s.  The Mie-Gr{\"u}neisen equation of 
           state should probably be replaced by a higher order equation of 
           state for extremely high rate processes.
    \item  The relations between the Rockwell C hardness and the model
           parameters that have been presented provide reasonable estimates
           of the parameters.  However, more data for the $R_c$ 30, 45, and
           49 tempers are needed for better estimates for intermediate 
           tempers.  There is an anomaly in the strain rate and temperature
           dependence of the yield strength for $R_c$ 50 and higher tempers
           of 4340 steel.  We would suggest that the values for $R_c$ 49
           steel be used for harder tempers.  For tempers below $R_c$ 30,
           the fits discussed earlier provide reasonable estimates of the
           yield stress.
     \item The strain hardening (Voce) rule in the MTS model may be a 
           major weakness of the model and should be replaced with a more
           physically based approach.  The experimental data used to 
           determine the strain hardening rate parameters appear to deviate
           significantly from Voce behavior is some cases.
     \item The determination of the values of $g_{0es}$ and $\sigma_{0es}$
           involves a Fisher type modified Arrhenius plot.  We have 
           observed that the experimental data for the $R_c$ 45 and $R_c$
           49 tempers do not tend to reflect an Arrhenius relationship.
           More experimental data (and information on the variation of the
           experimental data) are needed to confirm this anomaly.
  \end{enumerate}

\section*{Acknowledgments}
  This work was supported by the the U.S. Department of Energy through the 
  ASCI Center for the Simulation of Accidental Fires and Explosions, under grant 
  W-7405-ENG-48.

\appendix
\section{Fisher plot data for $\sigma_i$ and $g_{0i}$}\label{app:FisherSigi}
  Tables~\ref{tab:FisherSigi30}, \ref{tab:FisherSigi38}, 
  \ref{tab:FisherSigi45}, and \ref{tab:FisherSigi49} show the Fisher plot
  data used to calculate $g_{0i}$ and $\sigma_i$ for the four tempers of
  4340 steel.
  \begin{table}[p]
    \centering
    \caption{Fisher plot data used to calculate $g_{0i}$ and $\sigma_i$
             for 4340 steel of hardness $R_c$ 30.}
    \begin{tabular}{cccccc}
       \hline
       \hline
       $[k_b T/\mu b^3 \ln (\Epdot{0i}/\Epdot{})]^{1/q_i}$ &
       $[(\sigma_y - \sigma_a)/\mu]^{p_i}$ &
       $T$ (K) & $\Epdot{}$ (/s) & $\sigma_y$ (MPa) & $\mu$ (GPa) \\
       \hline
       0.08333    & 0.044658  & 298 & 0.002 & 802.577 & 79.745 \\
       0.0782424  & 0.0432816 & 298 & 0.009 & 768.052 & 79.745 \\
       0.0700974  & 0.0425718 & 298 & 0.1   & 750.461 & 79.745 \\
       0.0619864  & 0.0469729 & 298 & 1.1   & 861.843 & 79.745 \\
       0.0408444  & 0.0447093 & 298 & 570   & 803.874 & 79.745 \\
       0.0747152  & 0.0435192 & 500 & 604   & 710.861 & 72.793 \\
       0.122804   & 0.044074  & 735 & 650   & 648.71  & 64.706 \\
       \hline
    \end{tabular}
    \label{tab:FisherSigi30}
  \end{table}
  \begin{table}[p]
    \centering
    \caption{Fisher plot data used to calculate $g_{0i}$ and $\sigma_i$
             for 4340 steel of hardness $R_c$ 38.}
    \begin{tabular}{cccccc}
       \hline
       \hline
       $[k_b T/\mu b^3 \ln (\Epdot{0i}/\Epdot{})]^{1/q_i}$ &
       $[(\sigma_y - \sigma_a)/\mu]^{p_i}$ &
       $T$ (K) & $\Epdot{}$ (/s) & $\sigma_y$ (MPa) & $\mu$ (GPa) \\
       \hline
       0.0775492 & 0.0563156 & 258  & 0.0002 & 1134.12 & 81.121 \\
       0.0911186 & 0.0542514 & 298  & 0.0002 & 1057.67 & 79.745 \\
       0.0412876 & 0.0565499 & 298  & 500    & 1122.38 & 79.745 \\
       0.0375715 & 0.0576617 & 298  & 1500   & 1154.16 & 79.745 \\
       0.117866  & 0.0551309 & 373  & 0.0002 & 1048.86 & 77.164 \\
       0.0900788 & 0.0508976 & 573  & 500    & 857.017 & 70.281 \\
       0.0819713 & 0.0531021 & 573  & 1500   & 910.012 & 70.281 \\
       0.134713  & 0.0420956 & 773  & 500    & 597.559 & 63.398 \\
       0.11695   & 0.0461642 & 773  & 2500   & 678.83  & 63.398 \\
       0.173098  & 0.0367619 & 973  & 1500   & 448.348 & 56.515 \\
       0.165137  & 0.037095  & 973  & 2500   & 453.773 & 56.515 \\
       0.237617  & 0.0263176 & 1173 & 1500   & 261.902 & 49.632 \\
       0.226689  & 0.0287519 & 1173 & 2500   & 291.972 & 49.632 \\
       0.322911  & 0.0199969 & 1373 & 1500   & 170.886 & 42.75  \\
       0.30806   & 0.0220299 & 1373 & 2500   & 189.782 & 42.75  \\
       \hline
       & \\
       & \\
       & \\
       & \\
    \end{tabular}
    \label{tab:FisherSigi38}
  \end{table}
  \begin{table}[p]
    \centering
    \caption{Fisher plot data used to calculate $g_{0i}$ and $\sigma_i$
             for 4340 steel of hardness $R_c$ 45.}
    \begin{tabular}{cccccc}
       \hline
       \hline
       $[k_b T/\mu b^3 \ln (\Epdot{0i}/\Epdot{})]^{1/q_i}$ &
       $[(\sigma_y - \sigma_a)/\mu]^{p_i}$ &
       $T$ (K) & $\Epdot{}$ (/s) & $\sigma_y$ (MPa) & $\mu$ (GPa) \\
       \hline
       0.0514817 & 0.0645752 & 173 & 0.0001 & 1429.17 & 84.046 \\
       0.0214507 & 0.0679395 & 173 & 1000   & 1538.34 & 84.046 \\
       0.0934632 & 0.0611362 & 298 & 0.0001 & 1255.45 & 79.745 \\
       0.038943  & 0.0683132 & 298 & 1000   & 1473.83 & 79.745 \\
       0.120899  & 0.062664  & 373 & 0.0001 & 1260.43 & 77.164 \\
       0.0503745 & 0.0653759 & 373 & 1000   & 1339.85 & 77.164 \\
       \hline
    \end{tabular}
    \label{tab:FisherSigi45}
  \end{table}
  \begin{table}[p]
    \centering
    \caption{Fisher plot data used to calculate $g_{0i}$ and $\sigma_i$
             for 4340 steel of hardness $R_c$ 49.}
    \begin{tabular}{cccccc}
       \hline
       \hline
       $[k_b T/\mu b^3 \ln (\Epdot{0i}/\Epdot{})]^{1/q_i}$ &
       $[(\sigma_y - \sigma_a)/\mu]^{p_i}$ &
       $T$ (K) & $\Epdot{}$ (/s) & $\sigma_y$ (MPa) & $\mu$ (GPa) \\
       \hline
       0.0514817 & 0.0682425 & 173 & 0.0001 & 1548.31 & 84.046 \\
       0.0214507 & 0.0711498 & 173 & 1000   & 1645.07 & 84.046 \\
       0.0934632 & 0.0674349 & 298 & 0.0001 & 1446.46 & 79.745 \\
       0.038943  & 0.0710307 & 298 & 1000   & 1559.63 & 79.745 \\
       0.120899  & 0.0653628 & 373 & 0.0001 & 1339.46 & 77.164 \\
       0.0503745 & 0.0694326 & 373 & 1000   & 1461.75 & 77.164 \\
       \hline
    \end{tabular}
    \label{tab:FisherSigi49}
  \end{table}

\section{Fisher plot data for $\sigma_{0es}$ and $g_{0es}$}
   \label{app:FisherSige}
   The data used to compute the parameters $\sigma_{0es}$ and $g_{0es}$
   are shown in Tables~\ref{tab:FisherSige30}, \ref{tab:FisherSige38},
   \ref{tab:FisherSige45}, and \ref{tab:FisherSige49}.
  \begin{table}[p]
    \centering
    \caption{Fisher plot data used to calculate $g_{0es}$ and $\sigma_{0es}$
             for 4340 steel of hardness $R_c$ 30.}
    \begin{tabular}{ccccccc}
       \hline
       $[k_b T/\mu b^3 \ln (\Epdot{0es}/\Epdot{})]$ &
       $\ln(\sigma_{es})$ & $T_0$ (K) & $T_s$ (K) & $\Epdot{}$ (/s) & 
       $\sigma_{es}$ (MPa) & $\mu$ (GPa) \\
       \hline
       \hline
       \multicolumn{7}{l}{$R_c$ = 30} \\
       \hline
       0.075541 & 20.986 & 298 & 298 & 0.002 & 1300 & 79.745 \\
       0.070454 & 19.807 & 298 & 298 & 0.009 & 400  & 79.745 \\
       0.062309 & 19.989 & 298 & 298 & 0.1   & 480  & 79.745 \\
       0.054198 & 19.968 & 298 & 298 & 1.1   & 470  & 79.745 \\
       0.038728 & 20.986 & 298 & 344 & 570   & 1300 & 78.571 \\
       0.064987 & 20.125 & 500 & 532 & 604   & 550  & 71.984 \\
       0.10315  & 19.807 & 735 & 758 & 650   & 400  & 64.126 \\
       \hline
    \end{tabular}
    \label{tab:FisherSige30}
  \end{table}
  \begin{table}[p]
    \centering
    \caption{Fisher plot data used to calculate $g_{0es}$ and $\sigma_{0es}$
             for 4340 steel of hardness $R_c$ 38.}
    \begin{tabular}{ccccccc}
       \hline
       $[k_b T/\mu b^3 \ln (\Epdot{0es}/\Epdot{})]$ &
       $\ln(\sigma_{es})$ & $T_0$ (K) & $T_s$ (K) & $\Epdot{}$ (/s) & 
       $\sigma_{es}$ (MPa) & $\mu$ (GPa) \\
       \hline
       \hline
       \multicolumn{7}{l}{$R_c$ = 38} \\
       \hline
       0.07092  & 21.129 & 258  & 258  & 0.0002 & 1500 & 81.121 \\
       0.08333  & 20.986 & 298  & 298  & 0.0002 & 1300 & 79.745 \\
       0.10779  & 21.254 & 373  & 373  & 0.0002 & 1700 & 77.164 \\
       0.036227 & 20.212 & 298  & 320  & 500    & 600  & 79.183 \\
       0.075873 & 20.05  & 573  & 591  & 500    & 510  & 69.826 \\
       0.11153  & 19.552 & 773  & 785  & 500    & 310  & 63.096 \\
       0.037964 & 20.618 & 298  & 371  & 1500   & 900  & 77.886 \\
       0.070669 & 20.367 & 573  & 614  & 1500   & 700  & 69.246 \\
       0.14023  & 20.03  & 973  & 988  & 1500   & 500  & 56.154 \\
       0.098146 & 20.125 & 773  & 815  & 2500   & 550  & 62.342 \\
       0.13342  & 20.088 & 973  & 995  & 2500   & 530  & 55.989 \\
       0.19164  & 19.519 & 1173 & 1185 & 1500   & 300  & 49.282 \\
       0.25895  & 18.891 & 1373 & 1381 & 1500   & 160  & 42.505 \\
       0.18261  & 19.163 & 1173 & 1193 & 2500   & 210  & 49.047 \\
       0.24641  & 19.376 & 1373 & 1388 & 2500   & 260  & 42.289 \\
       \hline
    \end{tabular}
    \label{tab:FisherSige38}
  \end{table}
  \begin{table}[p]
    \centering
    \caption{Fisher plot data used to calculate $g_{0es}$ and $\sigma_{0es}$
             for 4340 steel of hardness $R_c$ 45.}
    \begin{tabular}{ccccccc}
       \hline
       $[k_b T/\mu b^3 \ln (\Epdot{0es}/\Epdot{})]$ &
       $\ln(\sigma_{es})$ & $T_0$ (K) & $T_s$ (K) & $\Epdot{}$ (/s) & 
       $\sigma_{es}$ (MPa) & $\mu$ (GPa) \\
       \hline
       \hline
       \multicolumn{7}{l}{$R_c$ = 45} \\
       \hline
       0.047192 & 19.414 & 173 & 173 & 0.0001 & 270 & 84.046 \\
       0.085675 & 17.034 & 298 & 298 & 0.0001 & 25  & 79.745 \\
       0.11082  & 19.715 & 373 & 373 & 0.0001 & 365 & 77.164 \\
       0.021177 & 19.139 & 173 & 211 & 1000   & 205 & 83.067 \\
       0.034507 & 18.683 & 298 & 327 & 1000   & 130 & 79.004 \\
       0.043234 & 18.826 & 373 & 397 & 1000   & 150 & 76.553 \\
       \hline
    \end{tabular}
    \label{tab:FisherSige45}
  \end{table}
  \begin{table}[p]
    \centering
    \caption{Fisher plot data used to calculate $g_{0es}$ and $\sigma_{0es}$
             for 4340 steel of hardness $R_c$ 49.}
    \begin{tabular}{ccccccc}
       \hline
       $[k_b T/\mu b^3 \ln (\Epdot{0es}/\Epdot{})]$ &
       $\ln(\sigma_{es})$ & $T_0$ (K) & $T_s$ (K) & $\Epdot{}$ (/s) & 
       $\sigma_{es}$ (MPa) & $\mu$ (GPa) \\
       \hline
       \hline
       \multicolumn{7}{l}{$R_c$ = 49} \\
       \hline
       0.047192 & 19.254 & 173 & 173 & 0.0001 & 230 & 84.046 \\
       0.085675 & 19.376 & 298 & 298 & 0.0001 & 260 & 79.745 \\
       0.11082  & 19.756 & 373 & 373 & 0.0001 & 380 & 77.164 \\
       0.02075  & 18.951 & 173 & 207 & 1000   & 170 & 83.17 \\
       0.035325 & 19.45  & 298 & 334 & 1000   & 280 & 78.826 \\
       0.043234 & 19.45  & 373 & 397 & 1000   & 280 & 76.553 \\
       \hline
    \end{tabular}
    \label{tab:FisherSige49}
  \end{table}

 \section{Johnson-Cook model and parameters} \label{app:JC}
  The Johnson-Cook (JC) model (\citet{Johnson83}) is purely empirical and 
  has the form 
  \begin{align}
    \sigma_y(\Ep,\Epdot{},T) & = 
    \sigma_0\left[1 + \cfrac{B}{\sigma_0} (\Ep)^n\right]
            \left[1 + C \ln(\Epdot{}^{*})\right]
            \left[1 - (T^*)^m\right] \\
    \Epdot{}^{*} & = \cfrac{\Epdot{}}{\Epdot{0}}; \quad
    T^* = \cfrac{(T-T_r)}{(T_m-T_r)}
  \end{align}
  where $\sigma_0$ is the yield stress at zero plastic strain, and
  $(B, C, n, m)$ are material constants, $\Epdot{0}$ is a reference strain 
  rate, and $T_r$ is a reference temperature.

  The value of $\sigma_0$ for 4340 steel in the Johnson-Cook model varies 
  with the temper of the steel.  We have fit the yield stress versus $R_c$
  hardness curve for 4340 steel from the ASM handbook~\cite{ASM78} to 
  determine the value of $\sigma_0$ for various tempers.  The equation
  for the fit is
  \begin{equation}
    \sigma_0 = \exp(A_1 R_c + A_2)~~\text{(MPa)}
  \end{equation}
  where $A_1 =$ 0.0355 ln(MPa), $A_2 =$ 5.5312 ln(MPa), and
  $R_c$ is the Rockwell-C hardness of the steel.  The value of 
  $B/\sigma_0 = $ 0.6339 is assumed to be a constant for all tempers.
  The strain hardening exponent ($n$) is  0.26 and the strain rate 
  dependence parameter ($C$) is 0.014, for all tempers.  The reference 
  strain rate $\Epdot{0}$ is 1 /s.  For temperatures less than 298 K, thermal
  softening is assumed to be linear and the parameter $m$ takes a value of
  1.  Above 298 K and lower than 1040 K, $m$ is assumed to be 1.03, and
  beyond 1040 K, $m$ is taken as 0.5 (\citet{Lee97}).  The reference 
  temperature ($T_r$) is 298 K and the melt temperature ($T_m$) is kept 
  fixed at 1793 K.  These parameters provide a reasonable fit to the 
  experimental data presented earlier in the context of the MTS model.

\bibliographystyle{harvard}
\bibliography{mybiblio}

\end{document}